\documentclass[10pt,article]{article}
\usepackage[latin1]{inputenc}
\usepackage{amsmath}
\usepackage{amsfonts}
\usepackage{amssymb}
\usepackage{pstricks}
\newcommand{\m}[1]{\mathcal{#1}}
\newcommand{\w}{\wedge}
\newcommand{\ph}[1]{\phantom{#1}}
\author{\\\\
Hans F. Westman$^1$\footnote{\texttt{westman@iff.csic.es}}\; and Tom G. Zlosnik$^2$\footnote{\texttt{t.zlosnik@imperial.ac.uk}}\\
{\small \it $(1)$ Instituto de F\'isica Fundamental, CSIC, Serrano 113-B, 28006 Madrid, Spain}\\
{\small \it $(2)$ Imperial College Theoretical Physics, Huxley Building, London, SW7 2AZ}
}
\date{\today}

\title{Cartan gravity, matter fields, and the gauge principle}
\begin{document}
\maketitle
\abstract{Gravity is commonly thought of as one of the four force fields in nature. However, in standard formulations its mathematical structure is rather different from the Yang-Mills fields of particle physics that govern the electromagnetic, weak, and strong interactions. This paper explores this dissonance with particular focus on how gravity couples to matter from the perspective of the Cartan-geometric formulation of gravity. There the gravitational field is represented by a pair of variables: 1) a `contact vector' $V^A$ which is geometrically visualized as the contact point between the spacetime manifold and a model spacetime being `rolled' on top of it, and 2) a gauge connection $A_{\mu}^{\ph\mu AB}$, here taken to be valued in the Lie algebra of $SO(2,3)$ or $SO(1,4)$, which mathematically determines how much the model spacetime is rotated when rolled. By insisting on two principles, the {\em gauge principle} and {\em polynomial simplicity}, we shall show how one can reformulate  matter field actions in a way that is harmonious with Cartan's geometric construction. This yields a formulation of all matter fields in terms of first order partial differential equations. We show in detail how the standard second order formulation can be recovered. Furthermore, the energy-momentum and spin-density three-forms are naturally combined into a single object here denoted the spin-energy-momentum three-form. Finally, we highlight a peculiarity in the mathematical structure of our first-order formulation of Yang-Mills fields. This suggests a way to unify a $U(1)$ gauge field with gravity into a $SO(1,5)$-valued gauge field using a natural generalization of Cartan geometry in which the larger symmetry group is spontaneously broken down to $SO(1,3)\times U(1)$. The coupling of this unified theory to matter fields and possible extensions to non-Abelian gauge fields are left as open questions.}
\newpage

\section{Introduction}\label{intr}
In the traditional metric formulation of General Relativity the metric $g_{\mu\nu}$ is the fundamental variable with the affine connection $\Gamma^\rho_{\mu\nu}$ viewed as a secondary object constructed from the metric (i.e. the identification of $\Gamma^{\rho}_{\mu\nu}$ with the Christoffel symbols). The corresponding field equations are of second order in spatio-temporal derivatives. Nevertheless, there is also a first order Palatini formulation \cite{d'Inverno:1992rk} in which the metric and affine connection are regarded as independent variables. The associated field equations are then first order partial differential equations with the relation between metric and affine connection enforced by the equations of motion assuming vanishing spacetime torsion $T^\rho_{\mu\nu}=0$.

However, neither of these approaches are suitable for the inclusion of fermionic fields which require the introduction of local Lorentz invariance as a gauge group. Within this first order Palatini approach the gravitational field is mathematically represented in terms of a pair of one-forms: the co-tetrad $e^I=e^I_\mu dx^\mu$ and the spin connection $\omega^I_{\ J}=\omega^{\ I}_{\mu\ J}dx^\mu$ where $I,J=0,\dots,3$ are $SO(1,3)$ indices. Under a local Lorentz transformation represented by a matrix $\Lambda^{I}_{\phantom{I}J}(x^{\mu})$, the gravitational fields transform as follows:
\begin{eqnarray}
\label{t1}\omega &\rightarrow & \Lambda \omega \Lambda^{-1}-d\Lambda \Lambda^{-1} \\
\label{t2} e &\rightarrow & \Lambda e
\end{eqnarray}
where indices have been suppressed for notational compactness. We see that the spin-connection $\omega^{IJ}$ behaves exactly as a standard type Yang-Mills gauge field under local gauge transformations. Specifically, it transforms inhomogeneously under a local Lorentz transformation. The co-tetrad, however, transforms {\em homogeneously} under this gauge transformation, highlighting the peculiar role of this field: the co-tetrad $e^I_\mu$ (or equivalently the metric tensor $g_{\mu\nu}=\eta_{IJ}e^I_\mu e^J_\nu$) appears to be the only fundamental field in physics which possesses a spacetime (Greek) index and yet is not a gauge field \cite{Zee:2003mt}. Thus, gravity cannot be viewed as a standard gauge field within this specific mathematical formulation, making gravity stand out compared to the other Yang-Mills fields.

One may construct an action $S_{P}$ (the Palatini action) which is polynomial in $\omega^I_{\ J}$ and $e^{I}$:
\begin{eqnarray}
\label{sp} S_{P}[\omega^{I}_{\ J},e^{I}] &=& \kappa \int d^{4}x \; \epsilon_{IJKL}\varepsilon^{\alpha\beta\gamma\delta}\left(\frac{1}{2}e^{I}_{\alpha}e^{J}_{\beta}R_{\gamma\delta}^{\ph{\gamma\delta}KL}-\frac{\Lambda}{6}e^{I}_{\alpha}e^{J}_{\beta}e^{K}_{\gamma}e^{L}_{\delta}\right) \\
\label{riman}
R_{\mu\nu}^{\ph{\mu\nu}KL} &\equiv & 2\partial_{[\mu}\omega_{\nu]}^{\ph\nu KL} + 2\eta_{IJ}\omega_{[\mu}^{\ph{[\mu}KI}\omega_{\nu]}^{\ph{\nu]}JL}
\end{eqnarray}
where the two notationally distinct Levi-Civita symbols $\epsilon$ and $\varepsilon$ are defined so that $\epsilon_{\mu\nu\rho\sigma}\varepsilon^{\mu\nu\rho\sigma}=+4!$ and $e\epsilon_{\mu\nu\rho\sigma}=e^I_\mu e^J_\nu e^K_\rho e^L_\sigma \epsilon_{IJKL}$ and $\epsilon_{IJKL}=\eta_{IM}\eta_{JN}\eta_{KO}\eta_{LP}\varepsilon^{MNOP}=det(\eta)\varepsilon_{IJKL}=-\varepsilon_{IJKL}$. This action is equivalent to General Relativity with a cosmological constant $\Lambda$. From the resulting field equations, one can look to solve algebraically for $\omega_\mu^{\ph\mu IJ}$ and so, upon substitution of this solution back into the action, recover an action which is a functional only of $e^{I}$. This is the Einstein-Hilbert action (with $e^I$ as fundamental variable rather than $g_{\mu\nu}$) which must be modified by adding a boundary term in order to compensate for the presence of second order derivatives in the action \cite{Wald:1984rg}. \footnote{See \cite{Sotiriou:2006gp} for an interesting different perspective on 
the role of the boundary term.} Alternatively, this action may be written more traditionally as a functional of a field  $g_{\mu\nu}\equiv\eta_{IJ}e^{I}_{\mu}e^{J}_{\nu}$ in terms of the metric tensor.

We can now highlight a second peculiarity. A notable feature is that the Einstein-Hilbert action is non-polynomial in the basic variable $e^I$ (or $g^{\mu\nu}$ within the traditional formulation) and is only recoverable from the first order Palatini formalism, when $e^{I}$ is \emph{invertible}, i.e. it contains the inverse $e^{\mu}_{I}$ of the co-tetrad. This inverse field is commonly referred to as the tetrad, or vierbein, and is defined as follows:
\begin{eqnarray}
\label{tetrad}
e^{\mu}_{I} &\equiv& 4\frac{\varepsilon^{\mu\nu\rho\sigma}\epsilon_{IJKL}e^{J}_{\nu}e^{K}_{\rho}e^{L}_{\sigma}}{\varepsilon^{\alpha\beta\gamma\delta}\epsilon_{MNOP}e^{M}_{\alpha}e^{N}_{\beta}e^{O}_{\gamma}e^{P}_{\delta}}
\end{eqnarray}
%
Within the first order Palatini approach one may regard the requirement of invertibility of $e^{I}$ to be unnecessarily restrictive \cite{Horowitz:1990qb}. One may furthermore find the non-polynomial structure of the Einstein-Hilbert action to be undesirable \cite{Pagels:1983pq}, and so hold the Palatini formulation of gravity to be the more fundamental and mathematically elegant one. Thus, the peculiar non-polynomial character of pure gravity can easily be cured by insisting on a first order Palatini formulation. However, with the notable exception of actions for fermionic fields which are of first order and already polynomial in all variables (see  \ref{standDirac}), the problem of the appearance of inverses and the related non-polynomial structure immediately reappears when one looks to couple matter fields to gravity. Specifically, if one attempts to formulate the second order actions for bosonic matter-fields, using only those fields and the Palatini gravitational fields $e^{I}$ and $\omega^{I}_{
\ J}$, one cannot avoid use of the tetrad which again enforces a non-polynomial structure of the actions and associated field equations. 

As a simple example consider the action for a real, massless scalar field $\phi$:
\begin{eqnarray}
\label{scalact}
S_{\phi}[\phi,e^{I}] = - \int d^{4}x\frac{1}{4!}\eta^{MN}e^{\mu}_{M}e^{\nu}_{N}\partial_{\mu}\phi \partial_{\nu}\phi\; \epsilon_{IJKL}\varepsilon^{\alpha\beta\gamma\delta}e^{I}_{\alpha} e^{J}_{\beta} e^{K}_{\gamma}  e^{L}_{\delta}
\end{eqnarray}
The compactness of notation in (\ref{scalact}) somewhat hides the non-polynomial structure and the complicated role of $e^{I}$ in the action as dictated by the definition of the tetrad \eqref{tetrad}. In the case of a Yang-Mills gauge connection one-form, the situation is compounded, with the appearance of four tetrads in the action. However, as shown in \ref{standDirac}, in the case of fermionic matter the action is polynomial in all variables \cite{Westman:2012xk}. Thus, gravity stands out from other fields by exhibiting non-polynomial structure in the bosonic matter actions.

A third notable peculiar feature of gravity is that, polynomial actions or not, the coupling of matter fields to gravity is different from the way a matter field is coupled to a Yang-Mills gauge field. For example, if we look to couple a scalar field $\phi^a$ valued in some representation of a Lie-algebra to the associated Yang-Mills field $B^a_{\ph a b}=B^{\ph \mu a}_{\mu\ph a b}dx^{\mu}$ we would proceed according to the {\em gauge prescription} which says that we should simply replace the partial derivative in the action with the gauge covariant derivative $\partial_\mu\phi^a \rightarrow \m D_\mu\phi^a =\partial_\mu\phi^a+ig B^{\ph \mu a}_{\mu\ph a b}\phi^b$. However, it is clear from the action \eqref{scalact} that the coupling to the gravitational field is not done according to such a gauge prescription. In particular, there is no Yang-Mills index associated with gravity in standard formulations.

Summarizing: there are three peculiarities with `standard' approaches to gravity and matter fields that makes the gravitational field different and stand out from other fields in nature: \footnote{Weinberg has pointed out an additional peculiarity of gravity. In \cite[p. 7]{Weinberg:1996kr} Weinberg writes: ``in General Relativity the affine connection is itself constructed from first derivatives of the metric tensor, while in gauge theories the gauge fields are not expressed in terms of any more fundamental fields.'' We note that this peculiarity is only present in the second order formulations, while being absent in the first order Palatini formulation.}
\begin{enumerate}
	\item the first order Palatini and second order Einstein-Hilbert formulations of gravity involve the co-tetrad field $e^{I}_\mu$ that has no natural analogue within standard gauge theory \cite{Zee:2003mt} and is the only fundamental field in modern physics with a spacetime index $\mu$ and yet does not behave like a gauge field
	\item even though the non-polynomial structure of the Einstein-Hilbert action can be avoided by adopting the first order Palatini action, the standard actions for bosonic matter actions still exhibit non-polynomial structure in the gravitational field variables.
	\item In standard formulations the coupling of matter fields to gravity does not mirror the coupling of matter to Yang-Mills fields.
\end{enumerate}
This paper aims to demonstrate that these peculiar features of gravity are absent within a Cartan-geometric formulation of gravity thus removing these dissimilarities between gravity and the Yang-Mills force fields of particle physics. As will be detailed in Section \ref{spontgravity}, the first peculiarity disappears if gravity is instead regarded as a spontaneously broken gauge theory of a group larger than the Lorentz group $SO(1,3)$. This alternative approach to gravity has traditionally been called the de Sitter/anti-de Sitter gauge theory of gravity \cite{MacDowell:1977jt,Stelle:1979va,Randono:2010cq}. More recently, its structure in relation to Cartan's approach to differential geometry has been illuminated \cite{Wise:2006sm,Wise:2009fu,Westman:2012xk}.

Given that the first peculiarity can readily be avoided by adapting the Cartan-geometric approach to gravity, it is only natural then to ask whether the second and third peculiarities of the standard approach to gravity can be avoided from the Cartan-geometric perspective. More specifically, can gravity be coupled to matter fields in way that is in accordance with the gauge prescription discussed above and in such a way that the gravitational field variables appear {\em polynomially} and so without the recourse to inverses? We shall find that this is indeed not only possible but lead to interesting perspectives regarding the geometric role of the Higgs field \cite{Wise:2011ab}, the possible role gravity plays in physics, and potential new avenues for unifying gravity with the other Yang-Mills gauge fields.

The structure of this paper is as follows: Section \ref{spontgravity} introduces the Cartan-geometric formulation of gravity  as a gauge theory with a spontaneously broken symmetry with the elegant geometric interpretation recalled in  \ref{cartangeometry}. Section \ref{dissatisfaction} provides the reader with the essentials of the formulation of standard matter actions using the language of differential forms. The language of differential forms is an ideal starting point for adapting a description of matter fields to the Cartan-geometric formalism and used extensively throughout this paper. The `tensor-minded' reader is referred to  \ref{diffpre} for additional background material. In Section \ref{rolling} we introduce two principles, the {\em gauge principle} and {\em polynomial simplicity}, which all subsequent Cartan-geometric matter actions will be required to satisfy. In Section \ref{matter} we proceed to show how to construct polynomial matter actions coupled to gravity consistent 
with the gauge principle and how the familiar second order field equations are recovered on-shell by imposing the equations of motion. We also briefly comment here on previous alternative ideas in the literature. In Section \ref{stressen} we introduce the spin-energy-momentum three-form which dictates the back-reaction of the matter fields on the gravitational field. We show how the canonical energy-momentum tensor is recovered on-shell and using these expressions we restrict the values of the parameters appearing in the Cartan-geometric matter actions. In Section \ref{unify} we note that enforcing the gauge principle and polynomial simplicity for Yang-Mills fields enforces a peculiar mathematical form of the gauge connections and an associated skewed gauge transformation property. This skewed mathematical structure hints that Yang-Mills fields and gravity should be collected into a single gauge field. We then proceed to detail how we can unify an $U(1)$ field with gravity using a natural generalization of 
Cartan geometry. This unification is is based on the gauge group $SO(1,5)$ which is spontaneously broken down to $SO(1,3)\times U(1)$. Section \ref{discus} ends with conclusions, discussion and outlook for future work.
\section{The first and second peculiarities and Cartan gravity}
We shall now review how one may address, within the gravitational sector, the first and second peculiarities mentioned in the introduction. This requires a different way of thinking about differential geometry developed by \'Elie Cartan \cite{SharpeCartan}. In Appendix \ref{cartangeometry} we provide a brief discussion of the geometric interpretation Cartan geometry in terms of contact vectors and the rolling models spaces. We note that the remainder of this paper will rely heavily on the calculus of forms. We shall follow closely the presentation \cite{Westman:2012xk} which explains in a pedagogical manner the geometrical machinery of Cartan geometry in terms of `idealized waywisers' with several appendices provided for `tensor-minded' readers detailing the mathematical techniques involved. Additionally, two excellent expositions of Cartan geometry can be found in \cite{Randono:2010cq,Wise:2006sm}.
\subsection{Gravitation and spontaneous symmetry breaking}\label{spontgravity}
For the purposes of this paper it will be instructive to introduce the Cartan-geometric formulation in the language of spontaneously broken gauge theories \cite{Stelle:1979va}. In fact, Cartan's approach to geometry is based on the idea of spontaneous symmetry breaking \cite{Wise:2011ab}, and appeared long before the idea of symmetry breaking made its way into condensed matter and particle physics.

Our starting point is the first order Palatini approach to gravity as realized in the action principle (\ref{sp}). Recall the discussion of Section \ref{intr}, wherein it was mentioned that the description of the gravitational field in the Palatini formalism involved a connection for the Lorentz group $\omega^{IJ}=\omega_{\mu}^{\ph\mu IJ}dx^\mu$
along with a field $e^I=e^{I}_{\mu}dx^\mu$ where $I$ and $J$ are $SO(1,3)$ indices. Together these amount to a set of ten one-form fields. This is precisely the number of independent one-form fields required to construct a gauge connection for the orthogonal groups $SO(p,q)$ with $p+q=5$. \footnote{The Poincar\'e group is of course also ten-dimensional with its translations, rotations, and boosts and can also be used to define Cartan geometry, also in this context called Poincar\'e gauge theory \cite{Kibble:1961ba,Gronwald:1995em,Ali:2009ee}.} We will refer to this connection as $A^{AB}=A_{\mu}^{\ph\mu AB} dx^{\mu}$, where capitalized Latin indices of the first part of the alphabet $A,B,C,\dots,H$ are taken to go from $0$ to $4$. On the other hand capitalized Latin letters from the middle part of the alphabet $I,J,\dots,P$ run from $0$ to $3$.

Given the coincidence between the number of one-form fields it is only natural to ask whether it is possible to understand the co-tetrad $e^I$ and the spin-connection $\omega^{IJ}$ as different parts of a $SO(p,q)$ connection. However, this idea immediately seems too simplistic since modern physical theories display local Lorentz $SO(1,3)$ symmetry but not $SO(p,q)$ invariance with $p+q=5$. Nevertheless, we may recall that the electroweak theory of particle physics \cite{Mandl:1985bg} also starts with a larger gauge group $SU(2)\times U(1)$ only to be spontaneously broken by a two-component Higgs field $\Phi$.\footnote{We note that it is more accurate to write $SU(2)_L\times U(1)$ indicating the parity violating left-handed character of the electroweak interactions in the fermionic sector. We have nevertheless dropped the subscript $L$ here since we are for the moment concerned with the Higgs field for which does not have chiral representations.} The $U(1)$ invariance  of electrodynamics, and its associated 
gauge connection $A_\mu$, correspond to the remnant subgroup of $SU(2)\times U(1)$-transformations that leaves the Higgs field $\Phi$ invariant. In this way the Higgs field is made electrically neutral by construction.

One can now try to implement the same symmetry breaking mechanism in gravity. We should then introduce a Higgs-type field so that the subgroup of $SO(p,q)$-transformations leaving this field invariant is precisely the Lorentz group $SO(1,3)$, the unbroken remnant symmetry of the first order Palatini formulation. Since only $SO(2,3)$ and $SO(1,4)$ contain $SO(1,3)$ as a subgroup our choice of orthogonal group is narrowed down to one of these two. Further, mimicking the electroweak symmetry breaking mechanism, we demand that the gravitational Higgs-type field should carry a $SO(2,3)$ or $SO(1,4)$ index, i.e. it should have the form $V^A(x)$ with $A$ an $SO(2,3)$ or $SO(1,4)$ index. 

In order for the Lorentz group to emerge as the subgroup of transformations leaving $V^A$ invariant, the vector $V^A$ must be time-like $V^AV^B\eta_{AB}\equiv V^{2}<0$ for $SO(2,3)$ and space-like $V^AV^B\eta_{AB}>0$ for $SO(1,4)$ \cite{Westman:2012xk}. Here $\eta_{AB}=diag(-1,1,1,1,\mp 1)$ is the $5$-dimensional Minkowski metric with the last component $\eta_{44}=-1$ for $SO(2,3)$ and $\eta_{44}=+1$ for $SO(1,4)$. Although a field $V^{A}$ with a single index may seem like an unfamiliar object, we will see in Section \ref{matter} that $V^A$ is nothing but a scalar field coupled to gravity as it would appear within a Cartan-geometric formulation. Secondly, we note that the field $V^A$ is a symmetry breaking field analogous to the Higgs field $\Phi$, although the gauge symmetries they break are different.

Thus, we may look to see if a realistic description of gravity can consist of the pair of field variables $\{V^A,A^{AB}\}$. To do so it is appropriate to ask how one may relate these variables to the Palatini variables $e^I$ and $\omega^{IJ}$. Given that the connections of orthogonal groups are anti-symmetric, one natural guess would be something like
\begin{eqnarray}
A^{AB}=\left(\begin{array}{cc}\omega^{IJ}&e^I\\-e^J&0 \end{array}\right)
\end{eqnarray}
where the decomposition is with respect to some internal direction defining the split between $I,J$, and $4$ index. This however cannot be the whole story. As stressed in the introduction, $e^I$ is neither a gauge field, nor part of any gauge field, since it does not transform inhomogeneously. 

We then proceed differently by noting that the simplest one-form object we can construct from the pair $\{V^A,A^{AB}\}$, that transforms homogeneously under $SO(2,3)/SO(1,4)$, is $DV^A$. That one-form has in general $5$ non-zero components. However, if we impose that the norm of $V^A$ is constant $V^AV^B\eta_{AB}=\mp\ell^2$ for some constant $\ell$, then the component along $V^A$ is zero, i.e. $V_ADV^A=0$, and the object $DV^A$ has only four independent non-zero components.\footnote{We note that restricting the norm to be constant may seem rather artificial a particle physics perspective. Indeed, no restriction on the norm of the Higgs field $|\Phi|$ is made at the fundamental level. This provides perhaps a hint that Cartan gravity should be generalized to include a dynamical symmetry breaking contact vector $V^A$. We shall return to this issue in Section \ref{discus}.} Adopting a special gauge in which $V^A\overset{*}{=}\ell\delta^A_4$ we can now make the identification
\begin{eqnarray}
DV^A=dV^A+A^A_{\ph AB}V^{B}\overset{*}{=}\mp\ell A^{A4}=(e^I,0)
\end{eqnarray}
If we now restrict the transformations to those that leave $V^A$ invariant we see that $e^I$ transforms according to \eqref{t2}. In this special gauge we can further make the identification $\omega^{IJ}\overset{*}{=}A^{IJ}$ and again verify that $A^{IJ}$ transforms inhomogeneously accoding to \eqref{t1}.

We can now fully understand the first peculiarity mentioned in the introduction: $e^I$ is indeed a one-form, but it is not a connection, nor part of any connection. In fact, the first peculiarity disappears once we accept that the co-tetrad $e^I$ is not a fundamental variable from a Cartan-geometric perspective. Rather the co-tetrad $e^I\overset{*}{=}DV^I$ is a compound object constructed from the more fundamental fields $V^A$ and $A^{AB}$. The situation is then rather trivial since one can always cook up compound objects with spacetime indices which are not gauge connections, e.g. $\m D\Phi$ where $\Phi$ is the Higgs field and $\m D$ the $SU(2)\times U(1)$ gauge covariant exterior derivative. In fact, within the Cartan-geometric formulation of gravity the only fundamental one-form around is $A^{AB}$ which indeed transforms inhomogeneously as a standard gauge connection.  

Thus, the first peculiarity disappears and pure gravity can be seen as a standard gauge theory with a spontaneously broken symmetry. We will call theories where the basic gravitational variables are the pair $\{V^A,A^{AB}\}$ {\em Cartan gravity}. The reason for this is that this pair is also the descriptor of geometry in the theory of Cartan geometry (see Appendix \ref{cartangeometry} or \cite{Westman:2012xk}).
\subsection{Polynomial actions for Cartan gravity}\label{polyaction}
Let us now ask: Can realistic polynomial gravitational actions be constructed from the variables $\{V^{A},A^{AB}\}$? In order to get a feel for the crucial role the field $V^A$ plays in Cartan gravity, it is instructive to consider actions constructed solely out of the the connection $A^{AB}$ and the structure associated with the gauge group $SO(2,3)$/$SO(1,4)$. In constructing gauge invariant polynomial actions one is then limited to three objects:
\begin{itemize}
\item the field strength of $A^{AB}$, defined as $F^{AB} \equiv DA^{AB}\equiv dA^{AB}+ A^{A}_{\phantom{A} C} \wedge A^{CB}$
\item the numerically invariant object $\eta_{AB}=diag(-1,1,1,1,\mp 1)$ and its inverse $\eta^{AB}$ where the final entry is equal to $-1$ for $SO(2,3)$ and $+1$ for $SO(1,4)$.
\item the five dimensional Levi-Civita symbol $\epsilon_{ABCDE}$ which is numerically invariant under local gauge transformations
\end{itemize}
Demanding the action to be gauge invariant and polynomial in the connection $A^{AB}$ and its gauge-covariant exterior derivatives $DA^{AB}$ is extremely restrictive. In fact, these conditions identify the following action as the only possible one:
\begin{eqnarray}
\label{tp1}
S = \int F^{AB}\wedge F_{AB}.
\end{eqnarray}
However, it may be shown that $\int F^{AB}\wedge F_{AB}= \int d(A^{AB}\wedge F_{AB}+\frac{1}{3}A^{AB}\wedge A_{A}^{\ph{A}D}\wedge A_{BD}) $ and so does not contribute to the equations of motion.\footnote{For the `tensor-minded' reader we note that the dual scalar density of an exterior derivative of a three-form $\Omega$ is $d\Omega\sim\frac{1}{3!}\partial_\mu(\varepsilon^{\mu\nu\rho\sigma}\Omega_{\nu\rho\sigma})$ and thus a divergence term. See Appendix \ref{duality} for more details.} We refer the reader to the appendices of \cite{Westman:2012xk} for an overview of the different mathematical techniques used throughout this paper.

In contrast, the space of polynomial gauge invariant actions for a spontaneously broken gauge theory is larger and much more interesting.  Consider then a hypothetical gravitational theory and a corresponding action constructed from the pair $\{V^A, A^{AB}\}$. Insisting on gauge invariance and that the action be polynomial in the basic variables and their derivatives we end up with a rather narrow but non-trivial class of possible gravitational actions:\footnote{A recent interesting suggestion for a non-polynomial action for gravity see \cite{Krasnov:2011pp}.}
\begin{eqnarray}\label{genaction}
S_{g}=\int a_{ABCD} F^{AB}\wedge F^{CD} + b_{ABCD} e^A\wedge e^B\wedge F^{CD}+c_{ABCD} e^A\wedge e^B\wedge e^C\wedge e^D
\label{actione}
\end{eqnarray}
where $e^A\equiv DV^A$ and
\begin{eqnarray}
a_{ABCD} &=& a_{1}\epsilon_{ABCDE}V^{E}+a_{2} V_{A}V_{C}\eta_{BD} +a_{3} \eta_{AC}\eta_{BD} \\
b_{ABCD} &=& b_{1}  \epsilon_{ABCDE}V^{E}+b_{2} V_{A}V_{C}\eta_{BD}+b_{3}\eta_{AC}\eta_{BD}\\
c_{ABCD} &=&  c_{1}\epsilon_{ABCDE}V^{E}
\end{eqnarray}
In general the quantities $a_{i},b_{i},c_{i}$ may depend on the scalar $V^{2}=V_{E}V^{E}$. We shall however restrict ourself to the case where they are just constants. Given this assumption, we note that the $a_{3}$ term is simply the action (\ref{tp1}) and hence topological. Furthermore, the $a_{2}$ and $b_{3}$ terms are topologically equivalent (i.e. identical up to a boundary term) \cite{Westman:2012xk}. Therefore, in this case only five of the $a_{i},b_{i},c_{i}$ independently contribute to the equations of motion, namely $a_1,a_2,b_1,b_2$, and  $c_1$. 

In order to recognize the familiar Palatini action within this polynomial class of actions we note that in the special gauge $V^A\overset{*}{=}\ell\delta^A_4 $ we have:
\begin{eqnarray}
F^{IJ}\equiv dA^{IJ}+\eta_{CD}A^{IC}\w A^{DJ} &\overset{*}{=}& d\omega^{IJ}+\omega^{I}_{\ph{I}K}\wedge \omega^{KJ} \pm \frac{1}{\ell^{2}}e^{I}\wedge e^{J}= R^{IJ} \pm \frac{1}{\ell^{2}}e^{I}\wedge e^{J} \\
F^{I4}\equiv dA^{I4}+\eta_{CD}A^{IC}\w A^{D4} &\overset{*}{=}& \mp \frac{1}{\ell}D^{(\omega)}e^{I} =\mp\frac{1}{\ell}T^{I}
\end{eqnarray}
where $D^{(\omega)}$ is the $SO(1,3)$ covariant exterior derivative operator, $R^{IJ}$ is the Riemannian curvature (\ref{riman}), and $T^{I}$ is the torsion two-form. Upon substituting the above expressions into (\ref{genaction}) we can immediately see that the Palatini action is directly present in the $b_{1}$ term and also in $a_{1}$ up to an additional boundary term \cite{Randono:2010cq}. Furthermore, the topologically equivalent $a_{2}$ and $b_{3}$ terms are seen to be equal to the so-called Holst term which does not alter the equations of motion if the spin density vanishes \cite{Freidel:2005sn,Westman:2012xk}. The $c_1$ term represents a contribution to the cosmological constant. The $b_2$ is interesting since it seems to allow for non-trivial dynamics of $V^2$.
\subsection{Dynamical versus non-dynamical approaches}\label{nondynapproach}
It is now important to ask whether the Higgs-type field $V^A$ should be regarded as a dynamical field, with the associated field equations obtained by varying the action with respect to it, or whether it should be regarded as a non-dynamical `absolute' object in very much the same way one regards the Minkowski metric in the special-relativistic Klein-Gordon action (see \cite{Westman:2012xk}). As noted in the previous subsection, in order to reproduce the Palatini formulation it is necessary that $V^2$ is constant. Within the non-dynamical approach this could simply be imposed {\em \`a priori}. However, within a dynamical approach this must somehow follow from the equations of motion or at least in some physical limit. The constancy of $V^2$ could simply be imposed by adding the Lagrange multiplier $\lambda(V^2\pm\ell^2)$ \cite{Stelle:1979va} but this is not always regarded as a satisfactory solution \cite{Randono:2010cq} and we shall regard it here as a variant of the non-dynamical approach. 

On the other hand, treating $V^A$ as a genuine dynamical object yields five additional equations and the possibility of an over-constrained solution space, perhaps ruling out physically important solutions or even creating inconsistencies. As noted in \cite{Westman:2012xk} the action corresponding to the sum of the $b_1$ and $c_1$ terms allow one to treat $V^A$ as a dynamical field and $V^2=const.$ is a consequence of the equations of motion without the need to resort to Lagrange multipliers. However, the inclusion of matter fields within this dynamical approach is not yet developed. In particular, it seems doubtful that the constancy of $V^2$ can be maintained in the presence of matter fields and this may require some modifications of Cartan gravity as defined in this paper. It should further be noted that this particular action yields an algebraic equation imposing the constancy of $V^2$. Specifically, derivatives of $V^{2}$ do not appear in the equations of motion in this formulation. Indeed, it may be 
shown that only the $b_{2}$ contribution to (\ref{genaction}) can result in terms involving the derivative of the magnitude of $V^{A}$. Moreover, the $b_2$ term yields non-trivial field equations only if the torsion two-form is non-vanishing.

Although one may suspect that the dynamical approach is ultimately the correct one, we shall for the sake of simplicity adopt the non-dynamical approach in this paper when constructing the Cartan-geometric actions for the matter fields. In the final section of the paper when nevertheless provide some remarks and comments regarding the dynamical approach .

\section{Coupling matter to gravity and polynomial simplicity}
In this section we will provide reasons for why the standard formulation of matter fields is not harmonious with the mathematical structure of Cartan gravity. We shall put fourth two principles, the {\em gauge principle} and {\em polynomial simplicity}, that the subsequent reformulated matter actions will be required to satisfy. The first principle will require that matter fields couple to the gravitational field in the same way they couple to other gauge fields. The second principle forbids the use of inverses and other non-polynomial structure. 

Before developing the Cartan-geometric reformulation of matter actions it will be helpful to review the standard formulation of the actions of the bosonic and fermionic fields using the language of differential forms. We refer the reader to Appendix \ref{diffpre} for a more thorough discussion aimed at `tensor-minded' physicists.
\subsection{Reasons to be dissatisfied from a Cartan gravity perspective}\label{dissatisfaction}
In the standard model of particle physics we find both fermionic and bosonic fields. The known bosonic fields are various Yang-Mills gauge fields $B=B_\mu dx^\mu$ valued in some Lie-algebra and the Higgs field $\Phi$. The known fermionic fields are various by Dirac spinors which are also representations of the relevant Lie-Algebra, i.e. they carry a Yang-Mills index. For the sake of notational compactness we shall suppress both spinor and gauge indices on all the fields.

The action of a symmetry breaking Higgs field $\Phi$ and a generic Yang-Mills field $B$ can be written very succinctly as
\begin{eqnarray}\label{standYMnKG}
S_{KG}=-\int \m D\Phi^\dagger\w *\m D\Phi+\epsilon_{IJKL}e^I\w e^J\w e^K\w e^L U(\Phi)\qquad S_{YM}=-\int Tr\frac{1}{2}G\w *G
\end{eqnarray}
where $U(\Phi)$ is some potential, e.g. $U(\Phi)=-m^2|\Phi|^2+\lambda |\Phi|^4$. In both cases the non-polynomial structure in the gravitational variables is not immediately visible but somewhat hidden in the Hodge dual (see Appendix \ref{Hodge}) defined in respective case by
\begin{eqnarray}
*\m D\Phi&\equiv&\frac{e}{3!}\epsilon_{\mu\nu\rho\sigma} g^{\mu\kappa}\m D_\kappa\Phi dx^\nu\w dx^\rho\w dx^\sigma=\frac{1}{3!}\epsilon_{IJKL}e^{I\mu}D_\mu\Phi e^J\w e^K\w e^L\\
*G&\equiv&\frac{e}{2!2!}\epsilon_{\mu\nu\rho\sigma}g^{\mu\kappa}g^{\nu\tau}G_{\kappa\tau}dx^\rho\w dx^\sigma=\frac{1}{4}\epsilon_{IJKL}e^{I\mu}e^{J\nu}G_{\mu\nu} e^K\w e^L.
\end{eqnarray}
This non-polynomial structure, evidenced by the presence of inverses $e^{I\mu}$ or $g^{\mu\nu}$, is characteristic for gravitational variables and not present for matter field variables which always appear polynomially. Furthermore, even though these two actions are, no doubt, mathematically elegant when written in terms of the Hodge dual, they look rather alien from a Cartan-geometric perspective. As discussed in  \ref{cartangeometry}, the action of gravity within Cartan's approach is fundamentally about `rolling'. Specifically, the gauge connection $A^{AB}$ corresponds to the action of infinitesimally `rolling' some object on the manifold. With this in mind one cannot help asking: what happened to this elegant geometric picture when coupling gravity to matter fields? If gravity is fundamentally about the action of `rolling', should it not be possible for matter fields to somehow be `rolled'?

We can sharpen these heuristic remarks if we look at the standard model fields and their interactions with Yang-Mills gauge fields. In particular, we see that coupling of a matter field to a gauge field follows the {\em  gauge prescription}: the coupling of a matter field carrying one or many Yang-Mills indices to a Yang-Mills field is done by replacing the exterior derivative with a gauge covariant exterior derivative. For example, in the case of coupling a $SU(2)$-valued scalar field $\Phi$ to an $SU(2)$-valued Yang-Mills field $B^a_{\ph ab}$ we simply replace the exterior derivative $d\Phi^a$ with the gauge covariant one $d\Phi^a\rightarrow \m D\Phi^a\equiv d\Phi^a+iB^a_{\ph ab}\Phi^b$. Here the Yang-Mills indices were written out explicit but as mentioned above we have suppressed them elsewhere for notational compactness.

In view of this, it is clear that the coupling of matter fields to gravity does not follow the gauge prescription. For example, in the case of coupling a scalar field to gravity we note that the $SO(2,3)/SO(1,4)$ Yang-Mills index is not even present in standard formulations. Instead we find that the Klein-Gordon field in a curved spacetime is described by the action \eqref{standYMnKG} which contains no gauge covariant derivative $D$ and only a scalar field $\phi$ without a `rolling' Yang-Mills index. We must then ask: why does not the coupling of matter fields to gravity follow the gauge prescription of Yang-Mills interactions? We are left with the idea that gravity is somehow the `odd man out' and it is not really a gauge field like the Yang-Mills fields.

Let us proceed to the fermionic fields and investigate if the situation is any different there. The action of a Dirac field coupled to gravity is given by
\begin{eqnarray}
S_D=\int \frac{1}{3!}\epsilon_{IJKL}e^J\w e^K\w e^L\w \frac{i}{2}(\bar\psi\gamma^I D^{(\omega)}\psi- D^{(\omega)}\bar\psi \gamma^I\psi)-\frac{1}{4!}\epsilon_{IJKL}e^I\w e^J\w e^K\w e^L m\bar\psi\psi
\end{eqnarray}
where the gauge covariant exterior derivative $D^{(\omega)}$ is defined by $D^{(\omega)}\psi=d\psi-\frac{i}{2}\omega^{IJ}S_{IJ}\psi$. We note that, in contradistinction to the bosonic fields, the action manifestly polynomial in the gravitational variables with no need for a Hodge dual, and we find that the exterior derivative $d\psi$ is replaced by a gauge covariant one $D^{(\omega)}\psi$. However, the action is still rather alien from a Cartan perspective. The Dirac field here is a (reducible) representation of $SL(2,\mathbb{C})$ and the action of `rolling' (or equivalently the action of the (anti-)de Sitter group) of $\psi$ is not defined. Specifically, we find the ordinary $SO(1,3)$ gauge connection $\omega^{IJ}$ in the gauge covariant exterior derivative $D^{(\omega)}\psi$ and not the rolling connection $A^{AB}$ of Cartan gravity. The elegant geometric picture of Cartan gravity is lost and we find again a reason from a Cartan gravity perspective to be dissatisfied with the way gravity is coupled to the 
fermionic fields.

The Yang-Mills fields of particle physics play distinguished roles as they are gauge connections themselves. Therefore, it is perhaps not clear that the gauge prescription should be applied in this case when coupling a Yang-Mills field to gravity. Nevertheless, we shall take the non-polynomial structure in the gravitational variables as enough a reason for dissatisfaction. 
\subsection{The gauge principle and polynomial simplicity}\label{rolling}
The main aim of this paper is to show how matter fields can be coupled to gravity in a straightforward polynomial fashion that is consistent with the geometric picture of Cartan geometry as well as the gauge prescription. In doing this we shall remove differences, and so lessen he dissonance, between the mathematical description of gravity and the Yang-Mills gauge fields of particle physics. Thus, in this paper we are going to include matter fields so that the two following principles are satisfied:
\begin{itemize}
\item[$\bullet$] {\bf Polynomial Simplicity:} All fundamental field variables (gravitational and non-gravitational) are represented exclusively by differential forms and all actions are integrals over four-forms constructed from these fields and their exterior derivatives, using only the wedge product and therefore rendering them manifestly polynomial in all variables.
\item[$\bullet$] {\bf Gauge Principle:} The coupling of a matter field to gravity is done exclusively through the gauge prescription. This requires us to attach a $SO(2,3)$/$SO(1,4)$ gauge index to the matter field, and couple this matter field to gravity by replacing the exterior derivative $d$ with the gauge covariant exterior derivative $D$, and write down a suitable gauge invariant action in terms of these objects. 
\end{itemize}
When it comes to building actions these two principles are very restrictive. When we multiply the various forms together we are restricted to only use the wedge-product and taking the derivative of a field must be done using the exterior derivative or a suitable gauge covariant version of it. The gauge principle enforces the idea that gravity should couple to matter fields in the same way matter fields are coupled to Yang-Mills fields.

One consequence of requiring gauge invariance and polynomial simplicity is that all field equations must be first order partial differential equations. This follows immediately since any number of gauge-covariant exterior derivatives $\m D$ applied to some field will never result in more derivatives than one in any of the field variables. As an example take a scalar field $\phi$ coupled to some Yang-Mills gauge field $B$ with $G\equiv dB+iB\w B$ denoting the associated curvature two-form with gauge indices suppressed. Then we have that $\m D\m D \phi= G \phi$ does not contain second order derivatives in either $\phi$ or $B$. It is no use to take another derivative since $\m D\m D\m D \phi= \m D(G \phi)=\m DG\phi+(-1)^{2}G\w\m D\phi=G\w \m D\phi$ due to the Bianchi identity $\m DG\equiv0$.

Thus, gauge invariance and polynomial simplicity forces all differential equations to be first order partial differential equations. We can now understand why both the Dirac equation and the MacDowell-Mansouri gravitational field equations (the field equations derived from the action (\ref{actione}) with only $a_{1}$ non-vanishing) are first order partial differential equations: They are both derived from gauge invariant actions satisfying the principle of polynomial simplicity. On the other hand, the standard Yang-Mills and scalar field actions are gauge invariant but are non-polynomial in the gravitational variables and therefore do not satisfy polynomial simplicity. Thus we see that it is the non-polynomial structure that is the key reason for why the resulting equations are second order partial differential equations.

As we shall see below, restoring the principle of polynomial simplicity in the bosonic sector can be done easily once we insist on the {\em gauge principle} which says that matter fields should be coupled to gravity through the gauge prescription. Since this requires us to attach a `rolling' index to the matter fields the number of dynamical fields is increased. However, this is expected if we are going from a second order formulation to a first order one. This will also remove a dissonance within the matter sector: the fermionic fields are subject to first order differential equations while the bosonic ones to second order differential equations.

We should also pointed out that it is by no means obvious that coupling Yang-Mills fields to gravity should follow the gauge standard prescription, the reason being that the Yang-Mills fields are gauge fields themselves. However, we shall press on enforcing the gauge principle even for Yang-Mills fields as it will lead us to a novel possibility of unifying Yang-Mills fields with gravity using a natural generalization of Cartan geometry. In Section \ref{unify} we detail such an attempt of unifying a $U(1)$ gauge field with gravity. Whether this $U(1)$ gauge field can play the role of either the Maxwell field or the $U(1)$ gauge field of the electroweak interactions require more analysis which we postpone to a possible future paper.
%
\section{Cartan-geometric matter actions}\label{matter}
Let us now turn the implementation of the gauge principle and polynomial simplicity for the typical fields of the standard model. We will work within the non-dynamical approach discussed in Section \ref{nondynapproach} (see also \cite{Westman:2012xk}) in which the contact vector $V^A$, required to satisfy $V^AV^B \eta_{AB}=const.$, is not subject to non-trivial equations of motion. As such the contact vector possesses no gauge invariant degrees of freedom.

In this section we shall systematically reformulate the various matter fields so that the gauge principle and polynomial simplicity are satisfied. As we shall see the fermionic fields require only a minimal change: instead of regarding the spinors as representation of the spin-half Lorentz group $spin(1,3)$ we regard them instead as representations of the spin-half versions of the de Sitter or anti-de Sitter rolling groups, i.e. $spin(1,4)$ or $spin(2,3)$. It is then clear how to `roll' a fermionic field and thus to minimally couple it to gravity through a gauge covariant derivative. From here it is then a simple task to find the appropriate actions that reduce to the standard Dirac spinor action. That the changes would be minimal could be anticipated since the Dirac equation is gauge covariant and already on polynomial form and thus a first order theory. 

On the other hand, the Higgs and Yang-Mills fields are subject to second order field equations and, as noted in Section \ref{rolling}, this is not consistent with the gauge principle and polynomial simplicity. Instead of replacing an already existing index we need to attach a new rolling index. In the case of the Higgs scalar field this is straightforward. However, Yang-Mills fields require a bit more care in regards to gauge transformations.

This section relies heavily on the variational calculus using differential forms and we refer the reader to Appendix \ref{diffpre} and the appendices of \cite{Westman:2007yx} for an exposition of the necessary mathematical techniques. Furthermore, the reader is referred to Appendix \ref{units} for a discussion on our conventions regarding the choice of units and dimensions of the various object and constants appearing in this paper. 

\subsection{Convenient notation}
Many of the terms of the Cartan-geometric matter actions below exhibit a recurring mathematical structure involving the five dimensional epsilon $\epsilon_{ABCDE}$, the contact vector $V^A$, and the covariant derivative $DV^{A}$ which we write as $e^A$ for notational compactness. Thus we introduce

\begin{eqnarray}
\star\Sigma&=&\frac{\mu}{4!} \epsilon_{ABCDE}V^E e^A\w e^B\w e^C\w e^D\\
\star\Sigma_D&=&\frac{\mu}{3!} \epsilon_{ABCDE}V^E e^A\w e^B\w e^C\\
\star\Sigma_{CD}&=&\frac{\mu}{2!} \epsilon_{ABCDE}V^E e^A\w e^B\\
\star\Sigma_{BCD}&=&\mu \epsilon_{ABCDE}V^E e^A
\end{eqnarray}
where the constant $\mu$ has dimensions of inverse length. The reason we introduce such a constant is to be able to write down dimensionless actions. We could naively guess that $\mu=\ell^{-1}$ and suggesting that no new constant in addition to $\ell$ is necessary. The appearance of $\ell^{-1}$ is acceptable if $\ell$ indeed is just a constant. However, if we believe that $V^A$ is ultimately a dynamical variable with $V^2(x)=\mp\ell^2(x)$ then $\ell^{-1}$ would have to be replaced by $(\mp V^2)^{-\frac{1}{2}}$ which violates polynomial simplicity. In addition, we will find in Section \ref{unify} where we show how to unify gravity with a $U(1)$ gauge field that $\mu$ is related to a different object. For these reasons we shall not equate $\mu$ with $\ell^{-1}$.

When we show how to reproduce the field equations of standard formulations we shall frequently adopt the gauge in which $V^A=\ell\delta^A_4$. In order to signal that we are using a particular $SO(2,3)$/$SO(1,4)$ gauge we shall use the notation $V^A\overset{*}{=}\ell\delta^A_4$.
\subsection{Scalar Fields}
Scalar fields play a pivotal role in modern physics and therefore must be considered when coupling gravity to matter fields.
We have seen that the action for a massless, real scalar field (\ref{scalact}) in the standard treatment of gravity involves non-polynomial terms in the gravitational field. We will demonstrate that this is no longer the case when one takes gravity to be described by the pair $\{V^{A},A^{AB}\}$. Although our results are of general applicability, for concreteness we will explicitly look at the case of the standard-model Higgs scalar field $\Phi^{a}$, where $a,b,..$ denote $SU(2)$ indices. In the spirit of the gauge principle, we may then expect that from the Cartan gravity perspective one should consider an object $\Phi^{a A}$. We may further define a covariant derivative ${\cal D}_{\mu}\Phi^{a A}$ as follows such that the field transforms homogeneously with respect to combined $SO(2,3)/SO(1,4)$ and $SU(2)\times U(1)$ gauge transformations:

\begin{eqnarray}
{\cal D}_{\mu}\Phi^{a A} \equiv d\Phi^{aA} + A^{\ph \mu A}_{\mu\ph A B}\Phi^{a B}+\frac{ig}{2}{\cal B}^{\ph \mu a}_{\mu \ph ab}\Phi^{b A}+ \frac{ig'}{2}{\cal U}_{\mu}\Phi^{a A}
\end{eqnarray}
where ${\cal B}^{a}_{\ph ab}$ and ${\cal U}$ are the $SU(2)$ and $U(1)$ gauge fields with $g$ and $g'$ their respective coupling constants. We now show that it is possible to construct an action which reduces to the familiar Lagrangian of the Higgs boson in the limit where $V^{2}= const$.. Following the idea of polynomial simplicity, we can look to construct actions from the following ingredients: 

\begin{itemize}
\item The invariant objects of $SO(2,3)/SO(1,4)$ and $SU(2)$: $\epsilon_{ABCDE}$, $\eta_{AB}$,
$\epsilon_{ab}$, $\delta_{ab}$
\item The fields $V^{A}$, $\Phi^{a A}$, and its conjugate $\Phi^{\dagger A}_{\ph\dagger a}$
\item The differential forms $e^{A}$, $F^{AB}$, ${\cal D}\Phi^{a A}$, and its conjugate ${\cal D}\Phi^{\dagger A}_{\ph\dagger a}$
\end{itemize}
Henceforth we will suppress all $SU(2)$ indices for notational compactness. In considering the object $\Phi^{A}$ we now have five $SU(2)$-valued scalar fields instead of one. The increase in the number of variables is expected since the equations of motion will be first order partial differential equations rather than the usual second order Klein-Gordon equation. We may suspect that the field $\Phi=V_A\Phi^A$ will play the role of the Higgs field in the standard second order formulation and the four components $\Phi^I$ orthogonal to $V^A$ will be related somehow to the four spacetime derivatives $\partial_\mu\Phi$. In fact, as we shall see such a relation will be enforced by the equations of motion given below.
\subsubsection{The Higgs-Cartan action}
It is now straightforward to write down an action for the Higgs field minimally coupled to the electroweak gauge fields.
Although more general actions may be constructed from the above ingredients, we restrict ourselves to obtaining an action that reduces to the familiar Klein-Gordon equations of motion when $V^2=const.$. The simple action, which we shall refer to as the {\em Higgs-Cartan action}, is as follows:
\begin{eqnarray}\label{HCaction}
S_{HC}=\kappa_{HC}\int \left(\star\Sigma_A\w(\Phi^{\dagger A}\m D\Phi^B+\m D\Phi^{\dagger B}\Phi^A)V_B+\frac{1}{2}\star\Sigma U(\Phi^{\dagger}\Phi)\right)
\end{eqnarray}
where $\kappa_{HC}$ is a dimensionless constant to be fixed, $U(\Phi^{\dagger}\Phi)=-m^2\Phi^\dagger\Phi+\lambda (\Phi^\dagger\Phi)^2$ is the `Mexican hat' potential and $\Phi\equiv V_A\Phi^A$. As usual the mass term has the `wrong' sign as is required for the vacuum expectation value of the Higgs field $\Phi$ to be non-zero.
\subsubsection{Equations of motion and recovery of the standard Klein-Gordon equation}
Let us now look at the equations of motion. Variation with respect to $\Phi^A$ yields
\begin{eqnarray}
\delta_\Phi S_{HC}&=&\kappa_{HC}\delta_\Phi \int \left(\star\Sigma_A\w(\Phi^{\dagger A}\m D\Phi^B+\m D\Phi^{\dagger B}\Phi^A)V_B+\frac{1}{2}\star\Sigma U\right)\nonumber\\
&=&\kappa_{HC}\int \star\Sigma_A\w(\delta\Phi^{\dagger A}\m D\Phi^B+\m D\delta\Phi^{\dagger B}\Phi^A+\Phi^{\dagger A}\m D\delta\Phi^B+\m D\Phi^{\dagger B}\delta\Phi^A)V_B\\
&+&\frac{1}{2}\star\Sigma\frac{\partial U}{\partial \Phi}\delta\Phi^BV_B+\frac{1}{2}\star\Sigma\delta\Phi^{\dagger B}V_B\frac{\partial U}{\partial \Phi^\dagger}\nonumber\\
&=&\kappa_{HC}\int \delta\Phi^{\dagger B}\left(\star\Sigma_B\w\m D\Phi^AV_A+\m D(\star\Sigma_A\Phi^AV_B)+\frac{1}{2}\star\Sigma V_B\frac{\partial U}{\partial \Phi^\dagger}\right)\nonumber\\
&+&\left(\star\Sigma_B\w\m D\Phi^{\dagger A}V_A+\m D(\star\Sigma_A\Phi^{\dagger A} V_B)+\frac{1}{2}\star\Sigma V_B\frac{\partial U}{\partial \Phi}\right)\delta\Phi^B\nonumber.
\end{eqnarray}
Requiring $\delta_\Phi S_{HC}$ to be zero for arbitrary variations $\delta\Phi^{A}$ then implies the four-form equations
\begin{eqnarray}
\star\Sigma_B\w\m D\Phi^AV_A+\m D(\star\Sigma_A\Phi^AV_B)+\frac{1}{2}\star\Sigma\frac{\partial U}{\partial \Phi^\dagger}V_B=0
\end{eqnarray}
which, using the identity $\star\Sigma_A\w e_B=\star\Sigma \eta_{AB}=\star\Sigma_B\w e_A$, can be rewritten as
\begin{eqnarray}
\star\Sigma_B\w(\m D\Phi-2 e_A\Phi^A)+\left(\m D(\star\Sigma_A\Phi^A)+\frac{1}{2}\star\Sigma\frac{\partial U}{\partial \Phi^\dagger}\right)V_B=0.
\end{eqnarray}
The first term is orthogonal to $V^A$ while the second normal to it and these terms must therefore be independently zero. Thus we have
\begin{eqnarray}\label{KGCeqs}
\star\Sigma_B\w(\m D\Phi-2 e_A\Phi^A)=0\qquad \m D(\star\Sigma_A\Phi^A)+\frac{1}{2}\star\Sigma\frac{\partial U}{\partial \Phi^\dagger}=0.
\end{eqnarray}
If the co-tetrad is non-degenerate, the left equation readily implies $\m D\Phi=2 e_A\Phi^A$, or in tensor notation $\m D_\mu\Phi=2 e_{\mu A}\Phi^A\overset{*}{=}2 e_{\mu J}\Phi^J$. Non-degeneracy also implies the existence of a unique inverse $e^\mu_I$ called the tetrad or vierbein. Multiplying both sides by $e^\mu_I$ yields 
$2\Phi^I=e^{I\mu}\m D_\mu\Phi$ and we see that $\Phi^I$ is related to the derivative $\m D_\mu\Phi$ through the equations of motion. We now have
\begin{eqnarray}
*\m D\Phi&=&\frac{e}{3!}\epsilon_{\mu\nu\rho\sigma}\m D^\mu\Phi dx^\nu\w dx^\rho\w dx^\sigma=\frac{1}{3!}\epsilon_{IJKL}e^I_\mu e^J_\nu e^K_\rho e^L_\sigma\m D^\mu\Phi dx^\nu\w dx^\rho\w dx^\sigma\\
&=&\frac{1}{3!}\epsilon_{IJKL}e^{I\mu} \m D_\mu\Phi e^J\w e^K\w e^L=-\frac{1}{\mu\ell}\star\Sigma_I (e^{I\mu}\m D_\mu\Phi)=-\frac{2}{\mu\ell}\star\Sigma_I\Phi^I \nonumber \\
& \overset{*}{=}&-\frac{2}{\mu\ell}\star\Sigma_A\Phi^A
\end{eqnarray}
so that $\star\Sigma_A\Phi^A=-\frac{\mu\ell}{2}*\m D\Phi$ which inserted in the right equation of \eqref{KGCeqs} yields
\begin{eqnarray}
\square\Phi-\frac{\partial U}{\partial \Phi^\dagger}=0
\end{eqnarray}
Let us now consider the dual scalar density (see \ref{duality})
\begin{eqnarray}
-\frac{\mu\ell}{2}\m D(*\m D\Phi)-\frac{1}{2}\star\Sigma\frac{\partial U}{\partial \Phi^\dagger}&=&-\frac{\mu\ell}{2}\m D_\mu \left(\frac{e}{3!}\epsilon_{\kappa\tau\rho\sigma}D^\kappa\Phi\right)dx^\mu\w dx^\tau\w dx^\rho\w dx^\sigma+\frac{1}{2}\star\Sigma\frac{\partial U}{\partial \Phi^\dagger}\\
&\sim& -\frac{\mu\ell}{2}\m D_\mu \left(\frac{e}{3!}\epsilon_{\kappa\tau\rho\sigma}D^\kappa\Phi\right)\varepsilon^{\mu\tau\rho\sigma}-\frac{\mu\ell}{2} e\frac{\partial U}{\partial \Phi^\dagger}\\
&=&-\frac{\mu\ell}{2}\m D_\mu \left(e D^\mu\Phi\right)-e\frac{\mu\ell}{2}\frac{\partial U}{\partial \Phi^\dagger}\\
&=&-e\frac{\mu\ell}{2}\left(\square\Phi-\frac{\partial U}{\partial \Phi^\dagger}\right)
\end{eqnarray}
where we made use of the identity $\epsilon_{\kappa\tau\rho\sigma}\varepsilon^{\mu\tau\rho\sigma}=3!\delta^\mu_\kappa$ and $\square \Phi \equiv \frac{1}{e}\m D_\mu(e g^{\mu\nu}\m D_\nu\Phi)$. If the potential $U$ is given by $U(\Phi^{\dagger}\Phi)=-m^2\Phi^\dagger\Phi+\lambda (\Phi^\dagger\Phi)^2$ so that $\frac{\partial U}{\partial \Phi^\dagger}=-m^2\Phi+2\lambda |\Phi|^2\Phi$ then we now recognize is nothing but the minimally coupled Klein-Gordon  equation with a negative mass term required for imposing a non-zero vacuum expectation value of the Higgs. Thus, our first order Higgs-Cartan equations are equivalent (on-shell) to the standard Higgs equation of motion. 
%
%
Therefore, as usual the $SU(2)\times U(1)$ symmetry of the model is spontaneously broken by the Higgs field $\Phi$ which attains a vacuum expectation value for low energies. It is also now clear that although the Cartan-geometric formulation makes use of more scalar fields (i.e. we have five $SU(2)$-valued scalars $\Phi^A$ rather than a single one $\Phi$) we nevertheless do not end up with more propagating degrees of freedom. This is due to the fact that the Cartan-geometric action necessarily yields upon extremization first order polynomial partial differential equations rather than second order non-polynomial ones.

Although we have focused on a description of the Higgs boson, the approach here is general and applicable to spacetime scalar fields in general. In equation (\ref{HCaction}) we have found a Cartan-geometric action which reproduces the familiar Klein-Gordon equations. This action is not the only action which is consistent with the requirement of polynomial simplicity. It was shown in Section \ref{rolling} that it should not be possible to construct actions for $\phi^{A}$ which result in equations of motion for $\phi^{A}V_{A}$ which are higher than second order. It is interesting to ask then whether polynomial simplicity naturally leads to so-called Galileon scalar-field theories for actions which are cubic or higher in the field $\phi^{A}$ \cite{Nicolis:2008in,Chow:2009fm,Appleby:2012ba}. These theories have the interesting property that the equations of motion are not higher than second-order in derivatives even though their Lagrangians may contain terms such as $g^{\mu\nu}g^{\alpha\beta}\nabla_{\mu}\pi\
\nabla_{\nu}\pi\nabla_{\alpha}\nabla_{\beta}\pi$. 
\subsubsection{The similar roles of the Higgs field and contact vector}\label{simroles}
It is interesting to compare the roles of the contact vector $V^A$ to the standard model's Higgs field $\Phi$ \cite{Wise:2011ab}. In the context of the standard model of particle physics, a role of the Higgs field is to spontaneously break the electroweak symmetry group $SU(2)\times U(1)$ so that only a $U(1)$ invariance, associated with the Maxwell field, remains. The symmetry is broken by partitioning the group into transformations that, on the one hand, change the Higgs field and those that leave it invariant on the other. The remnant unbroken symmetry corresponds of the subgroup of transformations that leave the Higgs field invariant. The group $SU(2)\times U(1)$ consists of all unitary $2\times 2$ matrices and it is easy to see that the subgroup that leaves some two-component complex field invariant is $U(1)$. The electromagnetic field $A$ can now be represented as a $2\times 2$ matrix so that $A\Phi=0$. 

The situation here is completely analogous to that in Cartan geometry. Here we consider a larger gauge group $SO(2,3)$/$SO(1,4)$ which is then partitioned into two parts: those that change the contact vector and those that leave it unchanged. The former transformations are generalized translations called transvections, and the latter the usual local Lorentz transformations which remains as an unbroken symmetry group. Thus we see that the object $V^A$, which seems unfamiliar and odd, plays the role of a Higgs-type field.

However, there is a significant difference in the way these objects are usually treated. While the Higgs field carries independent degrees of freedom which is of key importance for the renormalizability of the electroweak theory, antecedent work on Cartan gravity has often assumed that the contact vector field $V^A$ carries no gauge independent degrees of freedom of its own \cite{Petti:2006ue}. Specifically, the direction $V^A$ is a pure gauge degree of freedom due to the local $SO(2,3)$/$SO(1,4)$ gauge symmetry leaving the magnitude $V^2$ as the only gauge independent degree of freedom.  However, $V^2$ is typically restricted by hand to be constant \cite{Pagels:1983pq,Stelle:1979va}, corresponding to what we have called a non-dynamical approach.

The similar roles of the fields $V^A$ and $\Phi$ as symmetry breaking Higgs-type fields can be taken to signal the need to provide non-trivial dynamics for $V^A$ \cite{Randono:2010cq}. To understand the way matter couples to the contact vector $V^A$ it is necessary to have a fully Cartan-geometric formulation of the matter sector. This paper will thus provide the necessary tools to embark on such an enterprise in a more systematic fashion. We return to this issue in Section \ref{discus}.
\subsection{Yang-Mills fields}
\label{ymc}
We now turn to the coupling of gauge fields (here referred to in general as Yang-Mills fields whether Abelian or non-Abelian gauge fields) in the context of Cartan gravity. Consider then some Yang-Mills gauge field $B=B_{\mu}dx^{\mu}$ valued in the Lie algebra of a particular symmetry group (gauge indices suppressed). The gauge principle now requires us to attach a `rolling' index to the Yang-Mills field $B\rightarrow B^{A}=B^{\ph \mu A}_{\mu}dx^{\mu}$. In analogy to the scalar field case, we may anticipate that the original gauge field $B$ will be identified with $V_{A}B^{A}$ via the equations of motion, whereas the components of $B^{A}$ orthogonal to $V^{A}$ will contain information about derivatives of the gauge field $B=V_AB^A$. 

If indeed the gauge field $B$ is identified with $V_{A}B^{A}$ then one can define a field strength two form $G$ for $B$:
\begin{eqnarray}
G  \equiv dB - ig B\wedge B
\end{eqnarray}
where $g$ is the gauge coupling constant of the theory. This field strength clearly transforms homogeneously under Yang-Mills gauge transformations if $B_{\mu}$ transforms as usual under a gauge transformation represented by the matrix $U$:

\begin{eqnarray}
\label{btrans}
B &\rightarrow & U B U^{-1} - \frac{i}{g} dU U^{-1}
\end{eqnarray}
If we assume $V^{2}$ to be non-vanishing then we may generalize (\ref{btrans}) to a transformation law for $B^{A}$:

\begin{eqnarray}
\label{batrans}
B^{A} &\rightarrow & UB^{A}U^{-1} -\frac{i}{g}\frac{V^{A}}{V^{2}}dU U^{-1}.
\end{eqnarray}
which implies that $V_AB^A$ transforms according to \eqref{btrans}. We will now  present an action which is invariant under the transformation law (\ref{batrans}).
\subsubsection{The Yang-Mills-Cartan action}
Consider then the following Yang-Mills-Cartan action:
\begin{eqnarray}
\label{sym}
S_{YMC}=\kappa_{YMC} Tr\int\left(\star\Sigma_{AB}\w B^A\w B^B+e_A\w B^A\w G\right)
\end{eqnarray}
where $\kappa_{YMC}$ is a dimensionless constant to be fixed and  $B\equiv V_AB^A$. The invariance of (\ref{sym}) under (\ref{batrans}) can be seen as follows. Since $B^I$ transform homogeneously so must $\star\Sigma_{AB}\w B^{A} \w B^{B}\overset{*}{=}\star\Sigma_{IJ}\w B^{I} \w B^{J}$. As the constancy of $V^{2}$ has been assumed, we have that $e_{A}V^{A}=0$ and $e_{A}\wedge B^{A}\overset{*}{=}e_I\w B^I$ therefore transforms homogeneously under (\ref{batrans}). Additionally the two-form ${\cal D}B$ transforms homogeneously by construction.

We note however that the second term is invariant under \eqref{batrans} if and only if $V^{2}$ is constant. If $V^A$ is allowed to vary then $e^A$ will have a fifth non-zero component $V_Ae^A$ which then implies that $e_A\w B^A$ is not a gauge covariant term. We conclude that if $V^A$ is turned into a dynamical field then the gauge invariance of the Yang-Mills-Cartan action lost. Such a feature of this action might be considered a deficiency and we shall see in Section \ref{unify} how this can be cured by unification. 
\subsubsection{Equations of motion and the recovery of the standard Yang-Mills equation}
The equations of motion are obtained as usual by varying the action with respect to $B^A$ yielding
 the equations of motion
\begin{eqnarray}
\label{baeq}
2\star\Sigma_{AB}\w B^B+e_A\w G-V_A{\cal D}(e_B\w B^B)=0
\end{eqnarray}
As in the case of the field $\phi^{A}$, we may decompose (\ref{baeq}) into a part projected along $V^{A}$ and parts orthogonal to $V^{A}$:
\begin{eqnarray}\label{YMCeqs}
2\star\Sigma_{AB}\w B^B+ e_A\w G=0\qquad V_A{\cal D}(e_B\w B^B)=0
\end{eqnarray}
After some calculations detailed in Appendix \ref{YMCcalc} the first of the above equations can be shown to be equivalent to 

\begin{eqnarray}
\label{ebdb}
(\mu\ell) e^{A}\wedge B_{A}=  *G
\end{eqnarray}
Therefore, the remaining component of the equations of motion becomes

\begin{eqnarray}
\frac{1}{\mu\ell}{\cal D}(* G)=0
\end{eqnarray}
which is nothing but the Yang-Mills equations of motion written in the language of differential forms, as shown in equation (\ref{ymeq}).
\subsubsection{Gauge boson masses}
A key property of the electroweak theory is the presence of massive gauge bosons. Gauge boson masses can be included in the standard formulation by adding a gauge symmetry breaking mass term
\begin{eqnarray}
S_{Proca}&=&\int -Tr\frac{1}{2} G\w* G-m^2 B\w*B\nonumber
\end{eqnarray}
However, as is well-known, the Proca mass-term destroys renormalizability. The standard trick, which does note explicitly break gauge symmetry, is to introduce gauge field mass terms using a minimally coupled Higgs field
\begin{eqnarray}
S^{(kin)}_{H}&=&-\int \m D\Phi^\dagger\w*\m D\Phi=-\int d\Phi^\dagger \w *d\Phi+ig(d\Phi^\dagger\w *B\Phi-\Phi^\dagger B\w*d\Phi)+g^2\Phi^\dagger B\w *B\Phi\nonumber\\
&=&-\int d\Phi^\dagger \w *d\Phi+ig(d\Phi^\dagger\Phi-\Phi^\dagger d\Phi)\w*B+g^2\Phi^\dagger B\w *B\Phi\nonumber
\end{eqnarray}
where $\m D\Phi=d\Phi+ig B\Phi$. The first term is the usual kinetic term for the scalar field. The second term is proportional to $d\Phi^\dagger\Phi-\Phi^\dagger d\Phi$ which is pure gauge and is zero in the unitary gauge $\Phi\overset{*}{=}(0,v)$ with $v(x)$ is some real-valued scalar. The last term $g^2\Phi^\dagger B\w *B\Phi$ provides mass for the weak gauge fields but not the electromagnetic one $A$ since by construction $A\Phi=0$ (see Section \ref{simroles}).

Now that we have recapitulated how gauge bosons acquire mass in the standard formulation we turn to the Cartan-geometric formulation. We first note that the non-polynomial term $B\w *B$ violates both the gauge principle and polynomial simplicity. Secondly, it is not clear to us how to write down a polynomial mass term or how to make such terms appear on-shell, i.e. after imposing the equations of motions. Instead, it appears that the only way to generate gauge boson masses within the Cartan-geometric approach is to use the Higgs-Cartan field $\Phi^A$. 

Let us therefore expand the kinetic part of the Higgs-Cartan action minimally coupled to a Yang-Mills field $B$. This yields
\begin{eqnarray}
S^{(kin)}_{HC}&=&\kappa_{HC}\int \star\Sigma_A \w(\Phi^{A\dagger}\m D\Phi^B+\m D\Phi^{B\dagger}\Phi^A)V_B\nonumber\\
&=&\kappa_{HC}\int \star\Sigma_A \Phi^{A\dagger}\w(D\Phi^B+igB\Phi^B)V_B-(D\Phi^{B\dagger}-ig\Phi^{\dagger B} B)\w \star\Sigma_A \Phi^A V_B\nonumber\\
&=&\kappa_{HC}\int \star\Sigma_A \w(\Phi^{A\dagger} D\Phi^B+D\Phi^{B\dagger}\Phi^A)V_B+ig(d\Phi^\dagger\Phi-\Phi^\dagger d\Phi)\w*B+g^2\Phi^\dagger B\w *B\Phi\nonumber
\end{eqnarray}
and we see that we get the same two extra terms as above: one that is pure gauge and one that yields mass terms for the weak field but not the electromagnetic one. Therefore, the inclusion of gauge boson masses is straightforward in the Cartan-geometric formulation but only, as it seems, using a Higgs mechanism.
\subsection{Spinor fields}
Compared to scalar and gauge fields, the treatment of spinor fields from the Cartan gravity perspective is relatively straightforward, especially in the de Sitter $spin(1,4)$ case. This can already be suspected as the standard actions are already on polynomial and on a first order form. In fact, the necessary step consists only of replacing the Dirac $spin(1,3)$ index with a $spin(2,3)$ or $spin(1,4)$ index, and then write down an action which reduces to the standard Dirac action in the gauge $V^A\overset{*}{=}\ell\delta^A_4$.
\subsubsection{Anti-de Sitter and de Sitter spinors}
The anti-de Sitter and de Sitter Lie-algebras are given by
\begin{eqnarray}
[\m J_{AB},\m J_{CD}]=-i(\eta_{AC}\m J_{BD}-\eta_{AD}\m J_{BC}-\eta_{BC}\m J_{AD}+\eta_{BD}\m J_{AC})
\end{eqnarray}
where $\eta_{AB}=diag(-1,+1,+1,+1,-1)$ in the anti-de Sitter case and $\eta_{AB}=diag(-1,+1,+1,+1,+1)$ in the de Sitter case. Suitable choices for the Clifford algebra gamma-matrices are 
\begin{eqnarray}
\Gamma_A=(-i\gamma_5\gamma_I,\gamma_5)\ \text{for}\ spin(2,3)\qquad \Gamma_A=(\gamma_I,i\gamma_5)\ \text{for}\ spin(1,4)
\end{eqnarray}
satisfying $\{\Gamma_A,\Gamma_B\}=-2\eta_{AB}$ with the spin-$\frac{1}{2}$ generators in both cases given by $\m J_{AB}=-\frac{i}{4}[\Gamma_A,\Gamma_B]$. In accordance with the gauge principle, the coupling of this (anti-)de Sitter spinor field to gravity is done by replacing the exterior derivative with a $SO(2,3)$/$SO(1,4)$-gauge covariant derivative: 
\begin{eqnarray}
d\psi\rightarrow D\psi=d\psi-\frac{i}{2}A^{AB}\m J_{AB}\psi
\end{eqnarray}%
Coupling this spinor field to other gauge fields is done as usual by adding the relevant gauge connection $B$, i.e.
\begin{eqnarray}
D\psi\rightarrow \m D\psi=D\psi-ie B\psi=D\psi-ie V_AB^A\psi
\end{eqnarray}%
according to the usual gauge prescription.
\subsubsection{Dirac-Cartan action and equations of motion}
Before writing down an action it is necessary to understand how we can extract invariants from the spinor $\psi$. In the Lorentzian case $spin(1,3)$ we know that $\bar{\psi}\psi\equiv\psi^\dagger\gamma_0\psi$ is both real-valued and invariant under Lorentz transformations. In order to generalize this to the de Sitter and anti-de Sitter groups we make the ansatz 
\begin{eqnarray}
\tilde{\psi}\equiv \psi^\dagger \xi
\end{eqnarray}
and then require that $\tilde\psi\psi$ is real-valued and invariant under the relevant group, i.e. either $spin(2,3)$ or $spin(2,3)$. Real-valuedness of $\tilde{\psi}\psi$ immediately implies that $\xi^\dagger=\xi$. Invariance under the (anti-)de Sitter group can be imposed by considering an infinitesimal (anti-)de Sitter transformation
\begin{eqnarray}
\psi\rightarrow \psi + \frac{i}{2}\theta^{AB}\m J_{AB}\psi \qquad \psi^\dagger\rightarrow \psi^\dagger - \frac{i}{2}\theta^{AB}\psi^\dagger \m J_{AB}^\dagger
\end{eqnarray}
where $\theta^{AB}$ is anti-symmetric and real-valued parameters. The condition that $\tilde\psi\psi$ is invariant:
\begin{eqnarray}
\tilde\psi\psi=\psi^\dagger\xi\psi\rightarrow (\psi^\dagger - \frac{i}{2}\theta^{AB}\psi^\dagger \m J_{AB}^\dagger)\xi (\psi + \frac{i}{2}\theta^{AB}\m J_{AB}\psi)=\tilde\psi\psi+\frac{i}{2}\theta^{AB}\psi^\dagger(\xi \m J_{AB}-\m J_{AB}^\dagger\xi)\psi
\end{eqnarray}
now implies to first order in $\theta^{{AB}}$ that $\xi \m J_{AB}=\m J_{AB}^\dagger\xi$. In the case of $spin(1,4)$ it is easy to verify that $\xi=\gamma^0$ \footnote{We adopt the sloppy but harmless practice of writing $\xi=\gamma_0$ even though these two objects are members of distinct spaces. If we write $\psi\in \m W$ where $\m W$ is a four-dimensional complex vector space, then $\gamma_0\in \m W\otimes \m W^*$ and $\xi\in \m W^*\times \bar{\m W}^*$ where $\m W^*$ and $\bar{\m W}^*$ are the dual and conjugate dual spaces of $\m W$ \cite{Wald:1984rg}.} satisfies that condition and we have $\tilde\psi=\bar\psi=\psi^\dagger\gamma_0$. However, in the case of $spin(2,3)$, then $\xi=i\gamma_0\gamma_5$ is the appropriate choice \cite{Ikeda:2009xb,MirKasimov:1978ci} so that $\tilde\psi=\psi^\dagger i\gamma_0\gamma_5$.

Using the expressions
\begin{eqnarray}
spin(2,3):&&D\psi=d\psi-\frac{i}{2}A^{AB}\m J_{AB}\psi\overset{*}{=}D^{(\omega)}\psi+\frac{i}{2\ell}e^I\gamma_I\psi\\
spin(1,4):&&D\psi=d\psi-\frac{i}{2}A^{AB}\m J_{AB}\psi\overset{*}{=}D^{(\omega)}\psi-\frac{i}{2\ell}e^I\gamma_5\gamma_I\psi
\end{eqnarray}
where $D^{(\omega)}\psi\equiv d\psi-\frac{i}{2}\omega^{IJ}\m J_{IJ}\psi$ is the usual $spin(1,3)$ gauge covariant derivative, and introducing the symbol $\Gamma\equiv \frac{V^A}{\ell}\Gamma_A$ we can now verify that the following two Dirac-Cartan actions 
\begin{eqnarray}
S^{spin(2,3)}_{DC}&=& \kappa_{DC}^{(2,3)}\int\star\Sigma_{A}\w \frac{i}{2}(\tilde\psi\Gamma^AD\psi-D\tilde\psi\Gamma^A\psi)+i\star\Sigma (m-\frac{2}{\ell})\tilde\psi\Gamma\psi
\\
S^{spin(1,4)}_{DC}&=&\kappa_{DC}^{(1,4)}\int\star\Sigma_{A}\w \frac{i}{2}(\bar\psi\Gamma^AD\psi-D\bar\psi\Gamma^A\psi)-\star\Sigma m\bar\psi\psi
\end{eqnarray}
reduce to the standard Dirac action in the gauge $V^A\overset{*}{=}\ell\delta^A_4$ and for appropriate values of the dimensionless constants $\kappa_{DC}^{(2,3)}$ and  $\kappa_{DC}^{(1,4)}$. Due to the difference in Clifford algebra $\Gamma$-matrices and in $\xi$, the matter term of the action takes on a form that depends on whether the rolling group is $SO(2,3)$ or $SO(1,4)$. In particular, we note the presence of a `cosmological' mass $m_{cosm.}=\frac{2}{\ell}$ in the anti-de Sitter case that needs to be corrected for so that the action reduces to the standard Dirac one. Alternatively one could perhaps view this as a `cosmological' mass generation mechanism so that an anti-de Sitter fermion acquire mass not only from the electroweak Higgs field $\Phi$ but also from the gravitational Higgs field $V^A$.

Contrary to the bosonic actions there is no need to show that the corresponding equations of motion reduce to the standard Dirac equation. This follows immediately from the fact that these actions reduce to the standard Dirac action in the special gauge where $V^A\overset{*}{=}\ell \delta^A_4$.
\subsubsection{Contact vector, orientation, and parity violation}\label{chiral}
An interesting possible role of the contact vector $V^A$ is in relation to parity violation of the electroweak theory. First we  recall that a manifold which is {\em orientable} admits an everywhere non-vanishing four-form, 
\begin{eqnarray}
\m E(x)=\frac{1}{4!}\m E_{\mu\nu\rho\sigma}(x)dx^\mu\w dx^\nu\w dx^\rho\w dx^\sigma\neq0.
\end{eqnarray}
It induces an orientation since it attributes a sign $\pm$ to any collection of four ordered vectors $u_1^\mu$, $u_2^\mu$, and $u_3^\mu$, $u_4^\mu$ by 
\begin{eqnarray}
sign(\m E_{\mu\nu\rho\sigma}u_1^\mu u_2^\nu u_3^\rho u_4^\sigma).
\end{eqnarray}
We now see that the pair $\{V^A,A^{AB}\}$ implies such a natural four-form
\begin{eqnarray}
\m E=\star\Sigma=\frac{\mu}{4!}\epsilon_{ABCDE}V^E DV^A\w DV^B\w DV^C\w DV^D
\end{eqnarray}
at least as long as $V^A\neq 0\neq DV^A$. Since this four-form is odd in $V^A$ we see that the discrete transformation $V^A\rightarrow -V^A$ is associated with a change of orientation of the manifold. That the contact vector induces a natural orientation on the manifold is easily visualized if we imagine the manifold as embedded in a fifth dimension. There the contact vector $V^A$ is visualized as a normal to the manifold hyper surface (see \cite{Westman:2012xk}) and we see that the change $V^A\rightarrow -V^A$ is directly associated with a change of orientation.

The chiral asymmetry of the electroweak theory originates from the fact that the weak field $W$ couples only to the left part of the fermions. However, there are no faithful two-dimensional representations of $SO(2,3)$ or $SO(1,4)$ and thus the (anti-) de Sitter Dirac field does not split up into left- and right-handed representations. Thus, the Cartan-geometric formulation naturally starts from a chiral symmetric formalism employing four-component spinors. Nevertheless, the additional  structure that the contact vector $V^A$ provides can be used to partition the Dirac spinor $\psi$ into left- and right-handed components.

The left- and right-handed components of a Dirac field are usually defined using the chiral projector 
\begin{eqnarray}
P_{L,R}=\frac{1}{2}(1\mp \gamma_5)
\end{eqnarray}
with upper sign representing left projector $(L)$ and lower sign the right projector $(R)$. However, we see that within Cartan formulation of fermionic fields we can define the chiral projector as
\begin{eqnarray}
P^{SO(2,3)}_{L,R}=\frac{1}{2}(1\mp \Gamma)\qquad P^{SO(1,4)}_{L,R}=\frac{1}{2}(1\pm i\Gamma)
\end{eqnarray}
where $\Gamma\equiv \frac{V^A}{\ell}\Gamma_A$. Whether the gravitational Higgs-type field $V^A$ can play the role of breaking parity in the electroweak theory cannot be answered before a more systematic Cartan-geometric reformulation of the electroweak theory has been carried out.
\subsection{Alternative ideas}
It is now appropriate to make contact with some literature of similar but distinct pre-existing treatments of matter fields. Especially noteworthy is a paper by Pagels \cite{Pagels:1983pq} wherein actions consistent with polynomial simplicity were constructed that described matter for an $SO(5)$ Cartan gravity model. Despite the different choice of group, the results are readily applicable to the $SO(2,3)$ and $SO(1,4)$ groups considered here: scalar fields were found to be described by fields $\phi^{A}$ valued in the Lie algebra of $SO(5)$ whilst spinor fields were shown to be described by fields $\Psi^{\alpha}$ valued in the Lie algebra of $spin(5)\simeq sp(2)$. A point of difference is in the treatment of Yang-Mills fields, which were instead described by a pair $\{(Y^{AB})^{a}_{\phantom{a}b},{\cal B}^{a}_{\phantom{a}b}\}$. The field ${\cal B}$ is precisely the Yang-Mills gauge field $B$. The field $Y^{AB}$ meanwhile transforms homogeneously under $SO(5)$ and Yang-Mills transformations; it is assumed to 
satisfy the following properties:

\begin{eqnarray}
Y^{AB} &=&  Y^{[AB]}\\
Y^{AB}V_{B} &=& 0
\end{eqnarray}
Therefore $Y^{AB}$ has six independent components and in conjunction with $e^{A}$ is relatable to $*{\cal D}B$. This approach avoids the skewed transformation law \eqref{batrans} of $B^A$. As a consequence the constancy of $V^{2}$ need not be assumed in order to retain gauge invariance under Yang-Mills transformations.

More recently Wilczek \cite{Wilczek:1998ea} has considered the coupling of matter fields in the context of $SO(1,4)$ and $SO(2,3)$ Cartan gravity models. In discussing this approach it can be noted that the following spacetime vector may be defined:

\begin{eqnarray}
\label{ftetrad}
{\cal E}^{\mu}_{A} &\equiv& \frac{\varepsilon^{\mu\nu\delta\sigma}(\star\Sigma_{A})_{\nu\delta\sigma}}{\varepsilon^{\gamma\xi\chi\eta}(\star\Sigma)_{\gamma\xi\chi\eta}}
\end{eqnarray}
By inspection, when $V^{A}\overset{*}{=}\ell \delta^{A}_{4}$, the vector (\ref{ftetrad}) coincides with the tetrad of equation (\ref{tetrad}). Therefore one can construct $SO(2,3)/SO(1,4)$ invariant actions for fields $\phi$ and $B_{\mu}$ which correspond in the symmetry broken phase to the conventional Klein-Gordon and Yang-Mills actions. For example, it can be easily be checked that 

\begin{eqnarray}
\label{kgk}
-\frac{1}{\mu\ell}\int \eta^{AB}{\cal E}^{\mu}_{A}{\cal E}^{\nu}_{B}\partial_{\mu}\phi\partial_{\nu}\phi \star\Sigma
\end{eqnarray}
reduces to the Klein-Gordon action (\ref{scalact}). The price to pay for this is that the actions are non-polynomial in precisely the same way as the standard actions are. In \cite{Wilczek:1998ea} the following object is considered

\begin{eqnarray}
\label{ftetrad2}
\tilde{{\cal E}}^{\mu}_{A} &\equiv& \frac{\varepsilon^{\mu\nu\delta\sigma}(\star\Sigma_{A})_{\nu\delta\sigma}}{\varepsilon_{0}}
\end{eqnarray}
where ${\varepsilon}_{0}$ is a spacetime density introduced into the theory; it is assumed to be constant
and to not have its own equations of motion. Therefore, use of $\tilde{{\cal E}}^{\mu}_{A}$ instead of ${\cal E}^{\mu}_{A}$ in (\ref{kgk}) yields an action which is polynomial in dynamical fields but because of the existence of a constant density ${\varepsilon}_{0}$ in the action is not invariant under diffeomorphisms. A similar result applies to the construction of actions for a Yang-Mills field in this approach. Therefore one can avoid non-polynomial actions in the second order formalism in Cartan gravity if diffeomorphism invariance is broken. The construction of actions incorporating a constant spacetime density is reminiscent of the theory of unimodular gravity in the metric formalism (see for instance \cite{Coley:2011cd}).

An approach sharing features of the approach chosen in this paper and in \cite{Pagels:1983pq} is the Duffin-Kemmer-Petiau (DKP) formulation of Klein-Gordon and vector field equations \cite{PhysRev.54.1114,Kemmer:1939zz,Lunardi:1999jq,Kanatchikov:1999ut,Bogush:2007pw}. As is the case for field equations recovered from polynomially simple actions, the DKP field equations are first order in spacetime derivatives. Reminiscent of the present paper, the DKP formalism describes a scalar field as an $SO(1,4)$ vector $\phi^{A}$. In the present paper and in \cite{Pagels:1983pq}, the field $V^{A}$ sets a scale with which to decompose a field $\Phi^{A}$ into what is identified as the scalar field $\Phi\equiv V_{A}\Phi^{A}$ and its derivatives $\frac{1}{2}e^{\mu I}{\cal D}_{\mu}\Phi=\Phi^{I}$. In the DKP approach the scale arises not from $V^{A}$ but from the mass $m$ of the fields itself. In the notation of the present paper, the scalar field $\phi$ is defined as $\phi\equiv m^{-3/2}\delta^{4}_{A}\phi^{A}$ whereas $\
sqrt{m}e^{\mu I}\partial_{\mu}\phi= \phi^{I}$. Although $V^{A}$ is not explicitly referred to in the DKP formulation, its recovery of Lorentz invariant field equations from representations of $SO(1,4)$ assumes symmetry breaking. A similar decomposition exists for a field valued in the adjoint representation of $SO(1,4)$ into a spacetime vector field and its spacetime exterior derivative. However, due to the reliance on the scale $m$, the DKP approach is only applicable to massive fields and therefore requires modification in order to accommodate massless scalar and vector/one-form fields \cite{HarishChandra:1947zz,Casana:2002fu}.
%
\section{The Cartan gravitational field equations}\label{stressen}
Until now we have shown how to incorporate the effect of gravity on matter fields consistent with the gauge principle and polynomial simplicity. This yielded a formalism in which all fields, gravitational and matter fields, are subject to polynomial first order partial differential equations. However, in order to complete our Cartan-geometric reformulation it is also necessary to consider the back-reaction of the matter fields on the spacetime geometry described by the pair $\{V^{A},A^{AB}\}$. Since we have treated $V^A$ as a non-dynamical field here we shall restrict ourselves to how matter influences the dynamics of $A^{AB}$. 
\subsection{The spin-energy-momentum three-form}
Within the traditional metric formulation of General Relativity one introduces the canonical energy-momentum tensor $\m T_{\mu\nu}$ defined by
\begin{eqnarray}
\m T_{\mu\nu}\equiv-\frac{2}{\sqrt{-g}}\frac{\delta S_M}{\delta g^{\mu\nu}}.
\end{eqnarray}
However, in order to incorporate fermionic fields, which can induce spacetime torsion $T^I$, the new gravitational variables $e^I$ and $\omega^{IJ}$ are introduced. One then associates to these variables the three-forms $\m T_I$ and $\m S_{IJ}$ defined by
\begin{eqnarray}
\delta_e S_M\equiv\int \m T_I\w \delta e^I\qquad \delta_\omega S_M\equiv \int \m S_{IJ}\w\delta \omega^{IJ}
\end{eqnarray}
denoted the {\em energy-momentum} and {\em spin-density} three-forms respectively. However, from a Cartan-geometric perspective this is not particularly natural. Firstly, the co-tetrad $e^I$ is not a fundamental field and secondly the fundamental connection is  $A^{AB}$ and not $\omega^{IJ}$. Instead it is more natural from a Cartan-geometric perspective to collect the energy-momentum and spin-density three-forms into a single object $\m S_{AB}$ defined by
\begin{eqnarray}
\delta_A S_M\equiv\int \m S_{AB}\w \delta A^{AB}
\end{eqnarray}
which we denote the {\em spin-energy-momentum three-form}. The previous energy-momentum and spin-density three-forms are recovered in the gauge where $V^A\overset{*}=\ell\delta^A_4$ as follows
\begin{eqnarray}
\m S_{IJ}\overset{*}=\m S_{IJ}\qquad \m S_{I4}\overset{*}=(\mp\ell \m T_I,0)
\end{eqnarray}
Given that the spacetime geometry is described by a pair of fields $\{V^A,A^{AB}\}$ it is appropriate to consider the four-form $\m Q_A$ defined by
\begin{eqnarray}
\delta_V S_{G+M}\equiv\int \m Q_A \delta V^A
\end{eqnarray}
where $S_{G+M}$ is the combined gravitational and matter action. The equation $\m Q_A=0$ would presumably then provide the dynamics equations for $V^A$. We shall return to this issue in Section \ref{discus}.
\subsection{Spin-energy-momentum for matter fields}
In this section we shall showcase the spin-energy-momentum three-forms of our three typical standard model fields. The following relations
\begin{eqnarray*}
\delta_A \star\Sigma &=& V_{[D}\star\Sigma_{C]}\w \delta A^{CD}\\
\delta_A \star\Sigma_A &=& V_{[D}\star\Sigma_{C]A}\w \delta A^{CD}\\
\delta_A \star\Sigma_{AB} &=& V_{[D}\star\Sigma_{C]AB}\w \delta A^{CD}\\
\delta_A e^B&=&\delta A^{BC}V_C
\end{eqnarray*}
will be useful for carrying out the variations.
\subsubsection{Higgs-Cartan field}
In order to determine the spin-energy-momentum three-form we vary the Higgs-Cartan action with respect to $A^{AB}$. After some straightforward calculations we obtain 
\begin{eqnarray}
\delta_A S_{HC}=\int {\cal S}^{(HC)}_{CD}\w \delta A^{CD} 
\end{eqnarray}
with
\begin{eqnarray*}
\m S^{(HC)}_{CD}&=&\kappa_{HC}\left[-2V_{[D}\star\Sigma_{C]A}\w(\Phi^{\dagger A}\m D\Phi^B+\m D\Phi^{\dagger B}\Phi^A)V_B\right.\\
&&+\left.2\star\Sigma_A(\Phi^{\dagger A}V_{[C}\Phi_{D]}+V_{[C}\Phi^\dagger_{D]}\Phi_{A})+V_{[C}\star\Sigma_{D]}U(\Phi)\right]
\end{eqnarray*}
corresponding to the spin-energy-momentum three-form. First we note that, as expected, the spin-density ${\cal S}^{(HC)}_{IJ}$ is identically zero. In order to relate the components ${\cal S}^{(HC)}_{I4}$ to the canonical energy-momentum tensor we consider the dual vector density. This yields
\begin{eqnarray}
\frac{e}{3!}(\m S^{(HC)}_{I4})_{\mu\nu\rho}\varepsilon^{\mu\nu\rho\sigma}=\pm \frac{e}{4}\kappa_{HC}\mu\ell^2\left(\m D^\sigma\Phi^\dagger\m D_I \Phi+\m D_I\Phi^\dagger\m D^\sigma \Phi-e^\sigma_I(\m D^\mu\Phi^\dagger\m D_\mu\Phi+U)\right)
\end{eqnarray}
Clearly then, his corresponds to the usual stress-energy tensor of the Higgs field up to a pre-factor $\pm\frac{e}{4}\kappa_{HC}\mu\ell^2$.
\subsubsection{Yang-Mills-Cartan field}
Next we consider the spin-energy-momentum of the Yang-Mills-Cartan field which is obtained by varying the the Yang-Mills-Cartan action with respect to $A^{AB}$. This yields
\begin{eqnarray*}
\m S^{(YMC)}_{CD}=\kappa_{YMC}Tr\left[V_{[D}\star\Sigma_{C]AB}\w B^A\w B^B-V_{[D}B_{C]}\w G\right]
\end{eqnarray*}
Again we see that the spin-density three-form ${\cal S}_{IJ}^{(YMC)}$ is identically zero. As before we now consider the dual vector density of ${\cal S}^{(YMC)}_{I4}$ with the equations of motion imposed yields
\begin{eqnarray}
\m S^{(YMC)}_{I4}\sim\frac{e}{3!}(\m S^{(YMC)}_{I4})_{\mu\nu\rho}\varepsilon^{\mu\nu\rho\sigma}=\pm \frac{e\kappa_{YMC}}{2\mu}Tr(G_{IL}G^{\sigma L}-\frac{1}{4}e^\sigma_I G_{\mu\nu}G^{\mu\nu}).
\end{eqnarray}
This is the usual stress-energy tensor of a Yang-Mills field up to a pre-factor $\pm \frac{e\kappa_{YMC}}{2\mu}$.
\subsubsection{Dirac-Cartan fields} 
First let us consider the $spin(1,4)$ case.
Varying the $spin(1,4)$ Dirac-Cartan action with respect to $A^{AB}$ yields the spin-energy-momentum tensor
\begin{eqnarray}
\nonumber\m S^{(DC)}_{CD}&=&\kappa_{DC}\left[-V_{[D}\star\Sigma_{C]A}\w \frac{i}{2}(\bar\psi\Gamma^AD\psi-D\bar\psi\Gamma^A\psi)+\frac{1}{4}\star\Sigma_A \bar\psi \{\Gamma^A,S_{CD}\}\psi\right. \\
&& \left.-V_{[D}\star\Sigma_{C]}m\bar\psi\psi\right].
\end{eqnarray}
Adopting the gauge $V^A\overset{*}{=}\ell\delta^A_4$ we can identify a non-zero spin-density
\begin{eqnarray}
\m S^{(DC)}_{IJ}=\frac{\kappa_{DC}}{4}\star\Sigma_K \bar\psi \{\gamma^K,S_{IJ}\}\psi
\end{eqnarray}
and after some simplification from $\m S_{4I}$ we obtain the canonical energy-momentum tensor after the usual procedure of dualizing and lowering the indices
\begin{eqnarray}
\nonumber\m S^{(DC)}_{I4}&\sim&\frac{e}{3!}(\m S^{(DC)}_{I4})_{\mu\nu\rho}\varepsilon^{\mu\nu\rho\sigma}\\
&=&\mp e\frac{\kappa_{DC}\mu\ell^2}{2}\left(\frac{i}{2}(\bar\psi \gamma^\sigma D_I^{(\omega)}\psi-D_I^{(\omega)}\bar\psi \gamma^\sigma \psi)-e^\sigma_I(e^\mu_J\frac{i}{2}(\bar\psi \gamma^J D_\mu^{(\omega)}\psi-D_\mu^{(\omega)}\bar\psi \gamma^J\psi)-m\bar\psi\psi)\right) \nonumber
\end{eqnarray}
where $\gamma^\mu g_{\mu\nu}=\gamma_\nu\equiv e_\nu^I\gamma_I$. It may be checked that the results for the $spin(2,3)$ case are the same when evaluated explicitly up to a factor of $-1$ multiplying $S^{(DC)}_{I4}$ as given above. Thus, we obtain the standard energy-momentum tensor for a Dirac field up to the pre- factor $\mp e\frac{\kappa_{DC}\mu\ell^2}{2}$. We note that the energy-momentum tensor is not necessarily symmetric. This is consistent with the presence of torsion which makes the Einstein tensor non-symmetric as well. 

\subsection{Recovery of the Einstein equations}
As we have now seen, the familiar forms for the stress-energy tensor of matter fields may be recovered from the first order Cartan-geometric formulation. Recall the conventional form of the Einstein equations in the metric formalism:

\begin{eqnarray}
\label{eingg}
R_{\mu\nu}-\frac{1}{2}Rg_{\mu\nu}=8\pi{\cal G}{\cal T}_{\mu\nu}-\Lambda g_{\mu\nu}
\end{eqnarray}
where ${\cal G}$ is the gravitational constant, $\Lambda$ is the cosmological constant, and ${\cal T}_{\mu\nu}$ is the stress-energy tensor of matter fields. The following action leads the familiar gravitational parts of the equations (\ref{eingg}):

\begin{eqnarray}
\nonumber S_{\zeta} &=& \int \left( \zeta_{0}\star\Sigma_{ABCD}F^{AB}\w F^{CD}+\zeta_{1}\star \Sigma_{CD}\w F^{CD}\right)
\end{eqnarray}
Of course, either one of these two terms leads to the term of the form $R_{\mu\nu}-\frac{1}{2}Rg_{\mu\nu}+\Lambda g_{\mu\nu}$ but we have kept both terms to keep the discussion general. 

We now consider the action $S_{\zeta}+S_{DC}+S_{HC}+S_{YMC}$. After some calculation it may be shown that the Einstein field equations take the form: 

\begin{eqnarray}
\mp4\mu(\zeta_{0}\pm \zeta_{1}\ell^{2})\left(R_{\mu\nu}-\frac{1}{2}Rg_{\mu\nu}\right) &=&  \frac{\kappa_{HC}\mu\ell^2}{4}\left(D_\mu\Phi^\dagger D_\nu \Phi+D_\nu\Phi^\dagger D_\mu \Phi-g_{\mu\nu}(D^\rho\Phi^\dagger D_\rho\Phi+U)\right)\nonumber \\
&& + \frac{\kappa_{YMC}}{2\mu}Tr\left(G_{\mu\rho }G_\nu^{\ph \nu\rho}-\frac{1}{4}g_{\mu\nu} G_{\alpha\beta}G^{\alpha\beta}\right)\nonumber\\
&& -\frac{\kappa_{DC}\mu\ell^{2}}{2}\left[e_{ K\nu }\frac{i}{2}\left(\bar{\psi}\gamma^{K}D^{(\omega)}_{\mu}\psi-D^{(\omega)}_{ \mu}\bar{\psi}\gamma^{K}\gamma\right) \right.\nonumber \\
&& - \left. g_{\mu\nu}\left(\frac{i}{2}e^{\sigma}_{K}
\left(\bar{\psi}\gamma^{K}D^{(\omega)}_{\sigma}\psi-D_{\sigma}\bar{\psi}\gamma^{K}\psi\right)-m\bar{\psi}\psi\right)\right]\nonumber\\
&& - 12\mu(\zeta_{0}\pm 2\zeta_{1}\ell^{2})\frac{1}{\ell^{2}}g_{\mu\nu} \label{ggg}
\end{eqnarray}
In  \ref{convs} we detail our conventions for matter actions in the standard formalism for gravity i.e. where gravity is described by co-tetrad $e^{I}$ and spin-connection $\omega^{IJ}$. Requiring that equation (\ref{ggg}) is of identical form to  the Einstein field equations (\ref{convs2}) of \ref{convs} then fixes the values of $\Lambda$ and $8\pi {\m G}$ in terms of the seven constants  $\{\zeta_{0},\zeta_{1},\kappa_{HC},\kappa_{YMC},\kappa_{DC},\mu,\ell\}$ as well as places restrictions upon the relative values of these constants. By inspection we have that:

\begin{eqnarray}
\Lambda &=& \mp \frac{3}{\ell^{2}}\frac{(\zeta_{0}\pm 2\zeta_{1}\ell^{2})}{(\zeta_{0}\pm \zeta_{1}\ell^{2})} \quad\quad\quad 8\pi{\cal G} = \mp \frac{\kappa_{HC}\ell^{2}}{16(\zeta_{0}\pm\zeta_{1}\ell^{2})}= \mp \frac{\kappa_{YMC}}{8\mu^{2}(\zeta_{0}\pm\zeta_{1}\ell^{2})}
\end{eqnarray}
and that 

\begin{eqnarray}
\kappa_{HC} &=& -2\kappa_{DC} =  \left(\frac{2}{\mu^{2}\ell^{2}}\right)\kappa_{YMC}
\end{eqnarray}
We clearly have too many unknowns to be able to determine the constants $\{\zeta_{0},\zeta_{1},\kappa_{HC},\kappa_{YMC},\kappa_{DC},\mu,\ell\}$ uniquely. Of course, one parameter is an overall factor multiplying all the terms in the action and can easily be dropped. For example we may put the dimensionless constant $\zeta_0$ equal to one. We may also reduce the number of constants by putting either $\zeta_0$ or $\zeta_1$ equal to zero since both of them yields the Palatini term. However, we have deliberately kept things general.
%
\section{A unification of a $U(1)$ gauge field and gravity}\label{unify}
%
Let us now take a step back and consider the results of the previous sections. By insisting on two principles (the gauge principle and polynomial simplicity) we were able to formulate matter fields in a way that is harmonious with Cartan geometry where the spacetime geometry is in part represented by a standard gauge connection $A^{AB}$. One of the chief motivations for reformulating the matter actions was that gravity stood out from the Yang-Mills fields of particle physics both in the way it couples to matter (which did not follow the standard gauge prescription) and the presence of non-polynomial structure of matter actions. Although the reformulation of scalar and spinor fields proceeded in a straightforward manner, the adherence to these two principles enforced a peculiar mathematical structure of the Yang-Mills fields. In particular, the gauge principle required us to attach a rolling index to the gauge field $B\rightarrow B^A$. Thus, this object is no longer a standard Yang-Mills field especially 
since it does not transform as one under gauge transformations. Instead we found that $B^A$ had to transform in a skewed way under gauge transformations according to \eqref{batrans}. Furthermore, the Yang-Mills Cartan field $B^A$ has two (suppressed) Yang-Mills indices but only one rolling index. Thus, we see that requiring gravity to behave like a standard gauge field enforces a peculiar skewed mathematical representation of the standard Yang-Mills fields of particle physics. Furthermore, we see that the orthogonal components $B^I$ exhibits the same peculiarity as the co-tetrad $e^I$ (see Section \ref{intr}): $B^I$ is a one-form but does not transform as a gauge connection. Thus, the original problem seems to have mutated into a slightly different form plaguing instead the Yang-Mills fields. Of course, we could simply accept this as the appropriate mathematical structure of Yang-Mills fields within a Cartan-geometric formulation, but we are then back to the question original why gravity should behave in a 
different way than other Yang-Mills fields.

In this section we shall see that these peculiarities can be overcome in the case of a $U(1)$ gauge field coupled to gravity, by means of unification. As we shall now see, if we write both the Yang-Mills-Cartan and gravitational actions side to side, this total action has the mathematical structure of a spontaneously broken gauge theory with $SO(1,5)$ symmetry.
Whether the following approach may be generalized to the incorporation of non-Abelian gauge fields or whether coupling to fermions can be done using minimal coupling, are open questions. We will, for now, refrain from considering the coupling of this gauge theory to matter. For further details about spinor representations of $SL(2,\mathbb{H}) \simeq SO(1,5)$ see \cite{Morita:2007vc}.
\subsection{A generalization of Cartan geometry}
Let us now turn to the details of this speculative unification of gravity with a $U(1)$ gauge field. For the usual de Sitter or anti-de Sitter Cartan gravity we ask: what subgroup leaves the contact vector $V^{A}$ invariant? The answer is of course the Lorentz group $SO(1,3)$. On the other hand, it makes little sense to ask: what subgroup leaves an arbitrary tangent vector $u^A$ invariant? A tangent vector satisfies $V^Au_A=0$. Since any tangent vector $u^A=tT^A+xX^A+yY^A+zZ^A$ can be expanded in a basis $\{T^A,X^A,Y^A,Z^A\}$ orthogonal to $V^A $ this is equivalent to asking what subgroup leaves a set of four basis vectors  $\{T^A,X^A,Y^A,Z^A\}$ invariant. This subgroup is $SO(1)$ which has zero dimensions.

However, if the gauge group is enlarged the situation is different. Consider then $SO(1,5)$ as an extension beyond $SO(1,4)$ and $SO(3,3)$ beyond $SO(3,2)$. In order to break the $SO(3,3)/SO(1,5)$ group down to a residual subgroup $SO(1,3)$ we must again introduce symmetry breaking fields. We clearly need two so let us denote those $V^{\m A}$ and $W^{\m A}$, where Calligraphic indices $\m A,\m B,\dots$ go from $0$ to $5$. In the case of $SO(3,3)$ both contact vectors are assumed to be time-like and, in the case of $SO(1,5)$, spacelike. In addition these two contact vectors will be assumed to be orthogonal, i.e. $V^{\m A}W_{\m A}=0$. The subgroup of transformations leaving both contact vectors invariant is then the Lorentz group. However, since we have a larger group we can now also ask the reverse question: what subgroup of transformations leave an arbitrary tangent vector $u^A$ invariant? Again, this is the same question as asking what subgroup leave all four basis vectors $\{T^{\m A},X^{\m A},Y^{\m A},Z^{\
m A}\}$ invariant. This is clearly the group $SO(2)\simeq U(1)$.

Thus, we have the $U(1)$ group appearing as a subgroup of $SO(3,3)/SO(1,5)$. As noted above only $SO(1,3)$ transformations leave both contact vectors invariant thus also destroying the desired $U(1)$ invariance. However, as we shall see, if the contact vectors appear in the action together and anti-symmetrically, i.e. as $V_{[\m A}W_{\m B]}$, then the residual symmetry of the action upon symmetry breaking will be $SO(1,3)\times U(1)$.

Furthermore, the $SO(3,3)$/$SO(1,5)$ group is a fifteen dimensional Lie-group. This coincides exactly with the expected number of one-form fields we need. We have ten for Cartan gravity and five for Cartan-Maxwell theory (see Section \ref{ymc}), which adds up to fifteen \cite{Greenwald}. Naively we could have thought that a $U(1)$ field is represented by a single one-form connection $B$ and not five. However, as a consequence of the gauge principle a $U(1)$ field is represented by an object $B^A$ which indeed contains five one-forms. Thus the counting adds up and this raises the possibility that we can unify the $U(1)$ gauge field and gravitation into a single $SO(3,3)$/$SO(1,5)$ connection.

To explore this idea further let us work in a gauge where $V^{\m A}\overset{*}{=}(0,0,0,0,\ell,0)$ and $W^{\m A}\overset{*}{=}(0,0,0,0,0,\mu)$. Here $\mu$ is a constant with dimension of mass and $\ell$ is the constant appearing in the standard Cartan gravity based on the de Sitter group $SO(2,3)$ or $SO(1,4)$. We can then suspect that in the adapted gauge the components of the gravitational and electromagnetic fields can be organized as follows
\begin{eqnarray}\label{unifiedconnection}
\m A^{\m A\m B}\overset{*}{=}\left(\begin{array}{ccc}\omega^{IJ}&\frac{e^I}{\ell}&\frac{B^I}{\mu}\\-\frac{e^J}{\ell}&0&\frac{B}{\mu\ell}\\-\frac{B^J}{\mu}&-\frac{B}{\mu\ell}&0\end{array}\right)
\end{eqnarray}
Under a general $SO(3,3)$/$SO(1,5)$ we require the connection $\m A^{\m A\m B}$, which contains both the gravitational and $U(1)$ fields, to transform as a standard gauge connection, i.e.  
\begin{eqnarray*}
{\cal A}^{\m A}_{\phantom{\m A}\m B} &\rightarrow & U^{\m A}_{\phantom{\m A}\m C}{\cal A}^{\m C}_{\phantom{\m C} \m D}(U^{-1})^{\m D}_{\phantom{\m D}\m B} - dU^{\m A}_{\phantom{\m A}\m C} (U^{-1})^{\m C}_{\phantom{\m C}\m B}.
\end{eqnarray*}
This avoids the extravagant mathematical structure of the Yang-Mills-Cartan theory mentioned above as the unified field $\m A^{\m A\m B}$ can be regarded as a standard gauge field. Objects such as $V^{\m A}$ or $DW^{\m A}$ then transform homogeneously under gauge transformations.

We also note that the symmetric space (either $SO(3,3)$ or $SO(1,5)$) is a five-dimensional manifold, i.e. one more dimension than the spacetime manifold. This is in contrast to standard Cartan geometry in which the dimension of the model space we roll is the same as the manifold. Nevertheless, we shall here generalize the standard Cartan construction to allow for a higher dimensional model spacetime to be rolled on a manifold with smaller dimension.

This generalized Cartan geometry will automatically imply that there are gauge transformations that neither change the point of contact nor affect the tangent space. Such transformations would then possibly correspond to the gauge transformations of Yang-Mills theories, which we do not associate with any spacetime symmetry.
\subsection{Identifying $U(1)$ invariants}
In order to determine what kind objects might appear in the action without destroying $U(1)$ gauge invariance we will study infinitesimal gauge transformations. In the spin-$1$ representation we can write
\begin{eqnarray}
U^{\m A}_{\ph A\m B}=\delta^{\m A}_{\ph A\m B}-i\theta^{\m C\m D}(\m J_{\m C\m D})^{\m A}_{\ph A\m B}
\end{eqnarray}
The generators ${\cal J}_{\m A\m B}$ satisfy the usual algebra for an orthogonal groups
\begin{eqnarray}
[\m J_{\m A\m B},\m J_{\m C\m D}]=-i(\eta_{\m A\m C}\m J_{\m B\m D}-\eta_{\m A\m D}\m J_{\m B\m C}-\eta_{\m B\m C}\m J_{\m A\m D}+\eta_{\m B\m D}\m J_{\m A\m C})
\end{eqnarray}
and the spin-$1$ representation takes the form $(\m J_{\m A\m B})_{\m C}^{\ph C\m D}=i(\delta^{\m D}_{\m A}\eta_{\m B\m C}-\delta^{\m D}_{\m B}\eta_{\m A\m C})$. Here we shall isolate the generator $\m J_{45}$ to play the role of a $U(1)$ transformation which then takes on the form 
\begin{eqnarray}
(\m J_{45})_{\m C}^{\ph A\m D}=i(\delta^{\m D}_{4}\eta_{5\m C}-\delta^{\m D}_{5}\eta_{4\m C})
\end{eqnarray}
It is now straightforward to show that the connection $\m A^{\m A\m B}$ transforms under an infinitesimal $U(1)$ transformation as
\begin{eqnarray*}
\m A^{IJ}&\rightarrow& \m A^{IJ}\\
\m A^{I4}&\rightarrow& \m A^{I4}-\theta^{45} A^I_{\ph A 5}\\
\m A^{I5}&\rightarrow&\m A^{I4}+\theta^{45} A^I_{\ph A 4}\\
\m A^{45}&\rightarrow&\m A^{45}+d\theta^{45}\\
\end{eqnarray*}
and we see that the $A^{45}$ connection behaves like a $U(1)$ connection and $dA^{45}$ is a $U(1)$ invariant object. We also see that the components $\m A^{I4}$ and $\m A^{I5}$ rotate into each other under $U(1)$ transformations. Nevertheless, the objects $e^I=DV^I$ and $B^I=DW^I$ are not connections and are $U(1)$ invariants. In fact, any object which is a member of the tangent space (e.g. $V^I$, $e^I$, $R^{IJ}$,\dots), is invariant under $U(1)$ transformations by construction. Thus, a term in the action that contains the factor $\epsilon_{\m{ABCDEF}}V^{\m E}W^{\m F}$ is automatically a $U(1)$ invariant. However, since $(\m J_{45})^{\m A}_{\ph A\m B}V^{\m B}\neq 0\neq (\m J_{45})^{\m A}_{\ph A\m B}W^{\m B}$ we see the $U(1)$ symmetry is broken by the two contact vectors. However, it is easily shown that $V_{[\m A}W_{\m B]}$ is invariant under this $U(1)$ transformation and is therefore an allowed object in the action. \footnote{Alternatively we may take the symmetry breaking fields to consist of a contact 
point $V^{\m A}$ and a second object $U^{\m{AB}}$ which is a member of adjoint representation of the Lie algebra of $SO(1,5)$/$SO(3,3)$. The object $U^{\m{AB}}$ is then to be though of as the $U(1)$ generator which commutes with the Lorentz subgroup $SO(1,3)$ and also trivially with itself. The object $U^{\m{AB}}$ thus breaks the the group $SO(1,5)$/$SO(3,3)$ down to $SO(1,3)\times U(1)$}.

We can now better understand the skewed structure of the Yang-Mills-Cartan theory presented in Section \ref{ymc}. The Yang-Mills-Cartan field $B^A$ can be seen as the compound object $B^A=(DW^I,\frac{1}{\ell}\m A^{45})$. In particular, while the fifth component is part of the unified connection $\m A^{\m AB}$ and transforms inhomogeneously, the first four components are identified as the gauge-covariant derivative of part of the contact vector $W^{\m A}$. This mix of (parts of) connections $\m A^{45}$ and compound objects $DW^I$ seems to explain skewed transformation properties of the Yang-Mills-Cartan field $B^A$.
\subsection{Reproducing the standard Maxwell equations}
Now that we have identified $U(1)$ invariant objects we shall now show that we can reproduce the standard coupled Einstein-Maxwell equations by a suitable choice of action. The components of the curvature two-form $\m F^{\m A\m B}$ in this gauge are given by
\begin{eqnarray}
{\m F}^{IJ}&\overset{*}{=}&d\m A^{IJ}+\eta_{CD}\m A^{I\m C}\w A^{\m DJ}=d\m A^{IJ}+\m A^I_{\ph IK}\w A^{KJ}\mp\m A^{I4}\w A^{4J}\mp\m A^{I5}\w A^{5J}\nonumber\\
&=&R^{IJ}\pm\frac{1}{\ell^2} e^I\w e^J\pm\frac{1}{\mu^2}B^I\w B^J\\
{\m F}^{I4}&\overset{*}{=}&d\m A^{I4}+\eta_{CD}\m A^{I\m C}\w A^{\m D4}=d\m A^{I4}+\m A^I_{\ph IK}\w A^{K4}\mp\m A^{I5}\w A^{54}\nonumber\\
&=&\frac{1}{\ell}\left(D^{(\omega)}e^I\pm\frac{1}{\mu^2} B^I\w B\right)=\frac{1}{\ell}\left(T^I\pm\frac{1}{\mu^2} B^I\w B\right)\\
\nonumber{\m F}^{I5}&\overset{*}{=}&d\m A^{I5}+\eta_{CD}\m A^{I\m C}\w A^{\m D5}=d\m A^{I5}+\m A^I_{\ph IK}\w A^{K5}\mp\m A^{I4}\w A^{45}\\
&=&\frac{1}{\mu}\left(D^{(\omega)}B^I\pm\frac{1}{\ell^2} e^I\w B\right)=\frac{1}{\mu}\left(S^I\mp\frac{1}{\ell^2} e^I\w B\right)\\
{\m F}^{45}&\overset{*}{=}&d\m A^{45}+\eta_{CD}\m A^{4\m C}\w A^{\m D5}=d\m A^{45}+\m A^4_{\ph IK}\w A^{K5}=\frac{1}{\mu\ell}\left(dB-e^K\w B_K\right).
\end{eqnarray}
Note that whenever there is a $\pm$ or $\mp$, the top sign will correspond to the $SO(3,3)$ case whilst the bottom will correspond to the $SO(1,5)$ case. Let us then consider the following action
\begin{eqnarray}\label{standuni}
S_{1} &=& \xi \int \epsilon_{\m{ABCDEF}}V^{\m E} W^{\m F} e^{\m A}\w e^{\m B}\w {\cal F}^{\m C\m D} +\chi \int 
     V_{[\m A}W_{\m B ]}V_{[\m C}W_{\m D ]}{\cal F}^{\m A\m B}\w \m {\cal F}^{\m C\m D}
\end{eqnarray}
where the constant $\xi$ has dimensions of $L^{-2}$ and $\chi$ is dimensionless. If we write this action in the gauge $V^{\m A}\overset{*}{=}\ell\delta^{\m A}_4$ and $W^{\m A}\overset{*}{=}\mu\delta^{\m A}_5$ and make use of \eqref{unifiedconnection}, we obtain
\begin{eqnarray}\label{case1}
S_{1} &\overset{*}{=} &\int \xi\mu\ell\epsilon_{IJKL}e^{I}\w e^{J} \w {\cal F}^{KL}+\chi(\mu\ell)^{2} {\m F}_{45}\w {\m F}^{45} \\
      &=& (\mu\ell)\xi \int \epsilon_{IJKL}e^{I}\w e^{J} \w \left(R^{KL}\pm\frac{1}{\ell^{2}}e^{K}\w e^{L} \right)\nonumber \\
      && +\int\left(\pm\frac{\ell}{\mu}\xi\epsilon_{IJKL}e^{I}\w e^{J}\w B^{K} \w B^{L}-2\chi dB\w e_{J}\w B^{J}\right)\nonumber\\
       && +\chi\int\left( dB\w dB+e_{I}\w B^{I} \w e_{J}\w B^{J}\right).
\end{eqnarray}
By inspection the (\ref{case1}) indeed corresponds to the action for gravity coupled to a $U(1)$ gauge field up to the presence of two additional terms. The first term corresponds to the integral $\int dB \w dB$ which is a boundary term and may be omitted. The second term corresponds to $e_{J}\w B^{J} \w e_{I} \w B^{I}$. It might at first seem as this term would alter the dynamics of the $U(1)$ field $B$, but as we shall see below the effect of this term is simply to alter the relation between $e_{I}\w B^{I}$ and $*dB$ of equation (\ref{ebdb}) something. 

Let us now derive the field equations from the action \eqref{standuni} and show in detail how the Maxwell equations are reproduced. Varying the action with respect to the six-dimensional connection, which contains both the gravitational and $U(1)$ fields, yields the unified field equations:

\begin{eqnarray}
0&=&\xi V_{[\m D}\epsilon_{\m C]\m{ABGEF}}V^{\m E}W^{\m F}e^{\m A}\w \m F^{\m{BG}}\nonumber\\
&+&\xi\epsilon_{\m{ABCDEF}}(e^{\m E}W^{\m F}\w e^{\m A}\w e^{\m B}+V^{\m E}B^{\m F}\w e^{\m A}\w e^{\m B}+2V^{\m E}W^{\m F} T^{\m A}\w e^{\m B})\nonumber\\
&+&2\chi\left((e_{[\m A}W_{\m B]}+V_{[\m A}B_{\m B]})V_{[\m C}W_{\m D]}+V_{[\m A}W_{\m B]}(e_{[\m C}W_{\m D]}+V_{[\m C}B_{\m D]})\right)\w\m F^{\m{AB}}
\end{eqnarray}
where $T^{\m A}\equiv De^{\m A}$. As usual we can write these equations in the gauge where $V^{\m A}\overset{*}{=}\ell\delta^{\m A}_4$ and $W^{\m A}\overset{*}{=}\mu\delta^{\m A}_5$. This yields the following set of equations corresponding to $\{\m C=I,\m D=J\}$, $\{\m C=I,\m D=4\}$, $\{\m C=I,\m D=5\}$, and $\{\m C=4,\m D=5\}$,   
\begin{eqnarray}
0&=&-\xi\mu\ell\epsilon_{IJKL} T^K\w e^L\label{notorsion}\\
0&=&\pm\xi \mu\ell^2 \epsilon_{IJKL} e^J\w(R^{KL}\pm\frac{2}{\ell^2} e^K\w e^L)+\frac{\xi\ell^2}{\mu}\epsilon_{IJKL}e^J\w B^K\w B^L \nonumber\\
&&\mp\chi\ell B_I\w(G- e_J\w B^J)\label{einsteinfieldeq}\\
0&=&\xi \ell \epsilon_{IJKL} e^J\w e^K\w B^L\pm\chi\mu e_I\w G\mp\chi\mu e_I\w e_K\w B^K\label{BGrelation}\\
0&=&\pm \chi \m D(e_K\w B^K)\label{MWeq}
\end{eqnarray}
The first equation \eqref{notorsion} we recognize as the zero-torsion equation and apart from a different constant in front of it this is the same equation as in standard Cartan gravity. The second equation \eqref{einsteinfieldeq} we recognize as the Einstein field equations with some matter source with energy momentum three-form 
\begin{eqnarray}
{\cal S}_{I4}=\frac{\xi\ell^2}{\mu}\epsilon_{IJKL}e^J\w B^K\w B^L\mp\chi\ell B_I\w(G- e_J\w B^J)
\end{eqnarray}
The third equation \eqref{BGrelation} is identical to left equation of \eqref{YMCeqs} except for the term $e_I\w B^{I}\w e_J\w B^J$. However, as we shall now see, this term will not alter the standard second order Maxwell equations. If we consult  \ref{YMCcalc} we obtain the relation

\begin{eqnarray}\label{BGrelation2}
e_I\w B^I=\pm\frac{\chi\mu}{\xi^{2}\ell^{2}+\chi^{2}\mu^{2}}\left(\xi\ell*G\pm \chi\mu G\right)
\end{eqnarray}
which inserted in the fourth equation \eqref{MWeq} yields the standard Maxwell equations
\begin{eqnarray}
\pm\frac{\chi^{2}\mu}{\xi^{2}\ell^{2}+\chi^{2}\mu^{2}}\m D\left(\xi\ell*G\pm \chi\mu G\right)=\pm\frac{\chi^{2}\xi\mu\ell}{\xi^{2}\ell^{2}+\chi^{2}\mu^{2}}\m D*G=0
\end{eqnarray}
where we have used the identity $\m D G\equiv 0$. Thus, the effect of the extra term $e_{J}\w B^{J} \w e_{I} \w B^{I}$ only results in a rescaling of the Maxwell equation by a constant factor and not a change in the dynamics of $B$.
\subsection{Energy-momentum of the $U(1)$ field and the gravitational constant}
We have now seen that the $U(1)$ gauge field $B=V_AB^A$ indeed satisfies Maxwell's equations. Before we can be certain that we are dealing with a normal matter field we must check that this field has positive energy density. To do this we calculate the energy momentum tensor of the $U(1)$ field and check that comes with the right sign so that it is a field with positive energy density. We can immediately use the methods of Section \ref{stressen} to cast (\ref{einsteinfieldeq}) in the form 

\begin{equation}
\label{stanein}
R_{\mu\nu}-\frac{1}{2}Rg_{\mu\nu}= 8\pi {\m G} T^{U(1)}_{\mu\nu}-\Lambda g_{\mu\nu}
\end{equation}
where $\Lambda$ is the cosmological constant, ${\m G}$ is the effective gravitational constant, and $T^{U(1)}_{\mu\nu}$ is the stress energy tensor of the $U(1)$ gauge field. After a straightforward but quite tedious calculation it may be shown that equation (\ref{einsteinfieldeq}) is equivalent to (\ref{stanein}) with

\begin{eqnarray}
T^{U(1)}_{\mu\nu}&=&\left(G_{\mu\rho}G_{\nu}^{\phantom{\nu}\rho}-\frac{1}{4}G_{\alpha\beta}G^{\alpha\beta}g_{\mu\nu}\right)\\
\Lambda &=& \mp \frac{6}{\ell^{2}} \\
8\pi{\m G} &=& \mp \frac{1}{2\ell^{2}\left(\frac{\xi^{2}}{\chi^{2}}+\frac{\mu^{2}}{\ell^{2}}\right)}
\end{eqnarray}
Here we recognize the canonical energy-momentum tensor of a Yang-Mills field. We also see that the groups $SO(3,3)$ and $SO(1,5)$ are as usual associated with negative and positive cosmological constants. However, the group $SO(3,3)$ 
 is associated with a negative effective gravitational constant. This means that we are not dealing with a normal matter field, and in particular not a Maxwell field, since its energy density is negative. On the other hand the group $SO(1,5)$ implies a positive gravitational constant and the $U(1)$ field can be regarded as a proper matter field with positive energy density. Therefore, only the action (\ref{standuni}) with symmetry group $SO(1,5)$ may be regarded as a unification of gravity and a $U(1)$ matter gauge field.

To our knowledge, the unification, based on first order formulation, of a $U(1)$ field and gravity presented here has not yet been explored in the literature.  A more well explored idea is that of incorporating $\omega^{IJ}$ and $B$ into a single connection \cite{Nesti:2009kk,Percacci:2009ij,Nesti:2007ka,Lisi:2010td,Smolin:2007rx,TorresGomez:2010cd}. It would be interesting to see whether the first order unification could be extended beyond the $U(1)$ example considered here.

\subsection{A natural non-linear modification with non-minimal coupling}
In the pure gravitational action (\ref{actione}) the only difference between the action with only the $a_1$-term non-zero and an action with only the $b_1$-term non-zero, is the value of the cosmological constant which differs by a factor of two in the two cases. As we have seen above, an action rather similar to the $b_1$-action yields the Einstein field equations coupled to a $U(1)$ gauge field which reproduces Maxwell's equations. Consider then an action of $a_1$-type:
\begin{eqnarray}
S_{2} &=& \int \kappa_{{\m A}{\m B}{\m C}{\m D}}{\m F}^{{\m A}{\m B}}\w {\cal F}^{{\m C}{\m D}}             
\end{eqnarray}               
where
\begin{eqnarray}
\kappa_{\m A\m B\m C\m D} &=& \kappa_{0}\eta_{\m A\m C}\eta_{\m B\m D}+\kappa_{1}\epsilon_{\m A\m B\m C\m D\m E\m F}W^{[\m E}V^{\m F]}
  +\kappa_{2}\eta_{\m A\m D}V_{[ \m B}W_{\m C]}+\kappa_{3}V_{[\m A}W_{\m B]}V_{[\m C}W_{\m D]}\nonumber\\
  &&+\kappa_{4}V_{[\m A}W_{\m D]}V_{[ \m B}W_{\m C]}
\end{eqnarray}
This is the most general action quadratic in the curvature that is consistent with our assumption about the nature of the symmetry breaking. Given our conventions on units, it is seen that the constants $\kappa_{i}$ are dimensionless. We note that the action contains $V^{\m A}$ and $W^{\m A}$ only in the combination $V^{[\m A}W^{\m B]}$, therefore one may alternatively construct this action instead from an `internal two-form' $P^{\m A\m B}=-P^{\m B\m A}$ (no spacetime indices) along with the connection ${\m A}^{\m A\m B}$ as long as one may assume that $P^{\m A\m B}\overset{*}{\propto} \delta^{[\m A}_{4}\delta^{\m B]}_{5}$. Immediately we see that if $\kappa_{0}$ is a constant then it represents a  $Tr\int {\m F}\w {\m F}$ boundary term.  Now  It can be shown that given the assumed forms of $V^{\m A}$ and $W^{\m A}$ that only two of the $\kappa_{i}$ here have non-vanishing contributions. We take the remaining $\kappa_{i}$ to be constants and the action takes the form
\begin{eqnarray}
S_{2} &\overset{*}{=}& \int(\mu\ell)\kappa_{1}\epsilon_{IJKL}{\m F}^{IJ}\wedge {\m F}^{KL}+(\mu\ell)^{2}\kappa_{3}{\m F}^{45}\wedge {\m F}^{45} \\
&=& \pm\frac{\mu}{\ell}\kappa_{1}\int \epsilon_{IJKL}e^{I}\w e^{J}\w\left(2 R^{KL}\pm\frac{1}{\ell^{2}}e^{K}\wedge e^{L}\right)\nonumber \\
&&+\int\left(\frac{2\kappa_{1}}{(\mu\ell)}\epsilon_{IJKL}e^{I}\w e^{J}\w B^{K}\w B^{L}-2\kappa_{3}\left(e_{K}\wedge B^{K}\wedge dB -e^{K}\w B_{K}\w e^{L}\w B_{L}\right)\right)\nonumber \\
&& +(\mu\ell)\kappa_{1}\int\epsilon_{IJKL}\left(\pm 2 \frac{1}{\mu^{2}}R^{IJ}\wedge B^{K}\w B^{L}+\frac{1}{\mu^{4}}B^{I}\w B^{J}\w B^{K}\w B^{L}\right) \label{arg}
\end{eqnarray}
where the two boundary actions $\int \epsilon_{IJKL}R^{IJ}\w R^{KL}$ and $\int dB\w dB$ have been omitted in the final expression. We can see then from (\ref{arg}) that we may choose values of the constants such that the action (\ref{case1}) is recovered in the limit where the non-minimal coupling (i.e. the field $B^{I}$ coupling to curvature $R^{IJ}$) and quartic non-linear term  can be ignored. 

The counterpart of the Einstein equations yields the following result:

\begin{eqnarray}
\Lambda &=& \mp \frac{3}{\ell^{2}} \\
{\m G} &=& \mp \frac{1}{2}\frac{\ell^{2}}{\left(4\left(\frac{\kappa_{1}}{\kappa_{3}}\right)^{2}+\ell^{2}\mu^{2}\right)}
\end{eqnarray}
Therefore again it is the group $SO(3,3)$ which is disfavored. 
\section{Conclusions and discussion}\label{discus}
Gravity, although commonly considered as one of the four force-fields in nature, is within standard formulations notoriously different from the Yang-Mills gauge fields that govern the electromagnetic, weak, and strong interactions of particle physics. Chern pointed to this odd state of affairs when he wrote \cite[p. 437]{Baez:1995sj} 
\begin{quote}
{\em ``Electromagnetism is, as we have seen, a gauge field. That gravitation is a gauge field is universally accepted, although exactly how it is a gauge field is a matter still to be clarified."}
\end{quote} 
In this paper we began by exhibiting three key peculiar features of standard formulations of General Relativity that make it stand out from the other Yang-Mills fields. Then we set of to show how these peculiarities disappear once we adopt a Cartan-geometric formulation. We noted that the first peculiarity could be removed once we view gravity as gauge theory with a spontaneously broken gauge symmetry. The mathematical representation here consists of a $SO(2,3)$/$SO(1,4)$-valued gauge field $A^{AB}$ corresponding to the action of `rolling', and a symmetry breaking field $V^A$ visualized as the contact point between the manifold and the symmetric model spacetime we roll. The co-tetrad $e^I\overset{*}{=}DV^I$ was then seen as a compound object similar to $\m D\Phi$, and not connection, nor part of any connection, thus removing the first peculiarity. 

The second peculiarity of gravity stressed in the introduction is its non-polynomial structure: while matter fields appear in actions only polynomially, gravity commonly enters through determinants and inverses. While such non-polynomial structure can readily be removed in the pure gravitational sector by simply adopting a first order Palatini formulation based on the pair of one forms $\{e^I,\omega^{IJ}\}$, this non-polynomial structure reappears when coupling matter fields to gravity, with the notable exception of the fermionic fields. 

The bulk of this paper consisted of demonstrating how also this non-polynomial structure in the bosonic matter field actions can be removed by also removing a third peculiarity: the coupling of gravity to matter fields does not follow the gauge prescription and therefore stands out from the Yang-Mills fields of particle physics. However, when we insisted on that the coupling of gravity to matter fields should be done using the the gauge prescription, we saw that polynomial simplicity could be restored.\footnote{As is well-known, the appearance of non-polynomial structure in the Hamiltonian of phase space formulations complicates quantization \cite{Thiemann:2007zz}. We have achieved in this paper a manifestly polynomial formalism on configuration space which will easily carry over to a phase space formulation. However, the presence of second class constraints, which should be solved for a consistent 
quantization, might very well reintroduce non-polynomial structure and thus complicate quantization.} The Cartan-geometric matter actions we so obtained were fortunately very simple and could easily be obtained through some guess work. This is not too surprising since polynomial simplicity and gauge invariance limits the space of possible actions severely. A second important consequence of polynomial simplicity and gauge invariance is that all field equations must be first order partial differential equations. This puts all field equations on a simple cohesive  mathematical first order form with the second order equations reproduced only on-shell.

In order to consider the back-reaction of the matter fields onto gravity we introduced the spin-energy-momentum three-form $\m S_{AB}$ which collected the traditional energy-momentum tensor and the spin-density into a single object. It was then shown, upon adopting the gauge $V^A\overset{*}{=}\ell \delta^A_4$ that the spin-energy momentum of the bosonic fields was characterized by zero spin-density and the standard canonical energy-momentum tensor. Similarly for fermionic fields we found the usual spin-density and energy-momentum. This enabled us to fix the undetermined constants in front of the matter actions. 

It would then seem as if Cartan geometry clarifies, in a rather direct way, what kind of gauge field gravity is. Indeed, if we are to regard gravity as a standard Yang-Mills gauge field, then the pure gravitational field should correspond to the `rolling' connection $A^{AB}$. On such a view, the co-tetrad $e^I\overset{*}{=}DV^I$ does not purely represent the gravitational field as it also contains the symmetry breaking field $V^A$. We can compare this to the view adopted in the electroweak theory in which the electroweak field is thought of as a $SU(2)\times U(1)$ gauge connection and the Higgs field as a distinct physical field not to be thought of as an aspect of the electroweak force field. That gravity starts looking within a Cartan-geometric formulation much more like a Yang-Mills gauge field is, in hindsight, not too surprising. After all, Cartan geometry is based on precisely the same fiber-bundle structure (see e.g. \cite{Westman:2012xk}) that characterizes the mathematical structure of Yang-Mills 
fields. Therefore, if a common ground between gravitational and Yang-Mills actions is sough, then Cartan geometry provides a natural mathematical basis in terms of fiber bundles and spontaneous symmetry breaking.

This common mathematical ground naturally suggests that it might be fruitful to identify similarities and dissimilarities between Cartan gravity and the force fields of particle physics; in particular the electroweak theory. We note that both theories start off with a larger symmetry group, i.e. $SU(2)\times U(1)$ or $SO(2,3)$/$SO(1,4)$, only to be broken by the presence of a Higgs type field, i.e. $\Phi$ or $V^A$, leaving a remnant unbroken symmetry corresponding to the $U(1)$ symmetry of electrodynamics and the local Lorentz invariance of General Relativity. 

In this paper we have adopted for the sake of simplicity a non-dynamical approach in which the contact vector $V^A$ is not thought of as being subjected to non-trivial field equations. Instead, we regarded it as an {\em \`a priori} postulated object. Such an approach is also common in mathematical literature where the role of the symmetry breaking field is limited to defining a preferred section on the fiber bundle \cite{SharpeCartan,Wise:2011ab}. However, given the structural similarities between the contact vector $V^A$ and the  dynamical Higgs field $\Phi$ it is natural to suspect that a dynamical approach, with $V^A$ subject to a non-trivial equations of motion, is the more appropriate one. After all, if both $V^A$ and $\Phi$ play the role of symmetry breaking fields, it would be rather odd if one is dynamical and the other not.

In \cite{Westman:2012xk} we exhibited an action (corresponding to the $b_1$ and $c_1$ terms only from (\ref{actione}) being non-zero) for pure gravity which yielded consistent equations of motion for $V^A$ by varying the action with respect to it. It was shown that the equations corresponding to variation with respect to $A^{AB}$ and $V^A$ enforces {\em algebraically} that $V^{2}=\frac{3b_1}{c_1}$ under the assumption that $\epsilon_{ABCDE}V^{E}e^{A}\w e^{B}\w e^{C}\w e^{D}$ is non-vanishing. Although this shows that the constancy of $V^2$ can be established without the use of Lagrange multipliers it does not treat the contact vector $V^A$ as a genuine dynamical field as it was determined, not by a differential equation, but an algebraic equation.  

Consider then an action $S_{G+M}$ consisting of the sum of Yang-Mills, Higgs, fermionic, and a suitable polynomial gravitational action of the form \eqref{genaction}. Then the field equations $\m Q_A=0$ for $V^A$ are obtained by varying with respect to $V^A$:
\begin{eqnarray}
\delta_V S_{G+M}=\int \left[\frac{\partial \m L_{G+M}}{\partial V^A}-D\left(\frac{\partial \m L_{G+M}}{\partial e^A}\right)\right]\delta V^A\equiv\int \m Q_A \delta V^A
\end{eqnarray}
Since only the magnitude of $V^A$ is gauge independent we may take the dynamics of $V^{A}$ to be described by the field $V^{2}$. It may be expected that the dynamics of $V^{2}$ is determined by the projection of the equation of motion for $V^{A}$ along $V_{A}$ i.e. the equation ${\cal Q}_{A}V^{A}=0$. Schematically we may expect this equation to be of the following form:
\begin{eqnarray}
\label{vevo}
{\cal K}(V^{A},A^{AB},..)\w DV^{2} = {\cal J}(V^{A},A^{AB},..)
\end{eqnarray}
Where ${\cal K}$ and ${\cal J}$ are, respectively, a three-form and four-form constructed from the gravitational and matter fields and their covariant derivatives in a polynomial fashion. By way of example, the aforementioned action considered in \cite{Westman:2012xk} had ${\cal K}=0$ with the algebraic constraint on $V^{2}$ being enforced by ${\cal J}=0$. Clearly then dynamics of $V^{2}$, in the sense of being determined from a differential equation, may only follow when ${\cal K}\neq 0$. As mentioned above, if the contribution to ${\cal K}$ is purely from the gravitational sector, described by the action (\ref{actione}), it may be shown that the only potential contribution arises from the $b_{2}$ term and is proportional to the spacetime torsion $T^{I}$. Therefore in the absence of torsion, non-zero contributions to ${\cal K}$ must come from the matter sector. Even in the absence of non-vanishing ${\cal K}$, the matter sector may contribute significantly to the form of $V^{2}$ via contributions to ${\cal J}
$.\footnote{During the publication of this paper it has recently been shown in \cite{Dynamics} that by including a $b_2$ term we obtain a genuinely dynamical theory of $V^A$. Furthermore, if the $c_1$ term has the polynomial form $c_1=\gamma+\gamma_1 V^2$ then $V^2$ has a stable vacuum expectation value with its value determined by the constants $b_1,b_2,\gamma$ and $\gamma_1$. Matter fields were not included yet in this dynamical approach though. Nevertheless, the systematic reformulation carried out in this paper provides the appropriate mathematical prerequisites for including matter fields.}

We now comment briefly on the cosmological constant from the above perspective. In Cartan gravity the closest thing to an `integral over spacetime volume' $\int \sqrt{-g}d^{4}x$ comes via the $c_{1}$ term of (\ref{actione}) as well as contributions to the $a_{1}$ and $b_{1}$ actions. Each of these contributions are proportional to the familiar integral $\int \sqrt{-g}d^{4}x$ only when $V^{2}=const.$. In general we would however expect an epoch, or even spatially, dependent cosmological constant. It remains to see if it is possible to construct a Cartan-geometric theory in which $V^A$ is a genuine dynamical variable which is consistent with observations. In particular, such a dynamical theory must explain why $V^2$ is approximately constant at least where current observations support such a claim.

The implementation of the gauge principle and polynomial simplicity was achieved in a rather straightforward manner in this paper. We simply attached a rolling index on all matter fields, replaced the exterior derivatives with a gauge covariant ones, then proceeded to write down actions in terms of those quantities. However, we noted an peculiarity in the mathematical structure of Yang-Mills fields which is perhaps not too surprising since these are gauge connections themselves. The gauge transformation law of the Yang-Mills-Cartan field $B^A$ was `skewed' and contained the contact vector $V^A$. Secondly, the object $B^A$ is also a rather strange one: it contains two (suppressed) Yang-Mills indices but only one rolling index. Thirdly, it may be noted that gauge invariance is destroyed if $V^2$ is allowed to vary in spacetime. We then showed how these peculiarities could be removed by generalizing Cartan geometry in a natural way so as to accommodate a tentative unification of a $U(1)$ gauge field and the 
gravitational field. This required us to enlarge the gauge group from $SO(2,3)$/$SO(1,4)$ to $SO(3,3)$/$SO(1,5)$ and to employ two contact vectors $V^{\m A}$ and $W^{\m A}$ (or perhaps single anti-symmetric contact matrix $P^{\cal A\cal B}$). It was shown that actions exist such that the role of the contact fields was to break the gauge symmetry $SO(3,3)$/$SO(1,5)$, leaving the remnant symmetry $SO(1,3)\times U(1)$. By analyzing the stress-energy momentum tensor we found that the group $SO(3,3)$ yields a negative energy density for the $U(1)$ field and thereby rules it out as playing the role of the electromagnetic field which of course has positive energy density. Nevertheless, the group $SO(1,5)$ yields a positive energy density for the $U(1)$ field and seems therefore to be the more interesting group from a unification perspective. The group $SO(1,5)$ is however not usually associated with a unification of gravity and a $U(1)$ field. In fact, the group $SO(1,5)$ has $15$ generators while one would perhaps 
naively believe that such a unification would involve $10+1$ gauge fields. After all, within second order formulations we think of a $U(1)$ field as represented as a single one-form $B$. However, the Cartan-geometric formalism is with necessity a first order one in which a gauge field also carries a rolling index. This enlarged the number of one-form fields by a factor of five. Therefore, the counting matches: $10$ for the gravitational field and $5$ for the $U(1)$ gauge field.

We stress, that we cannot yet claim that this $U(1)$ field is either the electromagnetic one, or the $U(1)$ field of the electroweak theory associated with hypercharge, as we do not know how this field couples to the other fields of the standard model. It seems reasonable that a successful unification must involve a larger gauge group as to incorporate all the electroweak fields. Nevertheless, it should certainly be interesting to explore the new possibilities of unification that the Cartan-geometric formulation of matter fields opens up. 

On the conceptual level it seems important that we understand in more detail why there is an equivalence principle in the first place. Indeed, if gravity is but one type of Yang-Mills field, then why would we expect one of these fields to interact with matter fields in a universal manner? We shall leave a proper study of the equivalence principle within the Cartan-geometric formulation for future research.

The possibility of formulating both the gravitational and matter sectors without the use of inverses suggests the possibility that some singularities in General Relativity, associated with a degenerate co-tetrad, are an artifact of a specific non-polynomial formulation \cite{Horowitz:1990qb}. Indeed, within our formulation it seems plausible that a solution might be analytically continued through a hypersurface on which the co-tetrad is degenerate. However, singularities associated with curvatures `blowing up' will remain also within the first order Cartan-geometric formulation.

\section*{Acknowledgments}
We would like to thank Thomas Thiemann, Derek Wise, Florian Girelli, Ma\"it\'e Depuis, Juan Leon, and Friedrich Hehl for helpful discussions and encouragement. HW was supported by the Perimeter Institute-Australia Foundations (PIAF) program, the Australian Research Council grant DP0880860, and the CSIC JAE-DOC 2011 program.

\section{Bibliography}
\bibliographystyle{hunsrt}
\bibliography{references}

\begin{thebibliography}{10}

\bibitem{d'Inverno:1992rk}
R.~d'Inverno.
\newblock {Introducing Einstein's relativity}.
\newblock Book, Clarendon Press, 1992.

\bibitem{Zee:2003mt}
A.~Zee.
\newblock {Quantum field theory in a nutshell}.
\newblock Book, Princeton University Press, 2003.

\bibitem{Wald:1984rg}
Robert~M. Wald.
\newblock {General Relativity}.
\newblock Book, The University of Chicago Press, 1984.

\bibitem{Sotiriou:2006gp}
Thomas~P. Sotiriou and Stefano Liberati.
\newblock {Field equations from a surface term}.
\newblock {\em Phys.Rev.}, D74:044016, 2006, gr-qc/0603096.

\bibitem{Horowitz:1990qb}
Gary~T. Horowitz.
\newblock {Topology change in classical and quantum gravity}.
\newblock {\em Class.Quant.Grav.}, 8:587--602, 1991.

\bibitem{Pagels:1983pq}
Heinz~R. Pagels.
\newblock {Gravitational gauge fields and the cosmological constant}.
\newblock {\em Phys. Rev.}, D29:1690, 1984.

\bibitem{Westman:2012xk}
H.F. Westman and T.G. Zlosnik.
\newblock {Gravity, Cartan geometry, and idealized waywisers}.
\newblock 2012, 1203.5709.

\bibitem{Weinberg:1996kr}
Steven Weinberg.
\newblock {The quantum theory of fields. Vol. 2: Modern applications}.
\newblock 1996.

\bibitem{MacDowell:1977jt}
S.~W. MacDowell and F.~Mansouri.
\newblock {Unified Geometric Theory of Gravity and Supergravity}.
\newblock {\em Phys. Rev. Lett.}, 38:739, 1977.
\newblock [Erratum-ibid.38:1376,1977].

\bibitem{Stelle:1979va}
K.~S. Stelle and Peter~C. West.
\newblock {De Sitter gauge invariance and the geometry of the Einstein-Cartan
  theory}.
\newblock {\em J. Phys.}, A12:L205--L210, 1979.

\bibitem{Randono:2010cq}
Andrew Randono.
\newblock {Gauge Gravity: a forward-looking introduction}.
\newblock 2010, 1010.5822.

\bibitem{Wise:2006sm}
Derek~K. Wise.
\newblock {MacDowell-Mansouri gravity and Cartan geometry}.
\newblock {\em Class.Quant.Grav.}, 27:155010, 2010, gr-qc/0611154.

\bibitem{Wise:2009fu}
Derek~K. Wise.
\newblock {Symmetric space Cartan connections and gravity in three and four
  dimensions}.
\newblock {\em SIGMA}, 5:080, 2009, 0904.1738.

\bibitem{Wise:2011ab}
Derek~K. Wise.
\newblock {The geometric role of symmetry breaking in gravity}.
\newblock {\em J.Phys.Conf.Ser.}, 360:012017, 2012, 1112.2390.

\bibitem{SharpeCartan}
R.W Sharpe.
\newblock {Cartan's Generalization of Klein's Erlangen Program}.
\newblock 1997.
\newblock Book, Springer.

\bibitem{Kibble:1961ba}
T.W.B. Kibble.
\newblock {Lorentz invariance and the gravitational field}.
\newblock {\em J.Math.Phys.}, 2:212--221, 1961.

\bibitem{Gronwald:1995em}
Frank Gronwald and Friedrich~W. Hehl.
\newblock {On the gauge aspects of gravity}.
\newblock 1995, gr-qc/9602013.

\bibitem{Ali:2009ee}
S.A. Ali, C.~Cafaro, S.~Capozziello, and Ch. Corda.
\newblock {On the Poincare Gauge Theory of Gravitation}.
\newblock {\em Int.J.Theor.Phys.}, 48:3426--3448, 2009, 0907.0934.

\bibitem{Mandl:1985bg}
F.~Mandl and Graham Shaw.
\newblock {Quantum field theory}.
\newblock 1985.
\newblock Book, Wiley and Sons, 1984.

\bibitem{Krasnov:2011pp}
Kirill Krasnov.
\newblock {New Action Principle for General Relativity}.
\newblock {\em Phys. Rev. Lett.}, 106:251103, 2011, 1103.4498.

\bibitem{Freidel:2005sn}
Laurent Freidel, Djordje Minic, and Tatsu Takeuchi.
\newblock {Quantum gravity, torsion, parity violation and all that}.
\newblock {\em Phys.Rev.}, D72:104002, 2005, hep-th/0507253.

\bibitem{Westman:2007yx}
Hans Westman and Sebastiano Sonego.
\newblock {Coordinates, observables and symmetry in relativity}.
\newblock {\em Annals Phys.}, 324:1585--1611, 2009, 0711.2651.

\bibitem{Nicolis:2008in}
Alberto Nicolis, Riccardo Rattazzi, and Enrico Trincherini.
\newblock {The Galileon as a local modification of gravity}.
\newblock {\em Phys.Rev.}, D79:064036, 2009, 0811.2197.

\bibitem{Chow:2009fm}
Nathan Chow and Justin Khoury.
\newblock {Galileon Cosmology}.
\newblock {\em Phys.Rev.}, D80:024037, 2009, 0905.1325.

\bibitem{Appleby:2012ba}
Stephen~A. Appleby and Eric~V. Linder.
\newblock {Trial of Galileon gravity by cosmological expansion and growth
  observations}.
\newblock {\em JCAP}, 1208:026, 2012, 1204.4314.

\bibitem{Petti:2006ue}
R.~J. Petti.
\newblock {Translational spacetime symmetries in gravitational theories}.
\newblock {\em Class. Quant. Grav.}, 23:737--751, 2006.

\bibitem{Ikeda:2009xb}
Noriaki Ikeda and Takeshi Fukuyama.
\newblock {Fermions in (Anti) de Sitter Gravity in Four Dimensions}.
\newblock {\em Prog. Theor. Phys.}, 122:339--353, 2009, 0904.1936.

\bibitem{MirKasimov:1978ci}
R.M. Mir-Kasimov and I.P. Volobuev.
\newblock {Complex Quaternions and Spinor Representations of de Sitter Groups
  SO(4,1) and SO(3,2)}.
\newblock {\em Acta Phys.Polon.}, B9:91--105, 1978.

\bibitem{Wilczek:1998ea}
Frank Wilczek.
\newblock {Riemann-Einstein structure from volume and gauge symmetry}.
\newblock {\em Phys.Rev.Lett.}, 80:4851--4854, 1998, hep-th/9801184.

\bibitem{Coley:2011cd}
A.~Coley, J.~Brannlund, and J.~Latta.
\newblock {Unimodular Gravity and Averaging}.
\newblock 2011, 1102.3456.

\bibitem{PhysRev.54.1114}
R.~J. Duffin.
\newblock On the characteristic matrices of covariant systems.
\newblock {\em Phys. Rev.}, 54:1114--1114, Dec 1938.

\bibitem{Kemmer:1939zz}
N.~Kemmer.
\newblock {The particle aspect of meson theory}.
\newblock {\em Proc.Roy.Soc.Lond.}, A173:91--116, 1939.

\bibitem{Lunardi:1999jq}
J.~T. Lunardi, B.~M. Pimentel, and R.~G. Teixeira.
\newblock {Duffin-Kemmer-Petiau equation in Riemannian space-times}.
\newblock 1999, gr-qc/9909033.

\bibitem{Kanatchikov:1999ut}
Igor~V. Kanatchikov.
\newblock {On the Duffin-Kemmer-Petiau formulation of the covariant Hamiltonian
  dynamics in field theory}.
\newblock {\em Rept.Math.Phys.}, 46:107--112, 2000, hep-th/9911175.

\bibitem{Bogush:2007pw}
A.A. Bogush, V.V. Kisel, N.G. Tokarevskaya, and V.M. Red'kov.
\newblock {Duffin-Kemmer-Petiau formalism reexamined: Non-relativistic
  approximation for spin 0 and spin 1 particles in a Riemannian space-time}.
\newblock {\em Annales Fond.Broglie}, 32:355--381, 2007, 0710.1423.

\bibitem{HarishChandra:1947zz}
Harish-Chandra.
\newblock {On Relativistic Wave Equations}.
\newblock {\em Phys.Rev.}, 71:793--805, 1947.

\bibitem{Casana:2002fu}
R.~Casana, V.Y. Fainberg, J.T. Lunardi, B.M. Pimentel, and R.G. Teixeira.
\newblock {Massless DKP fields in Riemann-Cartan space-times}.
\newblock {\em Class.Quant.Grav.}, 20:2457, 2003, gr-qc/0209083.

\bibitem{Morita:2007vc}
Katsusada Morita.
\newblock {Quaternions, Lorentz Group and The Dirac Theory}.
\newblock 2007, hep-th/0701074.

\bibitem{Greenwald}
Susan~R. Greenwald.
\newblock {Two Plus Two is Not Five (Easy Methods to Learn Addition and
  Subtraction), 2006}.

\bibitem{Nesti:2009kk}
F.~Nesti and R.~Percacci.
\newblock {Chirality in unified theories of gravity}.
\newblock {\em Phys.Rev.}, D81:025010, 2010, 0909.4537.

\bibitem{Percacci:2009ij}
R.~Percacci.
\newblock {Gravity from a Particle Physicists' perspective}.
\newblock {\em PoS}, ISFTG2009:011, 2009, 0910.5167.

\bibitem{Nesti:2007ka}
Fabrizio Nesti and Roberto Percacci.
\newblock {Graviweak Unification}.
\newblock {\em J.Phys.A}, A41:075405, 2008, 0706.3307.

\bibitem{Lisi:2010td}
A.~Garrett Lisi, Lee Smolin, and Simone Speziale.
\newblock {Unification of gravity, gauge fields, and Higgs bosons}.
\newblock {\em J.Phys.A}, A43:445401, 2010, 1004.4866.

\bibitem{Smolin:2007rx}
Lee Smolin.
\newblock {The Plebanski action extended to a unification of gravity and
  Yang-Mills theory}.
\newblock {\em Phys.Rev.}, D80:124017, 2009, 0712.0977.

\bibitem{TorresGomez:2010cd}
Alexander Torres-Gomez, Kirill Krasnov, and Carlos Scarinci.
\newblock {A Unified Theory of Non-Linear Electrodynamics and Gravity}.
\newblock {\em Phys.Rev.}, D83:025023, 2011, 1011.3641.

\bibitem{Baez:1995sj}
J.~Baez and J.P. Muniain.
\newblock {Gauge fields, knots and gravity, 1995}.

\bibitem{Thiemann:2007zz}
Thomas Thiemann.
\newblock {Modern canonical quantum general relativity}.
\newblock 2001, gr-qc/0110034.

\bibitem{Dynamics}
H.F. Westman and T.G. Zlosnik.
\newblock {Gravity from dynamical symmetry breaking}.
\newblock 2013, 1302.1103.

\bibitem{audoin}
Claude Audoin and Bernard Guinot.
\newblock {The Measurement of Time: Time, Frequency and the Atomic Clock}.
\newblock Book, Cambridge University Press, 2001.

\bibitem{Nakahara:1990th}
M.~Nakahara.
\newblock {Geometry, topology and physics}.
\newblock Book, Adam Hilger, 1990.

\end{thebibliography}

\begin{appendix}
\section{Units and dimensions}\label{units}
In some of the literature the conventions regarding units and dimensions are not always clearly stated. However, as this will be important for the purposes of this paper, we provide here a discussion regarding the conventions we adopt in this paper. In particular, in order to write downs actions, which are required to have the same dimensions as $\hbar$, it is necessary to sort out the dimensions of the various objects, constants, and variables appearing in the actions. 

In this paper we choose to measure both length and duration in meters (i.e. how many meters a light ray has traveled during the time interval in question) so that the speed of light is dimensionless and numerically equal to one $c=1$. Similarly we will assume that Planck's constant is dimensionless and numerically equal to one $\hbar=1$ so that mass has the dimension of inverse length.  With $\hbar$ dimensionless we see that actions must be dimensionless too. The only dimension  remaining is that of length and we chose meter as the relevant unit.
\subsection{General philosophy regarding units and dimensions}
In pre-general relativistic theories it is standard practice to attach dimensions and units to Cartesian coordinates. For example, the time coordinate $t$ of special relativity has units of, say, seconds, and the spatial coordinates $(x,y,z)$ have units of metres. This is natural given the operational meaning these Cartesian coordinates enjoy in terms of spatio-temporal measurements. However, coordinates, in a theory which is invariant under general coordinate transformations (e.g. General Relativity), must necessarily be devoid of any operational meaning (we refer to \cite{Westman:2007yx} for a fuller account). In this paper we take the view that the units of length, duration, and mass are directly related to naturally occurring length-, or equivalently mass-, scales found in Nature. For example, all protons, at least as currently understood, have equal size, and the same is true for hydrogen atoms. Essentially, it is these naturally occurring units of length that enables us to theoretically define the 
notions of a standard ruler and a standard clock which engineers are trying to approximate \cite{audoin}. From this perspective it is clear that units and dimensions have nothing to do with the abstract labels of points in spacetime, the coordinates $x^\mu$. 

The position one takes on whether units should be attached to coordinates or not affects the conventions regarding the units and dimensions of the various fields appearing in gravitational physics. For example, it is commonplace  to assume that taking the derivative of an object decreases the length dimension with one unit. Specifically, if a tensor $T$ has dimension $L^\alpha$, then $\nabla T$ is often assumed to have dimension of $L^{\alpha-1}$. While that may be appropriate within special relativity it is not so within General Relativity. One reason is the above-mentioned lack of operational meaning of spacetime coordinates. A second reason, which complicates the issue even in special relativity, is that we often make use of angular coordinates which are dimensionless. For example, it is commonplace to attach dimensions of length to both the coordinates $t$ and $r$ in the Schwarzschild spherical coordinates while the angular coordinates $\theta$ and $\varphi$ are taken to be dimensionless. With such a 
`mixed' convention it is clear that $\nabla T$ cannot simply be assumed to have dimension of $L^{\alpha-1}$.

In this paper we shall insist on that coordinates in General Relativity are, without exception, dimensionless. This has the following consequences: Firstly, taking a derivative of some object can never change the dimensions of it; nor can integration. On the other hand, since proper time $d\tau$ along some worldline has immediate operational meaning in terms of the readings of ideal clocks, it is natural to attach the dimension $L^2$ to $d\tau^2=g_{\mu\nu}dx^\mu dx^\nu$. However, since the coordinates $x^\mu$ are dimensionless it is clear that the dimension of $g_{\mu\nu}$ is $L^2$. Furthermore, since $V^A$ has dimension $L$ (see \cite{Westman:2012xk}), and taking the derivative cannot change the dimension, then the co-tetrad $e^A$ must also have dimension $L$. This should be contrasted to different conventions adopted in the literature of Cartan geometry where the co-tetrad is sometimes taken to be dimensionless (see e.g. \cite{Randono:2010cq}). We also then see from the relation $g_{\mu\nu}=\eta_{AB}e^A_\mu e^B_\nu$ that $\eta_{AB}$ must be dimensionless. 

We note that this choice of length dimension also coincides with both the conformal weight of the metric and co-tetrad, i.e $g_{\mu\nu}\rightarrow e^{+2\Omega}g_{\mu\nu}$ and $e^I\rightarrow e^{+\Omega}e^I$. In fact, all fields variables in this paper are such that the objects conformal weight coincides with its dimension of length. For example, the Higgs field $\Phi$, which comes with dimensions of mass, i.e. it has dimension $L^{-1}$, has conformal weight $-1$. Furthermore, all connection fields, including both Yang-Mills fields $B$ and the gravitational rolling connection $A^{AB}$, have zero conformal weight and must therefore be dimensionless. Finally, spinor fields have the dimension $L^{-3/2}$ and also the same conformal weight.

\subsection{Tables of the dimensions of variables and constants}
The following table summarizes the dimensions of the various fields appearing in this paper
\begin{center}
\vspace{0.25cm}
\begin{tabular}{|l|l|}
\hline
{\bf Mathematical variable} & {\bf Dimension}\\
\hline $SO(2,3)$/$SO(1,4)$ contact vector: $V^A\overset{*}{=}\ell\delta^A_4$ & $+1$\\
\hline $SO(1,5)$ contact vector : $V^{\m A}\overset{*}{=}\ell\delta^A_4$ & $+1$\\
\hline $SO(1,5)$ contact vector : $W^{\m A}\overset{*}{=}\mu\delta^A_5$ & $-1$\\
\hline Co-tetrad: $e^A \equiv DV^A$ & $+1$\\
\hline Higgs field: $\Phi$ & $-1$\\
\hline Higgs-Cartan field: $\Phi^A$ & $-2$\\
\hline $SO(p,q)$ connections: $A^{AB}$,$\m A^{\m A\m B}$ & $\ \ 0$\\
\hline Yang-Mills field: $B$ & $\ \ 0$\\
\hline Yang-Mills-Cartan field: $B^A$ & $-1$\\
\hline Spinor field: $\psi$ & $-3/2$\\
\hline Exterior derivatives: $d,D,\m D$ & $\ \ 0$\\
\hline 
\end{tabular}
\vspace{0.25cm}
\end{center}
and the following table contains the dimensions of the various constants 
\begin{center}
\vspace{0.25cm}
\begin{tabular}{|l|l|}
\hline
{\bf Constant} & {\bf Dimension}\\
\hline Gravitational constant: $\m G$ & $+2$\\
\hline Planck's constant: $\hbar$ & $\ \ 0$\\
\hline Speed of light: $c$ & $\ \ 0$\\
\hline Size of model space: $\ell$& $+1$\\
\hline Cosmological constant: $\Lambda=\mp\frac{3}{\ell^2}$ & $-2$\\
\hline Cartan mass: $\mu$ & $-1$\\
\hline Action parameter: $a_1$ & $-1$\\
\hline Action parameter: $a_2$ & $-2$\\
\hline Action parameter: $b_1$ & $-3$\\
\hline Action parameter: $b_2$ & $-4$\\
\hline Action parameter: $c_1$ & $-5$\\
\hline Action parameter: $\zeta_0$ & $\ \ 0$\\
\hline Action parameter: $\zeta_1$ & $-2$\\
\hline Action parameter: $\xi$ & $-2$\\
\hline Action parameter: $\chi$ & $\ \ 0$\\
\hline Action parameter: $\kappa_i$ & $\ \ 0$\\
\hline 
\end{tabular}
\vspace{0.25cm}
\end{center}
\section{Conventions regarding Standard form of actions and energy-momentum tensors}
\label{convs}
In this paper we take the standard actions for the gravitational, Yang-Mills, complex Klein-Gordon, and Dirac fields to be
\begin{eqnarray}
S_{G}&=& \frac{1}{16\pi\m G}\int\left(R-2\Lambda\right)\sqrt{-g}d^{4}x\\
S_{YM}&=&-\frac{1}{4}Tr\int G_{\mu\nu}G^{\mu\nu}\sqrt{-g}d^{4}x\\
S_{KG}&=&-\int\left(g^{\mu\nu}\m D_\mu \Phi^\dagger \m D_\nu\Phi+U(|\Phi|^2)\right)\sqrt{-g}d^{4}x\\
S_{D}&=&\int \left(\frac{i}{2}e^\mu_I(\bar\psi \gamma^I \m D_\mu\psi-\m D_\mu\bar\psi \gamma^I \psi)-m \bar\psi\psi\right)e d^{4}x
\end{eqnarray}
with $e=det(e^I_\mu)=\sqrt{-det(g_{\mu\nu})}$. We stress that these actions are to be considered in the first order formalism and therefore the Ricci tensor $R_{\mu\nu}$ is to be considered as a function only of the spin-connection $\omega_{\mu}^{\phantom{\mu}IJ}$. We take the energy-momentum tensor to be defined by variation with respect to the tetrad, yielding the following field equations:

\begin{eqnarray}
R_{\mu\nu}-\frac{1}{2}g_{\mu\nu}R = 8\pi{\cal G}{\cal T}_{\mu\nu}-\Lambda g_{\mu\nu}
\label{convs1}
\end{eqnarray}
where
\begin{eqnarray}
{\cal T}_{\mu\nu} 
 &=& {\m D}_{\mu}\Phi^{\dagger}{\cal D}_{\nu}\Phi+{\m D}_{\nu}\Phi^{\dagger}{\m D}_{\mu}\Phi
 -\left(g^{\alpha\beta}{\m D}_{\alpha}\Phi^{\dagger}{\m D}_{\beta}\Phi+U\right)g_{\mu\nu} \nonumber \\
 && +Tr\left(G_{\mu\alpha}G_{\nu}^{\phantom{\nu}\alpha}-\frac{1}{4}g_{\mu\nu} G^{\alpha\beta}G_{\alpha\beta}\right) \nonumber \\
 && -\frac{i}{2}\left(\bar{\psi}\gamma^{I}{\cal D}_{\mu}\psi-{\cal D}_{\mu}\bar{\psi}\gamma^{I}\psi\right)e_{I\nu}
 +\left(\frac{i}{2}e^{\beta}_{I}\left(\bar{\psi}\gamma^{I}{\cal D}_{\beta}\psi-{\cal D}_{\beta}\bar{\psi}\gamma^{I}\psi\right)-m\bar{\psi}\psi\right)g_{\mu\nu} \label{convs2}
\end{eqnarray}
\section{Differential forms, the Hodge dual, and matter actions}\label{diffpre}
This paper rests heavily on the calculus of forms. This section will serve as a recapitulation of how matter actions can be expressed in the language of forms as well as fix the notation and conventions of this paper. Although we will use the concepts of co-tetrad $e^{I}_{\mu}$ and tetrad $e^{\mu}_{I}$ throughout this section it will be seen that the results suggest a path of approach in the context of Cartan geometry. For an introduction to the use of forms in Cartan geometry aimed at `tensor-minded' physicists see \cite{Westman:2012xk}.
\subsection{The Hodge dual}\label{Hodge}
The first useful concept is that of the Hodge dual. The basic idea is this: let $N$ be the dimension of the manifold and $0\leq p\leq N$, then the space $\Lambda^p$ of $p$-forms have the same dimension as the space $\Lambda^{N-p}$ of $(N-P)$-forms, i.e. the spaces have the dimension $\frac{N!}{p!(N-p)!}$. These spaces can therefore be regarded as dual to each other. Given an non-degenerate co-tetrad $e^I_\mu$ and a $p$-form $\Omega$ the Hodge dual $*\Omega$ is defined by:
\begin{eqnarray}
*\Omega=\frac{e}{p!(N-p)!}\epsilon_{\mu_1\mu_2\dots\mu_N}\Omega^{\mu_1\mu_2\dots\mu_p}dx^{p+1}\w \dots \w dx^{N} .
\end{eqnarray}
where the indices on $\Omega$ has been raised using the inverse metric $g^{\mu\nu}\equiv\eta^{IJ}e^\mu_Ie^\nu_J$. Using the Hodge dual we can also introduce a symmetric `inner-product' between two $p$-forms $\Omega_1$ and $\Omega_2$
\begin{eqnarray}
\langle \Omega_1|\Omega_2\rangle\equiv \Omega_1\wedge *\Omega_2=\Omega_2\wedge *\Omega_1
\end{eqnarray}
We note that the Hodge dual heavily rests on the existence of an inverse co-tetrad $e^\mu_I$ without which no natural isometry between $ \Lambda^p$ and $\Lambda^{N-p}$ exists. Furthermore, the Hodge dual is manifestly non-polynomial in the gravitational variables.
\subsection{Duality between forms and antisymmetric contravariant tensor densities}\label{duality}
There is however another form of duality which always exists: The completely antisymmetric Levi-Civita tensor density $\varepsilon^{\mu_1\mu_2\dots\mu_N}$ establishes an isometry between the space of $p$-forms and the space of completely antisymmetric $(N-p,0)$-rank tensor densities of weight $+1$. We will use the symbol $\sim$ to denote the dual quantity. Specifically, let $\Omega$ be some $p$-form, then the dual contravariant antisymmetric $+1$ tensor density $\Omega^{\mu_{p+1}\dots\mu_N}$ is defined as

\begin{eqnarray}
\Omega=\frac{1}{p!}\Omega_{\mu_1\dots\mu_p}dx^{\mu_1}\w \dots \w dx^{\mu_p}\sim\frac{1}{p!}\Omega_{\mu_1\dots\mu_p}\varepsilon^{\mu_1\dots\mu_p\dots\mu_N}
\end{eqnarray}
As a simple concrete example we can see that, in the case of four spacetime dimensions, the object dual to the four-form $\m E=\frac{1}{4!}\epsilon_{IJKL}e^I\w e^J\w e^K\w e^L$, is nothing but the usual scalar density volume element $e\equiv det(e^I_\mu)$, i.e. we have
\begin{eqnarray}
\m E=\frac{1}{4!}\epsilon_{IJKL}e^I\w e^J\w e^K\w e^L&=&\frac{1}{4!}\epsilon_{IJKL}e^I_\mu e^J_\nu e^K_\rho e^L_\sigma dx^\mu\w dx^\nu\w dx^\rho\w dx^\sigma\nonumber\\
&\sim& \varepsilon^{\mu\nu\rho\sigma}\frac{1}{4!}\epsilon_{IJKL}e^I_\mu e^J_\nu e^K_\rho e^L_\sigma\equiv det(e^I_\mu)=e
\end{eqnarray}
This duality between differential forms and contravariant antisymmetric tensor densities is useful since it allows us to translate between expressions written in differential forms forms and the more common tensorial notation which is more common within the physics community.
\subsection{Klein-Gordon field}
The action for a complex Klein-Gordon field with some `potential' $U(\phi)$, e.g. $U(\phi)=m^2|\phi|^2+\lambda |\phi|^4$, usually written as
\begin{eqnarray}
S_{KG}=-\int e\left(g^{\mu\nu}\partial_\mu\bar{\phi} \partial_\nu\phi +U(\phi)\right)d^4x
\end{eqnarray}
can be written as an integration over the four-form ($\m E\equiv \frac{1}{4!}\epsilon_{IJKL}e^I\w e^J\w e^K\w e^L$)
\begin{eqnarray}
S_{KG}=-\int \left(d\bar{\phi}\w *d\phi+\m EU(\phi)\right)=-\int \left(\langle d\bar{\phi}|d\phi\rangle+\m EU(\phi)\right)
\end{eqnarray}
where $\langle \Omega_1|\Omega_2\rangle\equiv \Omega_1\w *\Omega_2$ is the inner product associate with the Hodge dual \cite{Nakahara:1990th}.

To translate between the two actions we construct the scalar density dual to the four-form $d\bar{\phi}\w *d\phi+\m EU(\phi)$. We saw above that the scalar density dual to the volume form $\m E$ is the determinant $e$ and we only need to calculate the scalar density dual to $d\bar{\phi}\w *d\phi$:
\begin{eqnarray}
d\bar{\phi}\w *d\phi&\equiv & \partial_\mu\bar{\phi} dx^\mu\w \frac{e}{3!}\epsilon_{\kappa\nu\rho\sigma}g^{\kappa\tau}\partial_\tau\phi dx^\nu\w dx^\rho \w dx^\sigma=\frac{e}{3!}\epsilon_{\kappa\nu\rho\sigma}\partial_\mu\bar{\phi} g^{\kappa\tau}\partial_\tau\phi dx^\mu\w dx^\nu\w dx^\rho \w dx^\sigma\nonumber\\
&\sim&\frac{e}{3!}\epsilon_{\kappa\nu\rho\sigma}\partial_\mu\bar{\phi}  g^{\kappa\tau}\partial_\tau\phi\varepsilon^{\mu\nu\rho\sigma}=e\partial_\mu\bar{\phi} g^{\kappa\tau}\partial_\tau\phi\delta^\mu_\kappa =e g^{\mu\nu}\partial_\mu\bar{\phi}\partial_\nu\phi 
\end{eqnarray}
where we made use of the identity $\epsilon_{\kappa\nu\rho\sigma}\varepsilon^{\mu\nu\rho\sigma}=3!\delta^\mu_\kappa$. Thus, we have 
\begin{eqnarray}
d\bar{\phi}\w *d\phi+\m EU(\phi)\sim  e(g^{\mu\nu}\partial_\mu\bar{\phi}\partial_\nu\phi+U(\phi))
\end{eqnarray}
which is nothing but the usual Klein-Gordon Lagrangian density with potential $U(\phi)$.

Varying the Klein-Gordon action with respect to $\bar{\phi}$ and integrating by parts yields 
\begin{eqnarray}
\delta_{\bar{\phi}}S_{KG}=-\int \left(d\delta\bar{\phi}\w *d\phi+\m E\frac{\partial U(\phi)}{\partial \bar{\phi}}\delta\bar\phi\right)=-\int \delta\bar{\phi}\left(-d*d\phi+\m E\frac{\partial U(\phi)}{\partial \bar{\phi}}\right)
\end{eqnarray}
implying the four-form equation 
\begin{eqnarray}
-d*d\phi+\m E\frac{\partial U(\phi)}{\partial \bar{\phi}}=0
\end{eqnarray}
and varying with respect to $\phi$ using the identity $\langle \Omega_1|\Omega_2\rangle=\langle \Omega_2|\Omega_1\rangle$ yields the complex conjugate of that equation. Since 
\begin{eqnarray}
d*d\phi&=&d\left(\frac{e}{3!}\epsilon_{\kappa\nu\rho\sigma}g^{\kappa\tau}\partial_\tau\phi dx^\nu\w dx^\rho \w dx^\sigma\right)
=\partial_\mu\left(\frac{e}{3!}\epsilon_{\kappa\nu\rho\sigma}g^{\kappa\tau}\partial_\tau\phi\right)dx^\mu\w dx^\nu\w dx^\rho \w dx^\sigma\nonumber\\
&\sim&\partial_\mu\left(\frac{e}{3!}g^{\kappa\tau}\partial_\tau\phi\right)\epsilon_{\kappa\nu\rho\sigma}\varepsilon^{\mu\nu\rho\sigma}=\partial_\mu\left(e g^{\kappa\tau}\partial_\tau\phi\right)\delta^\mu_\kappa=\partial_\mu\left(e g^{\mu\nu}\partial_\nu\phi\right)=\square \phi
\end{eqnarray}
and $\frac{\partial U}{\partial \bar{\phi}}=m^2\phi+2\lambda |\phi|^2 \phi$, the above equation is just the Klein-Gordon equation $\square\phi-m^2\phi-2\lambda |\phi|^2 \phi=0$ with $\phi^4$ coupling.

\subsection{Yang-Mills field}
Let $B=B_{\mu}dx^\mu$ be some gauge connection with values in some Lie-algebra (e.g. $U(1)$ or $SU(2)$). For notational compactness these internal indices are suppressed. Let $G_{\mu\nu}=\partial_\mu B_\nu-\partial_\nu B_\mu+ig[B_\mu,B_\nu]$, or in the language of forms $G= dB+ig B\w B$, be the corresponding curvature two-form.\footnote{We note that if $B^\dagger=B$ then $(iB\w B)^\dagger=(iB_\mu B_\nu dx^\mu\w dx^\nu)^\dagger=-iB_\nu^\dagger B_\mu^\dagger dx^\mu\w dx^\nu=i B\w B$ so that $G^\dagger=G$.} The action for this gauge field, up to a constant, is commonly written as
\begin{eqnarray}
S_{YM}=-\int Tr\frac{e}{4} g^{\mu\rho}g^{\nu\sigma}G_{\mu\nu}G_{\rho\sigma}d^{4}x
\end{eqnarray}
\\
where $Tr$ denotes taking a trace over the internal indices.
Using the Hodge dual, this action can be written as
\begin{eqnarray}
S_{YM}=-\int Tr\frac{1}{2}G\w *G
\end{eqnarray}
As before the equivalence with the previous action can be established by considering the dual scalar density
\begin{eqnarray}
\frac{1}{2}G\w *G&=&\frac{1}{4}G_{\mu\nu}dx^\mu\w dx^\nu\w\frac{e}{2!2!}\epsilon_{\kappa\tau\rho\sigma}g^{\kappa\alpha}g^{\tau\beta}G_{\alpha\beta}dx^\rho\w dx^\sigma\nonumber\\
&=&\frac{e}{16}G_{\mu\nu}\epsilon_{\kappa\tau\rho\sigma}g^{\kappa\alpha}g^{\tau\beta}G_{\alpha\beta}dx^\mu\w dx^\nu\w dx^\rho\w dx^\sigma
\sim\frac{e}{16}G_{\mu\nu}\epsilon_{\kappa\tau\rho\sigma}g^{\kappa\alpha}g^{\tau\beta}G_{\alpha\beta}\varepsilon^{\mu\nu\rho\sigma}\nonumber\\
&=&\frac{e}{8}G_{\mu\nu}g^{\kappa\alpha}g^{\tau\beta}G_{\alpha\beta}(\delta^\mu_\kappa\delta^\nu_\tau-\delta^\mu_\tau\delta^\nu_\kappa)
=\frac{e}{4}g^{\mu\rho}g^{\nu\sigma}G_{\mu\nu}G_{\rho\sigma}
\end{eqnarray}
The equations of motion are obtained by varying with respect to the gauge field $B$, integrating by parts, and making use of the identity $\Omega_1\w *\Omega_2=\Omega_2\w *\Omega_1$ yielding
\begin{eqnarray}
\delta_B S_{YM}&=&-\delta_B \int Tr\frac{1}{2}\m G\w *\m G=-\int Tr\frac{1}{2}\left(\m D\delta B\w *G+G\w *\m D\delta B\right)\\
&=&-\int Tr\frac{1}{2}\left(\m D\delta B\w *G+\m D\delta B\w *G\right)=-\int Tr\delta B\w \m D*G 
\end{eqnarray}
which implies the equations of motion $\m D *G=0$. To translate between the forms to notation to the more common tensorial notation we consider the dual vector density and make use of the identity $\epsilon_{\mu\nu\rho\sigma} \varepsilon^{\kappa\rho\sigma\tau}=2!(\delta_\mu^\kappa\delta_\nu^\tau-\delta_\mu^\tau\delta_\nu^\kappa)$
\begin{eqnarray}
\m D* G&=&\m D(\frac{e}{2!2!}\epsilon_{\mu\nu\rho\sigma}g^{\mu\alpha}g^{\nu\beta}G_{\alpha\beta}dx^\rho\w dx^\sigma)=\frac{1}{4}\m D_\kappa(e\epsilon_{\mu\nu\rho\sigma}g^{\mu\alpha}g^{\nu\beta}G_{\alpha\beta})dx^\kappa\w dx^\rho\w dx^\sigma\nonumber\\
&\sim &\frac{1}{4}\m D_\kappa(eg^{\mu\alpha}g^{\nu\beta}G_{\alpha\beta})\epsilon_{\mu\nu\rho\sigma} \varepsilon^{\kappa\rho\sigma\tau}=\frac{1}{2}(\delta_\mu^\kappa\delta_\nu^\tau-\delta_\mu^\tau\delta_\nu^\kappa)\m D_\kappa(eg^{\mu\alpha}g^{\nu\beta}G_{\alpha\beta})\nonumber\\&=&\m D_\kappa G^{e\kappa\tau}=0 \label{ymeq}
\end{eqnarray}
and we see that the three-form equation $\m D*G=0$ is nothing but the standard Maxwell-Yang-Mills equations $\m D_\kappa G^{\kappa\tau}=0$. 
\subsection{Dirac field}\label{standDirac}
The standard action for a Dirac field in tensorial notation is given by
\begin{eqnarray}
S_D=\int \left[\frac{i}{2}e e^\mu_I (\bar\psi\gamma^I\m D_\mu\psi-\m D_\mu\bar\psi\gamma^I\psi)-em\bar\psi\psi \right]d^4x
\end{eqnarray}
where $\m D_\mu\psi=\partial_\mu\psi-\frac{i}{2}\omega_\mu^{\ph{\mu} IJ}\m J_{IJ}\psi-g B_\mu\psi$ with $\omega_\mu^{\ph\mu IJ}$ the spin-connection, $B_\mu$ some suitable gauge field and $\m J_{IJ}=-\frac{i}{4}[\gamma_I,\gamma_J]$. The presence of an inverse $e^\mu_I$ naively suggests that the Dirac action is non-polynomial in the gravitational variables just as the Klein-Gordon and Yang-Mills actions. However, this is not the case which can be readily seen by rewriting it using he definition of the co-tetrad  inverse \eqref{tetrad}
\begin{eqnarray}
S_D=\int \left[\frac{1}{3!}\varepsilon^{\mu\nu\rho\sigma}\epsilon_{IJKL}e^J_\nu e^K_\rho e^L_\sigma\frac{i}{2} (\bar\psi\gamma^I\m D_{\mu}\psi-\m D_{\mu}\bar\psi\gamma^I\psi)-em\bar\psi\psi\right]d^4x
\end{eqnarray}
Thus, since the determinant $e$ is manifestly polynomial, we see that the co-tetrad enters only polynomially in the Dirac action. Secondly, We may also note that no Hodge dual is present which in fact was the source of the non-polynomial structure of both the Klein-Gordon and Yang-Mills actions. Thirdly, we note that the Dirac equation is a first order partial differential equation in contrast to the standard Klein-Gordon and Yang-Mills equations. In fact, as is shown in Section \ref{rolling}, the gauge principle and polynomial simplicity forces all equations to be of first order.

Using the symbol for the volume element $\m E=\frac{1}{4!}\epsilon_{IJKL} e^I\w e^J\w e^K\w e^L$ the dual four-form to the Dirac Lagrangian scalar density is easy to read off and is given by
\begin{eqnarray}
S_D=\int -\frac{1}{3!}\epsilon_{IJKL}e^J\w e^K\w e^L\w \frac{i}{2}(\bar\psi\gamma^I\m D\psi-\m D\bar\psi \gamma^I\psi)-\m E m\bar\psi\psi
\end{eqnarray}
in which the mathematically elegant polynomial structure of the Dirac action is manifest.

The four-form equations of motion are obtained by varying with respect to $\bar\psi$ (and the complex conjugate equations with respect to $\psi$) 
\begin{eqnarray}
\delta_{\bar\psi}S_D&=&\delta_{\bar\psi}\int-\frac{1}{3!}\epsilon_{IJKL}e^J\w e^K\w e^L\w \frac{i}{2}(\bar\psi\gamma^I\m D\psi-\m D\bar\psi \gamma^I\psi)-\m Em\bar\psi\psi\nonumber\\
&=&\int-\frac{1}{3!}\epsilon_{IJKL}e^J\w e^K\w e^L\w \frac{i}{2}(\delta\bar\psi\gamma^I\m D\psi-\m D\delta\bar\psi \gamma^I\psi)-\m Em\delta\bar\psi\psi\nonumber\\
&=&\int\delta\bar\psi\left[-\frac{1}{3!}\epsilon_{IJKL}e^J\w e^K\w e^L\w \frac{i}{2}\gamma^I\m D\psi+\frac{1}{3!}\epsilon_{IJKL}\m D(e^J\w e^K\w e^L\w \frac{i}{2}\gamma^I\psi)-\m Em\psi\right]\nonumber\\
&=&\int\delta\bar\psi\left[-\frac{1}{3!}\epsilon_{IJKL}e^J\w e^K\w e^L\w i\gamma^I\m D\psi-\m Em\psi+\frac{i}{4}\epsilon_{IJKL}T^J\w e^K\w e^L\gamma^I\psi\right]\nonumber
\end{eqnarray}
which implies the equations of motion
\begin{eqnarray}
\frac{i}{3!}\epsilon_{IJKL}e^J\w e^K\w e^L\w \gamma^I\m D\psi+\m Em\psi=\frac{i}{4}\epsilon_{IJKL}T^J\w e^K\w e^L\gamma^I\psi.
\end{eqnarray}
Whenever spacetime torsion is zero $T^I=0$ this equation is the usual Dirac equation in curved spacetimes which can be checked by constructing the dual scalar density
\begin{eqnarray}
\frac{i}{3!}\epsilon_{IJKL}e^J\w e^K\w e^L\w \gamma^I\m D\psi+\m E m\psi\sim e(ie^\mu_I\gamma^I\m D_\mu\psi-m\psi)=0.
\end{eqnarray}
However, whenever spacetime torsion, which is induced by spin-density $\m S_{IJ}$, is non-zero, then the Dirac equation is modified by the extra term $\frac{i}{4}\epsilon_{IJKL}T^J\w e^K\w e^L\gamma^I\psi\sim e\frac{i}{2}T^J_{IJ}\gamma^I\psi$ with $T^K_{IJ}=e^\mu_Ie^\nu_JT^K_{\mu\nu}$. For a `non-minimal coupling' generalization we refer to \cite{Freidel:2005sn}.

\section{Relation between $G$ and $e_A\w B^A$}\label{YMCcalc}
The Yang-Mills-Cartan action \eqref{sym} and the unified action \eqref{standuni} yields for the Yang-Mills field an equation of the form
\begin{eqnarray}\label{geneq}
a \epsilon_{IJKL} B^J\w e^K\w e^L+b e_I\w G+c e_I\w e^J\w B_J=0
\end{eqnarray}
imposing a relation between the curvature two-form $G$ and $e_I\w B^I$ which plays an important role when reproducing the standard second order formalism. This appendix provides the necessary calculations to establish the exact form of that relation.

First we rewrite equation \eqref{geneq} by defining $B^J=B_M^{\ph MJ}e^M$ and $G=\frac{1}{2}G_{KL}e^K\w e^L$
\begin{eqnarray}
a \epsilon_{IJKL} B_M^{\ph MJ}e^M\w e^K\w e^L+\frac{b}{2}G_{JK} e_I\w e^J\w e^K+c e_I\w e^J\w e^K B_{KJ}=0.
\end{eqnarray}
Then we construct the dual vector density using $e^I\w e^J\w e^K\sim e \varepsilon^{IJKL}e^\sigma_L$ yielding the equation
\begin{eqnarray}
a \epsilon_{IJKL} B_M^{\ph MJ}\varepsilon^{MKLN}e^\sigma_N+\frac{b}{2}G_{JK} \varepsilon_I^{\ph I JKL}e^\sigma_L+c \varepsilon_I^{\ph IJKL}e^\sigma_L B_{KJ}=0\end{eqnarray}
which using the identity $\epsilon_{IJKL}\varepsilon^{MKL}=2(\delta_I^M\delta^N_J-\delta_J^M\delta^N_I)$ simplifies to 
\begin{eqnarray}\label{BIJeq}
2a B_{IJ}-c\varepsilon_{IJKL}B^{KL}+\frac{b}{2}\varepsilon_{IJKL}G^{KL}=0.
\end{eqnarray}
In order to solve for $B_{IJ}$ we dualize this equation with $\varepsilon_{MNIJ}$ which yields
\begin{eqnarray}
2a\varepsilon_{MNIJ}B^{IJ}+\frac{b}{2}\varepsilon_{MN}^{\ph{MN}IJ}\varepsilon_{IJ}^{\ph{IJ}KL}G_{KL}-c\varepsilon_{MN}^{\ph{MN}IJ}\varepsilon_{IJ}^{\ph{IJ}KL}B_{KL}=0.
\end{eqnarray}
By reusing \eqref{BIJeq} and making use of the identity $\varepsilon_{IJKL}\varepsilon^{MNKL}=-2(\delta_I^M\delta^N_J-\delta_J^M\delta^N_I)$ we arrive at 
\begin{eqnarray}
B_{IJ}=-b\frac{a\varepsilon_{IJKL}G^{KL}-2c G_{IJ}}{4(a^2+c^2)}
\end{eqnarray}
or equivalently
\begin{eqnarray}
e^I\w B_I=b\frac{-a*G-cG}{a^2+c^2}.
\end{eqnarray}
\end{appendix}

\section{Geometric interpretation of Cartan gravity}\label{cartangeometry}
We have discussed formulations of gravitation where the gravitational field is described by the pair $\{V^{A},A^{AB}\}$,
and recovery of familiar gravitational physics occurs when $V^{2}= \mp \ell^{2}=const.$. Furthermore, we have emphasized that the first order Palatini formulation is recovered only when the $SO(2,3)/SO(1,4)$ symmetry is spontaneously broken i.e. $V^{2}(x^{\mu})\neq 0$. 

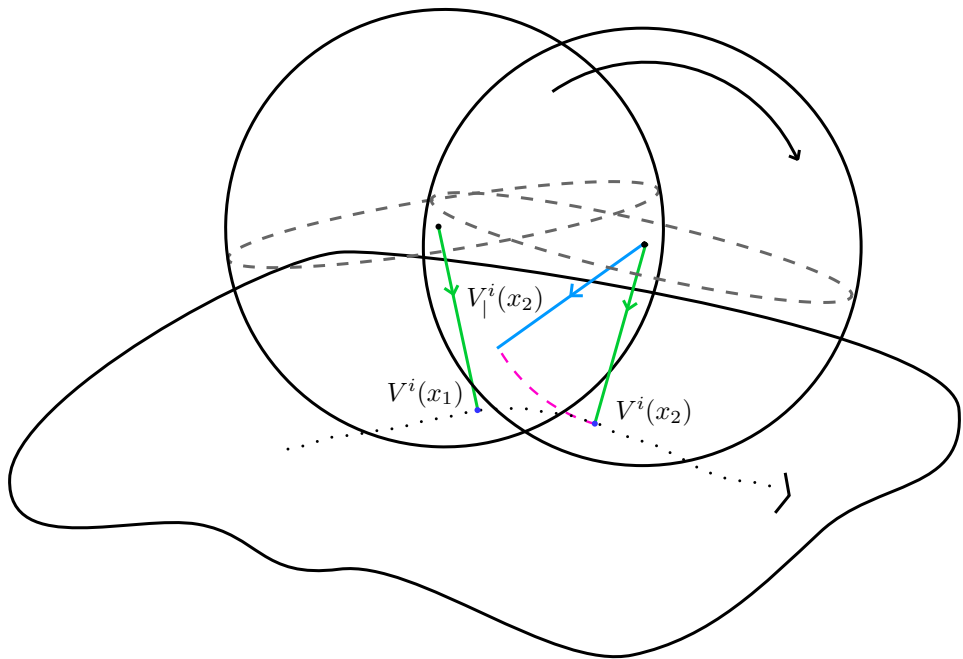
\begin{figure}[!h]
\centering
\begin{pspicture}(0,-4.5304008)(12.778019,4.367478)
\definecolor{color318}{rgb}{1.0,0.0,0.8}
\definecolor{color337}{rgb}{0.4,0.4,0.4}
\definecolor{color349}{rgb}{0.0,0.8,0.2}
\definecolor{color356}{rgb}{0.0,0.6,1.0}
\definecolor{color368}{rgb}{0.2,0.2,1.0}
\rput{122.863335}(13.677005,-5.75182){\psarc[linewidth=0.032,linecolor=color318,linestyle=dashed,dash=0.16cm 0.16cm](8.404438,0.8476607){2.1445203}{88.6461}{131.90227}}
\rput{-45.99371}(1.5889405,6.543079){\psarc[linewidth=0.04](8.502911,1.3995991){2.2}{69.9506}{170.73167}}
\psline[linewidth=0.04cm](10.512911,2.309599)(10.53291,2.429599)
\psline[linewidth=0.04cm](10.49291,2.309599)(10.39291,2.3495991)
\psbezier[linewidth=0.04](12.65291,-1.0186095)(12.547802,0.101329386)(5.4304876,1.1391683)(4.4329104,1.069599)(3.4353333,1.0000298)(0.0658209,-0.8830862)(0.03291045,-1.950401)(0.0,-3.0177157)(1.7129104,-2.4104009)(2.5329103,-2.550401)(3.3529105,-2.690401)(3.3729105,-3.2819734)(4.4129105,-3.1488345)(5.4529104,-3.0156953)(7.259678,-4.510401)(8.31291,-4.290401)(9.366143,-4.070401)(10.060142,-3.3511217)(10.81291,-2.6504009)(11.565679,-1.9496803)(12.7580185,-2.1385484)(12.65291,-1.0186095)
\usefont{T1}{ptm}{m}{n}
\rput(8.60291,-1.045401){$V^{i}(x_{2})$}
\usefont{T1}{ptm}{m}{n}
\rput(5.5629106,-0.80540097){$V^{i}(x_{1})$}
\rput{9.062611}(0.2988462,-0.8956818){\psellipse[linewidth=0.04,linecolor=color337,linestyle=dashed,dash=0.16cm 0.16cm,dimen=outer](5.800306,1.4375886)(2.9180002,0.37)}
\psdots[dotsize=0.06](8.472911,1.169599)
\usefont{T1}{ptm}{m}{n}
\rput(6.6229105,0.41459903){$V^{i}_|(x_{2})$}
\psdots[dotsize=0.06](8.49291,1.169599)
\psdots[dotsize=0.06](7.8129106,-1.2104009)
\pscustom[linewidth=0.04,linestyle=dotted,dotsep=0.16cm]
{
\newpath
\moveto(3.7329104,-1.550401)
\lineto(3.8429105,-1.530401)
\curveto(3.8979104,-1.520401)(4.0079103,-1.4904009)(4.0629106,-1.4704009)
\curveto(4.1179104,-1.450401)(4.2029104,-1.420401)(4.2329106,-1.410401)
\curveto(4.2629104,-1.400401)(4.3529105,-1.385401)(4.4129105,-1.380401)
\curveto(4.4729104,-1.3754009)(4.5729103,-1.3654009)(4.6129103,-1.3604009)
\curveto(4.65291,-1.3554009)(4.7329106,-1.3404009)(4.7729106,-1.330401)
\curveto(4.8129106,-1.320401)(4.9079103,-1.305401)(4.9629107,-1.300401)
\curveto(5.0179105,-1.295401)(5.1279106,-1.280401)(5.1829104,-1.270401)
\curveto(5.2379103,-1.260401)(5.3179107,-1.2404009)(5.3429103,-1.2304009)
\curveto(5.3679104,-1.2204009)(5.4379106,-1.200401)(5.4829106,-1.190401)
\curveto(5.52791,-1.180401)(5.6029105,-1.165401)(5.6329103,-1.160401)
\curveto(5.6629105,-1.155401)(5.7179103,-1.145401)(5.7429104,-1.140401)
\curveto(5.7679105,-1.135401)(5.8279104,-1.1254009)(5.8629103,-1.1204009)
\curveto(5.8979106,-1.1154009)(5.9679103,-1.1004009)(6.0029106,-1.0904009)
\curveto(6.0379105,-1.080401)(6.0979104,-1.065401)(6.1229105,-1.060401)
\curveto(6.1479106,-1.055401)(6.1979103,-1.045401)(6.2229104,-1.040401)
\curveto(6.2479105,-1.035401)(6.3029103,-1.030401)(6.3329105,-1.030401)
\curveto(6.3629103,-1.030401)(6.4379106,-1.025401)(6.4829106,-1.020401)
\curveto(6.52791,-1.015401)(6.6479106,-1.010401)(6.7229104,-1.010401)
\curveto(6.79791,-1.010401)(6.92291,-1.015401)(6.9729104,-1.020401)
\curveto(7.0229106,-1.025401)(7.0879107,-1.040401)(7.1029105,-1.050401)
\curveto(7.1179104,-1.060401)(7.1679106,-1.070401)(7.2029104,-1.070401)
\curveto(7.2379103,-1.070401)(7.3079104,-1.075401)(7.3429103,-1.080401)
\curveto(7.3779106,-1.0854009)(7.4429107,-1.0954009)(7.4729104,-1.1004009)
\curveto(7.5029106,-1.1054009)(7.54791,-1.1204009)(7.5629106,-1.130401)
\curveto(7.5779104,-1.140401)(7.6129103,-1.155401)(7.6329103,-1.160401)
\curveto(7.65291,-1.165401)(7.6979103,-1.170401)(7.7229104,-1.170401)
\curveto(7.7479105,-1.170401)(7.7929106,-1.180401)(7.8129106,-1.190401)
\curveto(7.8329105,-1.200401)(7.8729105,-1.2204009)(7.8929105,-1.2304009)
\curveto(7.9129105,-1.2404009)(7.9529104,-1.265401)(7.9729104,-1.280401)
\curveto(7.9929104,-1.295401)(8.1229105,-1.3454009)(8.23291,-1.380401)
\curveto(8.342911,-1.415401)(8.48791,-1.4604009)(8.52291,-1.4704009)
\curveto(8.557911,-1.4804009)(8.61791,-1.5004009)(8.64291,-1.510401)
\curveto(8.667911,-1.520401)(8.72791,-1.550401)(8.762911,-1.570401)
\curveto(8.797911,-1.5904009)(8.8729105,-1.630401)(8.91291,-1.650401)
\curveto(8.95291,-1.670401)(9.02291,-1.705401)(9.052911,-1.7204009)
\curveto(9.082911,-1.7354009)(9.14791,-1.775401)(9.182911,-1.800401)
\curveto(9.217911,-1.825401)(9.28291,-1.8554009)(9.31291,-1.8604009)
\curveto(9.342911,-1.8654009)(9.40291,-1.885401)(9.432911,-1.900401)
\curveto(9.462911,-1.915401)(9.52791,-1.935401)(9.56291,-1.940401)
\curveto(9.597911,-1.945401)(9.65791,-1.955401)(9.682911,-1.9604009)
\curveto(9.707911,-1.9654009)(9.752911,-1.9754009)(9.77291,-1.9804009)
\curveto(9.792911,-1.9854009)(9.847911,-1.9904009)(9.882911,-1.9904009)
\curveto(9.917911,-1.9904009)(9.99291,-2.000401)(10.03291,-2.010401)
\curveto(10.07291,-2.020401)(10.137911,-2.040401)(10.16291,-2.050401)
\curveto(10.18791,-2.060401)(10.222911,-2.085401)(10.23291,-2.100401)
\curveto(10.24291,-2.115401)(10.252911,-2.140401)(10.252911,-2.1704009)
}
\psline[linewidth=0.04](10.34343,-1.8664153)(10.396755,-2.1674976)(10.210434,-2.3992493)
\psline[linewidth=0.04cm,linecolor=color349](6.2529106,-1.010401)(5.7329106,1.4095991)
\psdots[dotsize=0.04](8.472911,1.189599)
\psdots[dotsize=0.04](8.472911,1.189599)
\psdots[dotsize=0.08](8.472911,1.169599)
\psdots[dotsize=0.08,linecolor=white](2.4529104,-0.83040094)
\psdots[dotsize=0.08,linecolor=white](2.4329104,-1.450401)
\psdots[dotsize=0.025999999,linecolor=white](1.7929105,0.10959904)
\psline[linewidth=0.04cm,linecolor=color356](8.472911,1.189599)(6.5129104,-0.21040095)
\psdots[dotsize=0.025999999,linecolor=white](3.0129104,-1.050401)
\psdots[dotsize=0.08](8.472911,1.169599)
\rput{-1.7559003}(-0.039849814,0.17876847){\pscircle[linewidth=0.04,dimen=outer](5.8129106,1.3895991){2.93}}
\psline[linewidth=0.04cm,linecolor=color349](8.49291,1.189599)(7.8129106,-1.2104009)
\pscircle[linewidth=0.04,dimen=outer](8.44291,1.1395991){2.93}
\psdots[dotsize=0.04](8.472911,1.169599)
\psdots[dotsize=0.08](8.472911,1.169599)
\rput{-12.908506}(-0.0392304,1.9124159){\psellipse[linewidth=0.04,linecolor=color337,linestyle=dashed,dash=0.16cm 0.16cm,dimen=outer](8.432911,1.1295991)(2.8880005,0.38)}
\psdots[dotsize=0.08,linecolor=color368](7.8129106,-1.2104009)
\psdots[dotsize=0.08,linecolor=color368](6.2529106,-1.030401)
\psdots[dotsize=0.08](5.7329106,1.4095991)
\psline[linewidth=0.04,linecolor=color349](5.8129106,0.589599)(5.9329104,0.50959903)(5.9929104,0.62959903)
\psline[linewidth=0.04,linecolor=color356](7.5129104,0.649599)(7.4929104,0.48959905)(7.65291,0.48959905)
\psline[linewidth=0.04,linecolor=color349](8.19291,0.42959905)(8.252911,0.30959904)(8.3729105,0.38959906)
\end{pspicture} 

\caption{This figure acts acts as an illustration of concepts in Cartan geometry for the case of two dimensional spatial geometries, where the `rolling' group is $SO(3)$. Here we imagine a sphere $X_{i}X^{i}=\ell^{2}$ (where $X^{i}$ are Cartesian coordinates in a three dimensional flat Euclidean space) being rolled from point $x_1$ to $x_2$ on the manifold. The contact vectors $V^i(x_1)$ and $V^i(x_2)$ at $x_1$ and $x_2$ respectively can be visualized as having their origin (black dots) in the center of the corresponding spheres and pointing towards the `point of contact' (the blue dots) between the sphere and the two-dimensional surface. The figure shows how the contact point $V^i(x_1)$ at $x_1$ is `rolled' to $x_2$ yielding $V_|^i(x_2)$ (light blue line). The distance between $x_1$ and $x_2$ is identified as the difference between the rolled $V^i_|(x_2)$ and the contact point $V^i(x_2)$ at $x_2$, i.e. $ds^2=dx^adx^bD_aV^iD_bV^j\delta_{ij}$.}
\label{cartangeometryfig}
\end{figure}

As clearly pointed out in \cite{Wise:2006sm}, this structure, which admits an elegant geometric interpretation, is that of Cartan geometry introduced by \'Elie Cartan in 1923 \cite{SharpeCartan}. To introduce this formulation, recall that $SO(2,3)/SO(1,4)$ symmetry could be broken down to SO(1,3) symmetry by an object $X^{A}$ satisfying $\eta_{AB}X^{A}X^{B}= \mp\ell^{2}(x^{\mu})$. Note however that there are infinite number of possible solutions to this equation. The solutions $X^{A}$ in fact correspond to coordinates describing a de-Sitter ($X^{2}=l^{2}$) or anti de-Sitter ($X^{2}=-l^{2}$) space of radius $\ell$ embedded in a five dimensional space with metric $\eta_{AB}$. Therefore we may reach the following interpretation of the field $V^{A}(x^{\mu})$: Consider at each point $x^{\mu}$ on a manifold an internal de-Sitter or anti de-Sitter space, which we denote as $X^{A}(x^{\mu})$. The field $V^{A}(x^{\mu})\in X^{A}(x^{\mu})$ represents a point in this space. A helpful visual of this field as a `point of 
contact' vector in a lower dimensional example is given in Figure \ref{cartangeometryfig}. 

The additional presence of a connection field $A_{\mu}^{\phantom{\mu}AB}$ allows one to parallel transport vectors such as $V^{A}$ as solutions to the `parallel transport' equation along a curve $x^{\mu}(\lambda)$:

\begin{eqnarray}
\label{pt}
\frac{dx^{\mu}}{d\lambda}D_{\mu}V^{A}=\frac{dx^{\mu}}{d\lambda}\left(\partial_{\mu}V^{A}+A^{AB}_{\mu}V_{B}\right)=0
\end{eqnarray}
Can familiar geometrical objects be recovered then from the pair $\{V^{A},A^{AB}\}$? Consider an internal space $X^{A}(x_{1})$ at a spacetime point $x_{1}$, with preferred vector $V^{A}(x_{1})$ . We may parallel transport all $X^{A}(x_{1})$ to a nearby second point $x_{2}$ according to the $SO(1,4)$/$SO(2,3)$ parallel transport equation
(\ref{pt}) applied to a vector $X^{A}$. At $x_{2}$, the $X^{A}$ will have changed by a transformation that preserves $X^{A}X_{A}=\mp \ell^{2}$ as well as the projection $\eta_{AB}X^{A}_{1}X^{B}_{2}$ between any two coordinate vectors $X^{A}_{1}$ and $X^{A}_{2}$. Therefore in parallel transporting the set $\{X^{A}\}$ from $x_{1}$ to $x_{2}$ we have in essence \emph{rolled} the internal anti-de Sitter/de Sitter space. Of course at $x_{2}$ we expect another preferred vector $V^{A}(x_{2})$. What information is contained in the comparison between the $V^{A}$ rolled from $x_{1}$ to $x_{2}$ (which we will call $V^{A}_{|}(x_{2})$) and $V^{A}(x_{2})$? Performing a Taylor expansion and recalling the definition of the covariant derivative in the parallel transport equation we have:
\begin{eqnarray}
V^{A}(x_{2}) &\simeq & V^{A}(x_{1}) + \partial_{\mu}V^{A}(x_{1})\delta x^{\mu} \\
V_{|}^{A}(x_{2}) &=& V^{A}(x_{1})- A^{AB}_{\mu}(x_{1})V_{B}(x_{1}) \delta x^{\mu}
\end{eqnarray}
Therefore in rolling from $x_{1}$ to $x_{2}$ one may identify the infinitesimal distance 
\begin{eqnarray}
ds^{2} &\equiv &\eta_{AB}(V^{A}(x_{2})-V_{|}^{A}(x_{2}))(V^{B}(x_{2})-V_{|}^{B}(x_{2}))\\
&=&\eta_{AB}D_{\mu}V^{A}(x_{1})D_{\nu}V^{B}(x_{1})\delta x^{\mu}\delta x^{\nu}\\
&\equiv & g_{\mu\nu}\delta x^{\mu}\delta x^{\nu}
\end{eqnarray}
If the radii of internal spaces are identical at different points then $V_{A}D_{\mu}V^{A}=0$ and $ds^{2}$ is simply a meaure of the distance being `traversed' on the internal de-Sitter/anti-de Sitter space in the process of rolling from $x_{1}$ to $x_{2}$. Much like a waywiser probing a two dimensional surface, this traversal gives information about physical distances on the surface itself. Therefore we can see that the identification of $DV^{I}$ with $e^{I}$ in the previous section is recovered. Furthermore, it may be shown under this assumption that the traditional objects of differential geometry (for instance Riemannian curvature, affine connection, torsion) can be recovered in the language of Cartan geometry \cite{Westman:2012xk}.

The possibility $V_{A}D_{\mu}V^{A}\neq 0$ represents a generalization of sorts; the quantity $ds^{2}$ as defined above additionally allows for a distance effect due to change of size of the internal space as it is rolled from from $x_{1}$ to $x_{2}$. By way of example, consider a situation where $V^{2}=\mp e^{2\phi(x^{\mu})}\ell^{2}$ where $\ell^{2}$ is a fixed scale. It follows that
\begin{eqnarray}\label{disf}
ds^{2} &=& e^{2\phi}\left(\eta_{IJ}e^{I}_{\mu}e^{J}_{\nu}\mp \ell^{2}\partial_{\mu}\phi\partial_{\nu}\phi\right)\delta x^{\mu}\delta x^{\nu}
\end{eqnarray}
We have argued that the field $V^{A}$ can be considered as a physical field in the description of gravitation. As such, it may be expected that its magnitude isn't constant over all of spacetime and it should be expected that the line element $ds^{2}$ should exhibit the disformal dependence upon gradients of $V^{2}$ apparent in (\ref{disf}).

\end{document}